\documentclass[LRconference]{IEEEtran}
\usepackage{cite}
\usepackage{amsmath,amssymb,amsfonts}
\usepackage{algorithmic}
\usepackage{graphicx}
\usepackage{textcomp}
\usepackage{xcolor}
\usepackage{url}
\usepackage{afterpage}
\usepackage{longtable}
\usepackage{booktabs}
\usepackage{tabularx}
\usepackage{geometry} 
\usepackage{caption}
\usepackage{float}
\usepackage{enumitem}
\usepackage{hyperref}

\def\BibTeX{{\rm B\kern-.05em{\sc i\kern-.025em b}\kern-.08em
    T\kern-.1667em\lower.7ex\hbox{E}\kern-.125emX}}
\begin{document}

\title{Scalability and Maintainability Challenges and Solutions in Machine Learning:SLR}

\author{\IEEEauthorblockN{Karthik Shivashankar}
\IEEEauthorblockA{\textit{Department of Informatics},
\textit{University of Oslo, Norway}\\
karths@ifi.uio.no}
\and

\IEEEauthorblockN{Ghadi S. Al Hajj}
\IEEEauthorblockA{\textit{Department of Informatics},
\textit{University of Oslo, Norway}\\
ghadia@uio.no}

\IEEEauthorblockN{Antonio Martini}
\IEEEauthorblockA{\textit{Department of Informatics},
\textit{University of Oslo, Norway}\\
antonima@ifi.uio.no}
}

\maketitle

\begin{abstract}
Background: The rapid advancement of Machine Learning (ML) across various domains has led to its widespread adoption in academia and industry. However, ML systems present unique maintainability and scalability challenges not encountered in conventional software projects. Addressing these challenges and possible solutions ensures long-term value and prevents ML performance decline.

Objective: This research aims to identify and consolidate the maintainability and scalability challenges and solutions at different stages of the ML workflow, discern the interdependencies between these stages, and investigate and identify potential tradeoffs and solutions to overcome these challenges.

Methodology: We conducted a systematic literature review, initially screening over 17,000 papers and selecting 124 papers to be included in this study.

Contributions: Our study presents (i) a catalogue of maintainability and scalability challenges and solutions in various stages of Data Engineering and Model Engineering workflows, as well as  difficulties in building ML systems or applications in the current ecosystem of framework and tools ; (ii)  Identified and consolidated 41 maintainability challenges and 13 Scalability challenges with some potential solutions, tools,  and recommendations  (iii) mapping of 13  interdependent  maintainability challenges  and a mapping 3 interdependent scalability challenges  impacting different stages of ML that influence the overall workflow is also synthesised; (IV) Synthesised a list of tradeoff and  strategies for striking a balance in achieving Maintainability and Scalability  and finally  (V) proposed solutions and insights to assist ML tool developers and researchers 

Conclusions: Through this study, practitioners and organisations can better understand the maintainability and scalability challenges and available solutions. Moreover, the identified tradeoffs and challenges can serve as a foundation for future research, further enhancing our comprehension of ML system maintainability and scalability in different stages of the ML workflow. This knowledge will empower them to circumvent potential pitfalls and facilitate the development of maintainable and scalable ML systems and applications. 

\end{abstract}

\section{Introduction}

Modern software applications depend heavily on Machine Learning (ML) systems to derive valuable insights from ever-expanding and evolving data sources. Numerous businesses have integrated ML into their service offerings and have realised its benefits. As ML adoption grows, new data and model engineering challenges have emerged [3,4,5,6].

Machine Learning systems are fundamentally anchored in the data they are trained on, which means that every nuance, pattern, and anomaly within the dataset can influence the behaviour and outcome of the model. The subtleties of a dataset, from its features to its biases, dictate how the model perceives and interprets information. As such, there exists a complex and intertwined relationship between the dataset's characteristics and the model's resulting performance. Consequently, these systems are prone to model staleness and training-serving skew, potentially diminishing their effectiveness without appropriate mitigation strategies [1]. Moreover, data dependency tends to be more costly than code dependency, making ML systems susceptible to technical debt and incurring significant ongoing maintenance costs compared to conventional software projects [2,7].

In machine learning, as models become increasingly sophisticated and datasets grow, two primary challenges arise scalability and maintainability [13].
Scalability is concerned with efficiently training ML models with growing datasets, managing storage requirements, and allocating computational resources. Large datasets can significantly prolong training times, particularly for Deep Learning (DL)  models that require extensive training to achieve desired performance levels. This prolonged training strains computational resources and can increase financial costs, mainly when utilising cloud-based infrastructures. Hence, scalable ML solutions often employ distributed training techniques, dividing computations across multiple machines or GPUs [13, 14]. Maintainability in the context of machine learning goes beyond just ensuring code readability or modular design. It directly addresses the evolving nature of data and the dynamic requirements of real-world applications. As data drifts and models degrade over time, a maintainable system allows for seamless updates and recalibrations, which ensures that models remain relevant and effective in their predictive capabilities. Moreover, by emphasising maintainability, teams can swiftly adapt to changes, whether integrating a new data source, tweaking model parameters, or handling unforeseen challenges, ensuring ML deployments' long-term viability and resilience [9, 10, 11].

The challenges and tradeoffs between maintainability and scalability need to be discussed because they are two crucial aspects of building an ML system and often have an inverse relationship. Complex ML systems with high computational demands and distributed components can be challenging to maintain and coordinate. Therefore, it is crucial to identify the balance between scalability and maintainability when designing systems that can handle growth while being effectively managed and updated. As an ML system grows in complexity, such as transitioning from a small experimental model to a  production-scale model, it may require more resources and infrastructure to handle the increased computational demands. However, managing and maintaining this distributed system becomes more challenging with scale. The more components there are, the more potential points of failure and the greater the complexity of coordinating and synchronising their operations [13].

For instance, when it comes to artefact management. As the number of ML models in the system increases, monitoring, updating, reproducing these models becomes more intricate. Manual monitoring and updates may suffice with a single model, but automation becomes necessary with a hundred models. Managing inferences, tracking versions, and ensuring reproducibility across multiple models require robust systems and practices to maintain clarity and efficiency. To illustrate the importance of considering these tradeoffs, let us take the example of a rapidly growing startup that provides ML services to enterprise customers. Initially, the startup serves one enterprise customer with a single ML model. However, as the customer base expands, the startup must deploy one or more models per customer or client, leading to managing thousands of models. Ensuring the scalability of resources and artefact management becomes crucial to meet the growing demands [13]. Failure to address these scalability challenges can decrease maintainability, making it difficult for different contributors (e.g., ML engineers, DevOps engineers, and  Data scientists) to collaborate effectively, identify issues, and implement solutions.

Therefore, it is essential to have a comprehensive understanding of challenges, solutions, and tradeoffs and to strike a balance between scalability and maintainability by designing systems that can handle growth while effectively managing and updating to build successful and sustainable ML systems.

In addition, ML systems often entangle signals, making it challenging to isolate improvements and leading to the "Changing Anything Changes Everything" (CACE) principle. Correction cascades can arise when improving one component decreases the system's accuracy, resulting in improvement deadlocks. Undeclared consumers and data dependencies further complicate maintenance, increasing the cost and difficulty of making changes. ML systems also exhibit unique feedback loops, both direct and hidden, that influence their behaviour over time [2]. 

Both scalability and maintainability challenges can impact the effectiveness and reliability of ML models, making it crucial for developers and researchers to successfully address these issues to implement machine learning systems in real-world applications. Therefore, organisations and practitioners must comprehend how to develop maintainable and Scalable ML systems and ensure long-term value delivery. The first step in addressing this issue is to understand the maintainability and scalability challenges faced when developing ML systems and how different stages in the workflow impact their quality. 

Regrettably, despite their significance, no systematic literature review, to our knowledge, has fully explored maintainability and scalability challenges,  solutions,  and tradeoffs and found a strategy to balance these tradeoffs when developing an ML system. Except for our previous SLR, which only covers Maintainability challenges and interdependencies between them [34], we will discuss later this  earlier work in the Related Works Section.

ML development usually comprises two primary workflows: Data Engineering and Model engineering. Data Engineering workflow in ML is concerned with designing and managing the infrastructure for acquiring, processing, and storing data. On the other hand, Model Engineering workflow in ML is concerned with developing and maintaining machine learning models by designing, training, evaluating, and deploying them for accurate predictions or decision-making [15]. Besides these two main workflows,  Building ML systems and applications in the Current Ecosystem is a multifaceted endeavour that extends beyond algorithm design and model training. At its core, orchestrating data and training pipelines demands a synergy between robust infrastructure and distributed systems. Infrastructure ensures that the underlying hardware and network resources can accommodate ML workloads' often intensive computational needs. Distributed systems, on the other hand, facilitate parallel processing and data handling, allowing for faster training and inferencing times. Equally crucial is the software development aspect, where best practices and principles are employed to guarantee the ML solution's sustainability, scalability, and maintainability. As ML becomes deeply integrated into various domains, the importance of harmonising these elements cannot be overstated, as it directly impacts the effectiveness and efficiency of the end-to-end ML application.

\subsection{Research Questions:
}
The research questions (RQs) listed below aim to investigate and understand the specific challenges and solutions related to maintainability and scalability in ML/DL systems and identify  the interdependence challenges  and tradeoffs with some potential strategies to overcome them.

\textbf{RQ1: What are the Data Engineering Maintainability challenges and solutions? } - This question focuses on the challenges faced in maintaining ML systems related to data engineering tasks, such as data preprocessing, feature engineering, and data storage, as well as the potential solutions to overcome those challenges. 

\textbf{RQ2: What are the Model Engineering Maintainability challenges and solutions?} - This question explores the challenges associated with maintaining ML models, including issues like model versioning, retraining, and incorporating updates or improvements, along with the corresponding solutions.

\textbf{RQ3: What are the current maintainability challenges and solutions when building ML systems and applications?} - This question aims to identify the overall maintainability challenges faced in the development and deployment of ML systems in the current ecosystem, considering both data engineering and model engineering and other related aspects like Software engineering, networks,  infrastructure, hardware and software considerations and cloud-native landscape.

\textbf{RQ4:What are the  Data Engineering Scalability challenges and solutions?} - This question delves into the challenges related to scaling ML systems in data engineering, such as handling large datasets, data distribution, parallel processing, and investigating potential solutions.

\textbf{RQ5:What are the  Model Engineering Scalability challenges and solutions? }- This question explores the challenges in scaling ML models, including issues like model size, training time, and inference speed, and seeks solutions to address these challenges.

\textbf{RQ6: What are the current scalability challenges and solutions when building ML systems and applications? }- This question investigates the broader scalability challenges in developing and deploying ML systems or applications in the current ecosystem, considering both data engineering and model engineering and other related aspects like software engineering, networks,  infrastructure, hardware and software considerations and cloud-native landscape.

\textbf{RQ7: What are the Interdependence challenges in maintainability?} - This question delves into how interdependence between various components or stages in ML system development can lead to challenges. It aims to explore the complex interactions where an issue in one ML stage affects another and subsequently impacts the system's overall maintainability.

\textbf{RQ8: What are the Interdependence challenges in scalability?} - This question delves into how interdependence between various components or stages in ML system development can lead to challenges. It aims to explore the complex interactions where an issue in one ML component affects another and subsequently impacts the system's overall scalability.

\textbf{RQ9: What are the tradeoffs between Scalability and Maintainability?} - This question aims to uncover the tradeoffs between scalability and maintainability in ML systems, where improving one aspect may come at the expense of the other.

\textbf{RQ10: What are the strategies to strike a  balance against the tradeoffs between Scalability and Maintainability?} - This question explores strategies and approaches to find a balance between scalability and maintainability in ML systems, enabling practitioners to make informed decisions when faced with these tradeoffs.

Overall, this research provides valuable insights into the challenges and solutions related to maintainability and scalability in ML systems, contributing to developing more robust and efficient machine learning applications.

\subsection{Contributions of the Study}

In this SLR, we contribute by
providing a comprehensive catalogue of maintainability and scalability challenges and solutions across various stages of Data Engineering, Model Engineering workflows and when Building ML systems or applications within the current framework and tool ecosystem;
Consolidated 41 maintainability and 13 scalability challenges and offered potential solutions and recommendations;
Synthesised 13 interdependent maintainability challenges and 3 interdependent scalability challenges that impact different stages of ML workflows;
Identified the tradeoffs between maintainability and scalability while exploring potential strategies for striking a balance against this tradeoff; and
We also offer valuable insights and implications to aid ML tool developers and researchers.

In summary, scalability and maintainability are vital concerns in ML. As datasets and models increase in size and complexity, training, storing, and operating them in production environments become more challenging. Nonetheless, various solutions can help address these challenges. By understanding these issues and solutions, practitioners and researchers can make better-informed decisions and tradeoffs when designing and building ML systems.

\section{Background}

\subsection{Maintainability of Software Systems and ML }

Maintainability is a crucial aspect of software engineering that refers to the ease with which a software system can be modified, updated, or debugged. It measures how well a system can be understood, changed, and tested and how well it can adapt to changing requirements and environments [9]. Maintainability is vital because software systems are often used for long periods, and they must be able to evolve and adapt as technology and business requirements change [10,11].

Maintainability is typically achieved by designing software systems that are modular, well-documented, and easy to understand. This includes using clear naming conventions, consistent coding style, and appropriate abstraction levels. Additionally, automated testing and continuous integration can be used to ensure that changes to the system do not break existing functionality [11, 13].

 Maintainability in the context of machine learning goes beyond just ensuring code readability or modular design. It directly addresses the evolving nature of data and the dynamic requirements of real-world applications. As data drifts and models degrade over time, a maintainable system allows for seamless updates and recalibrations. This ensures that representatives remain relevant and effective in their predictive capabilities. Moreover, by emphasising maintainability, teams can swiftly adapt to changes, whether integrating a new data source, tweaking model parameters, or handling unforeseen challenges, ensuring ML deployments' long-term viability and resilience [9, 10, 11]. Which requires continuous monitoring to detect issues such as data drift (when the distribution of data changes over time), errors, and technical debt [12,13].

In addition to monitoring, maintainability also focuses on managing multiple versions of the model and the data and configurations used to train them [8]. This enables easy rollback to prior versions in case of issues with the latest version and makes comparing performance between versions easier. By prioritising version management and monitoring, ML systems can achieve high maintainability and ensure they continue to perform effectively over time [12].

The reason for conducting an SLR on the maintainability of  ML software systems is to gain a comprehensive understanding of best practices, challenges, and strategies involved in effectively managing and improving the maintainability of ML systems over time. It helps researchers and practitioners identify gaps, trends, and opportunities for further advancements in this field.

\subsection{Scalability of Software Systems and ML 
}
Scalability is a crucial aspect of software engineering that refers to the ability of a software system to handle increasing amounts of work or traffic. A system can function efficiently and effectively as the workload or number of users increases. Scalability is important because many users often use software systems and must be able to handle a large amount of traffic and data [14].

Software systems' scalability is typically achieved using horizontal and vertical scaling patterns. Horizontal scaling refers to adding more machines to a system to handle more traffic. In contrast, vertical scaling refers to adding more resources to a single machine to handle more traffic. Additionally, caching and load balancing can be used to distribute traffic across multiple machines and improve scalability [14].

However, scalability in machine learning (ML) systems differs slightly from scalability in software systems. In software systems, scalability is focused on the command of the system to handle increasing numbers of users or traffic. In ML systems, scalability is concerned with efficiently training ML models with growing datasets, managing storage requirements, and allocating computational resources. Large datasets can significantly prolong training times, particularly for Deep Learning (DL)  models that require extensive training to achieve desired performance levels. This prolonged training strains computational resources and can increase financial costs, mainly when utilising cloud-based infrastructures. Hence, scalable ML solutions often employ distributed training techniques, dividing computations across multiple machines or GPUs [13, 14]. Additionally, cloud services such as Amazon Web Services, Microsoft Azure, and Google Cloud can be used to provide access to large amounts of computational resources at a reasonable cost. Nevertheless, during the ML inference or serving, both the scalability of ML systems and other Software Systems convergence focuses on the system's ability to handle increasing numbers of users or traffic [ 13,14].

The reason for conducting an SLR on the scalability of software systems and ML models is to understand how these systems can effectively handle large-scale datasets and DNN training. It helps identify design patterns, techniques, and technologies used to achieve scalability and ensure efficient performance as the system grows in size and complexity. Additionally, it explores specific considerations for scalability in ML systems, such as handling large datasets and computational requirements, parallel processing, distributed computing, model compression, and leveraging cloud services for cost-effective scalability.

\subsection{Data and Model Engineering Workflow  in ML }

Data engineering refers to the process of acquiring, preparing, and managing data for use in ML models, which is a critical aspect of ML development, as the quality of data directly affects the accuracy of the models built. Data engineering pipelines involve a series of operations on data from various sources, aiming to create high-quality training and testing datasets. The process typically includes the following stages [15]:(i) Data acquisition and exploration: Gathering data from different sources and exploring it to understand its structure, patterns, and potential issues. (ii)Data processing: Cleaning, transforming, and preprocessing the data to make it suitable for ML models. (iii)Data validation and management: Ensuring the data meets quality standards and managing it effectively throughout the ML workflow [15].

Model engineering, on the other hand, focuses on building, optimising, and deploying ML models based on processed data. This process involves several operations, such as:(i) Model training: Teaching the ML algorithm to recognise patterns in the data. (ii) Hyper-Parameter Optimization (HPO): Fine-tuning the model's parameters to achieve optimal performance.(iii)Model governance: Establishing guidelines and best practices for model development and use. (iv)Model monitoring: Continuously track the model's performance to ensure it remains accurate and relevant.(v)Model testing: Assessing the model's performance using various evaluation metrics.(vi) Model drift: Monitoring changes in the data or the model's performance over time and adjusting the model accordingly.(vii)Model deployment: Integrating the trained model into production systems for real-world use [15].

Data engineering focuses on acquiring, processing, and managing high-quality data, while model engineering involves building, optimising, and deploying ML models. Both processes are crucial to the success of ML projects, as they ensure that the models built can accurately and effectively solve the tasks they are designed for.

The coverage of maintainability and scalability challenges and solutions in the SLR is necessary to address the specific issues related to ML data engineering and model engineering. Understanding these challenges and solutions helps researchers and practitioners identify effective strategies, techniques, and best practices for managing the maintainability and scalability aspects of data and model engineering workflows. By exploring the existing literature, valuable knowledge and recommendations can be gathered to enhance the overall effectiveness and efficiency of data and model engineering workflows in ML projects.

\section{Related Works }

Limited research has been conducted to delve into the maintainability and scalability challenges and solutions of Machine Learning (ML) systems. To the best of our knowledge, we could only find our previous work, as detailed in the paper [34], exclusively addressed the maintainability challenges, their solutions, and the interdependencies of these challenges in maintainability. This literature review expands upon our previous work by introducing seven additional research questions (RQs). This SLR encompasses maintainability and delves into scalability challenges and solutions. Moreover, we analyse the interdependencies, tradeoffs, and potential strategies to over these tradeoffs. 

Nascimento et al. [31]  focused on understanding the development process of ML systems and identifying the challenges developers face. They proposed checklists to support developers in essential development tasks. While this study provides valuable insights into the challenges of ML system development, it must explicitly address maintainability and scalability challenges.

Nahar et al. [32] explored the collaboration challenges teams face building ML systems. It emphasised the importance of effective communication, documentation, engineering practices, and process alignment. While collaboration challenges are essential for maintainability and scalability, this study does not explicitly delve into those aspects.

Serban et al. [33] focused on adopting software engineering best practices for ML applications. They identified 29 engineering best practices and assessed their adoption and perceived effects. 

Shivashankar and  Martini [34] specifically investigated the maintainability challenges in different stages of the ML workflow and identified and synthesised maintainability challenges and their interdependencies, providing insights for building maintainable ML systems. While maintainability challenges are central to this study, scalability challenges and solutions, tradeoffs and finding potential solutions against these tradeoffs are not explicitly addressed in our previous study.

Kolltveit and  Li [35] systematically reviewed techniques, tools, and infrastructures required for operationalising ML models. Although operationalisation is not directly synonymous with maintainability and scalability, it contributes to ML systems' long-term value and performance.

Liu et al. [36] focused on the reproducibility and replicability of DL studies in SE tasks. While reproducibility and replicability are essential for maintaining and scaling ML systems, this study must address the challenges and solutions associated with maintainability and scalability.

Giray [37] conducted a systematic literature review to analyse the state of SE research for engineering ML systems. The study identified the non-deterministic nature of ML systems and the need for more mature tools and techniques across various SE aspects. While this study provides insights into the overall challenges of ML system engineering, the specific focus on maintainability and scalability is not explicitly emphasised.

Wang et al. [21] comprehensively reviewed ML/DL-related SE papers. It examined the complexity of applying ML/DL solutions to SE problems and highlighted issues concerning reproducibility and replicability. While this study touches on maintainability and scalability challenges, it needs to analyse these aspects in-depth.

In this SLR, we provide a catalogue of challenges and solutions for scalability and maintainability, map interdependencies in various stages in ML workflow, discuss tradeoffs and potential solutions, and discuss insights and recommendations. Our study aims to address these gaps in understanding and provide a foundation for further research and development of maintainable and scalable ML systems and applications.

\section{Research Methodology}

The research methodology adopted in this study is a Systematic Literature Review (SLR), which follows a rigorous, well-defined, and transparent process to identify, evaluate, and synthesise relevant literature in a reproducible manner. This section provides an in-depth description of the step-by-step approach employed in conducting the SLR [16].

An SLR is a type of study that collects and analyses the results from multiple primary studies. An SLR aims to provide an overview of a specific field for researchers and practitioners and identify gaps in the existing literature, according to Wohhlin et al. [27]. 

For this review, we followed established guidelines for conducting an SLR, as described by Kitchenham. The primary purpose of conducting a systematic review is to bring together and analyse existing work. To ensure integrity and high research quality, we followed a  search strategy that can be evaluated. The essential advantage of using this method is that it provides robust and transferable evidence, allowing for further examination of sources of variation [16].

To ensure that our study was rigorous and could be replicated, we adhered to a review protocol consisting of multiple phases:

(a) We clearly defined the research questions that guided our study.
(b) We established a well-defined search process to identify relevant literature.
(c) We determined specific criteria for including or excluding studies from our review.
(d) We used established and solid methods for assessing the quality of the included studies.
(e) We outlined a systematic approach for collecting the necessary data.
(f) We analyse and synthesise the collected data from our RQ.

By following this comprehensive review protocol, we aimed to maintain high rigour and ensure that others could replicate our study.

Our SLR study was conducted as depicted in Fig. 1, which provides an overview of the process. We started by searching the database using a predefined set of search terms and queries to obtain papers that address our Research questions. The papers were then iteratively evaluated based on inclusion and exclusion criteria and Quality assessment to ensure that only the most relevant and high-quality papers were included in our review. 

\subsection{Step 1: Search Strategy}

We employed a comprehensive search strategy to identify relevant literature from various databases and digital libraries to conduct our SLR. The selected databases for our search included IEEE Xplore, ACM Digital Library, Web of Science, Google Scholar, and Scopus. These databases were chosen due to their widespread usage and relevance in computer science, software engineering, and machine learning.

To ensure the identification of the most pertinent literature, we developed a well-crafted search query that encompassed keywords present in the titles, abstracts, and index terms of the papers. The search query utilised a combination of primary keywords derived from our research questions, alternative spellings, synonyms, and additional keywords obtained from pilot searches and related papers. This approach aimed to maximise the coverage of relevant literature across different databases.

The generic query used for our SLR was as follows:

\begin{itemize}
\item \textbf{TITLE-ABS-KEY} (machine \textbf{AND} learning \textbf{AND} software \textbf{AND} (ml \textbf{OR} ai \textbf{OR} dl \textbf{OR} neural \textbf{OR} intelligence \textbf{OR} learning) \textbf{AND} (adapt* \textbf{OR} maintain* \textbf{OR} scal* \textbf{OR} exten* \textbf{OR} evol* \textbf{OR} flex*) \textbf{AND} (system \textbf{OR} architect* \textbf{OR} design \textbf{OR} build \textbf{OR} application \textbf{OR} engineering \textbf{OR} test) \textbf{AND} (data \textbf{OR} algorithm \textbf{OR} debt \textbf{OR} pattern \textbf{OR} code))
\end{itemize}

This search query was designed to capture the essential terms related to our research topic, ensuring that the retrieved papers would address machine learning, software, system architecture, and associated concepts such as adaptation, maintainability, scalability, evolution, and flexibility. A more detailed search query for each digital library is shown in Table I. 

To focus our search on recent and relevant publications, we limited the search timeframe to articles published between January 1, 2014, and February 15, 2023. This period was deemed crucial for studying the maintainability and scalability of machine learning systems, as it coincided with the emergence and development of numerous machine learning projects and libraries [26]. Due to platform limitations, our search was restricted to the first 1000 papers from Google Scholar and the first 2000 papers from Scopus, sorted by relevance.

To summarise, we comprehensively searched multiple databases and digital libraries using a carefully crafted search query. The aim was to identify relevant literature published within a specific timeframe about machine learning, software, and associated topics. The search process was designed to maximise the coverage of relevant publications and ensure the inclusiveness of significant research in our SLR.

\begin{table*}[htbp]
  \centering
  \begin{tabular}{p{2cm} p{8cm} p{3cm} p{3cm}}
    \toprule
   \textbf{Database} &\textbf{Search query} &\textbf{Filter} &\textbf{No of Results obtained} \\
    \midrule

IEEE Xplore &("All Metadata": Machine learning AND Software AND (ML OR AI OR DL OR Neural OR intelligence OR learning) AND ( scal* OR exten* OR adapt* OR maintain* OR evol* OR flex*) AND (system OR architect* OR design OR build OR application OR engineering OR test) AND (data OR algorithm OR debt OR pattern OR code)) &2014 -2023 Search in (title , keyword and  abstract ) &4819 Results 
\\

Scopus &TITLE-ABS-KEY ( machine AND learning AND software AND ( ml OR ai OR dl OR neural OR intelligence OR learning ) AND ( scal* OR exten* OR adapt* OR maintain* OR evol* OR flex* ) AND ( system OR architect* OR design OR build OR application OR engineering OR test ) AND ( data OR algorithm OR debt OR pattern OR code ) ) &Year GT 2013 Langage English ( title , abstract and keyword) & 7285 results Only the first 2000 results sorted by relevance was available for review because of limitation in Scopus
\\

Web of Science &TS=(Machine learning AND Software AND (ML OR AI OR DL OR Neural OR intelligence OR learning) AND ( scal* OR exten* OR adapt* OR maintain* OR evol* OR flex*) AND (system OR architect* OR design OR build OR application OR engineering OR test) AND (data OR algorithm OR debt OR pattern OR code)) &2014-2023 English (Title, abstract, author keywords and keyword plus) &2588 Results 
\\

ACM &Machine learning AND Software AND (ML OR AI OR DL OR Neural OR intelligence OR learning) AND ( scal* OR exten* OR adapt* OR maintain* OR evol* OR flex*) AND (system OR architect* OR design OR build OR application OR engineering OR test) AND (data OR algorithm OR debt OR pattern OR code) &2014 - 2023 (Abstract only) &1053 Results 
\\

Google Scholar &design system engineering architect* application algorithm pattern* code data build test scal* OR maintain* OR adapt* OR evol* OR exten* OR flex* OR debt OR AI OR ML OR DL OR Neural OR intelligence OR learning "machine learning software" &2014 -2023 Search anywhere in the article &2191 Results Only the first 1000 results sorted by relevance was available for review because of limitation in Google Scholar. 
\\
\bottomrule
  
  \end{tabular}
  \\
  \caption{Search query used for different databases and other metadata}
  \label{tab:example1113}
\end{table*}

\subsection{Step 2: Snowballing}

To enhance the comprehensiveness of our Systematic Literature Review (SLR), we supplemented our initial search with the snowballing method. This iterative process, as outlined by Wohlin [29], comprised two approaches: backwards and forward snowballing.

Backwards snowballing involved scrutinising the reference lists of selected publications to identify relevant papers that our initial search might have overlooked [29]. Conversely, forward snowballing focused on locating studies that cited our selected publications, enabling us to discover more recent research built upon the initially identified works [29].

Both snowballing techniques adhered to the same selection criteria employed in our initial search phase. This process entailed evaluating papers based on their titles, abstracts, and full texts, ensuring alignment with our Research Question (RQ) and predefined inclusion criteria [28].

The snowballing process began with the 145 articles retained after applying inclusion and exclusion criteria upon analysing their full texts, as shown in Figure 1. These articles served as the foundation for identifying additional related literature through their references [28, 29].

This complementary snowballing step broadened our review's scope, ensuring the inclusion of pertinent literature that might have been missed in the initial search. Through this process, we incorporated an additional 20 research articles into our study, further enriching our analysis [28, 29].

\subsection{Step 3: Selection Criteria}

To ensure the quality and relevance of the papers included in our SLR, we applied a set of inclusion and exclusion criteria from Table II in three iterations, as shown in Fig. 1, following the guideline from  Kitchenham [16].

 In machine learning and Software Engineering (SE), our search queries produced extensive search results, reflecting our aim to prioritise inclusivity and capture as many relevant papers as possible. When studying maintainability and scalability in ML and SE, one often needs help without a standardised set of clearly defined terms [20]. This issue becomes more pronounced when conducting research. S. Wang et al. [21], in their SLR concerning software engineering and machine learning, also faced a significant number of search results during their search process.

 In the first stage, we analysed the paper's title and abstract to filter out irrelevant papers which do not meet our criteria, as shown in Table II. In the second and third stages, we conducted a more thorough analysis of the paper's full text to make a final decision on its inclusion or exclusion criteria. This study did not include any paper that satisfied the exclusion criteria and was not included in the quality assessment stage, as shown in Fig 1 below.

\begin{table}[htbp]
\centering
\caption{Inclusion and Exclusion Criteria}
\label{tab:criteria}
\begin{tabular}{@{}lp{0.7\columnwidth}@{}}
\hline
\textbf{Criteria} & \textbf{Description} \\
\hline
\textbf{Inclusion } & \\
\textbf{IC1} & The paper discusses Data Engineering Challenges in both Maintainability and Scalability Contexts. \\
\textbf{IC2} & The paper discusses Model Engineering Challenges in both Maintainability and Scalability Contexts. \\
\textbf{IC3} & The paper discusses Data Engineering Solutions in both Maintainability and Scalability Contexts. \\
\textbf{IC4} & The paper discusses Model Engineering Solutions in both Maintainability and Scalability Contexts. \\
\textbf{IC5} & The paper discusses Current Challenges and Solutions when Building an ML System. \\
\hline
\textbf{Exclusion} & \\
\textbf{EC1} & Non-English papers. \\
\textbf{EC2} & Publication for which the full text is not available. \\
\textbf{EC3} & Grey literature (Not Peer reviewed). \\
\textbf{EC4} & Duplicate papers and shorter versions of already included publications. \\
\hline
\end{tabular}
\end{table}

\begin{figure}[htbp]
\centerline{\includegraphics[scale=0.75]{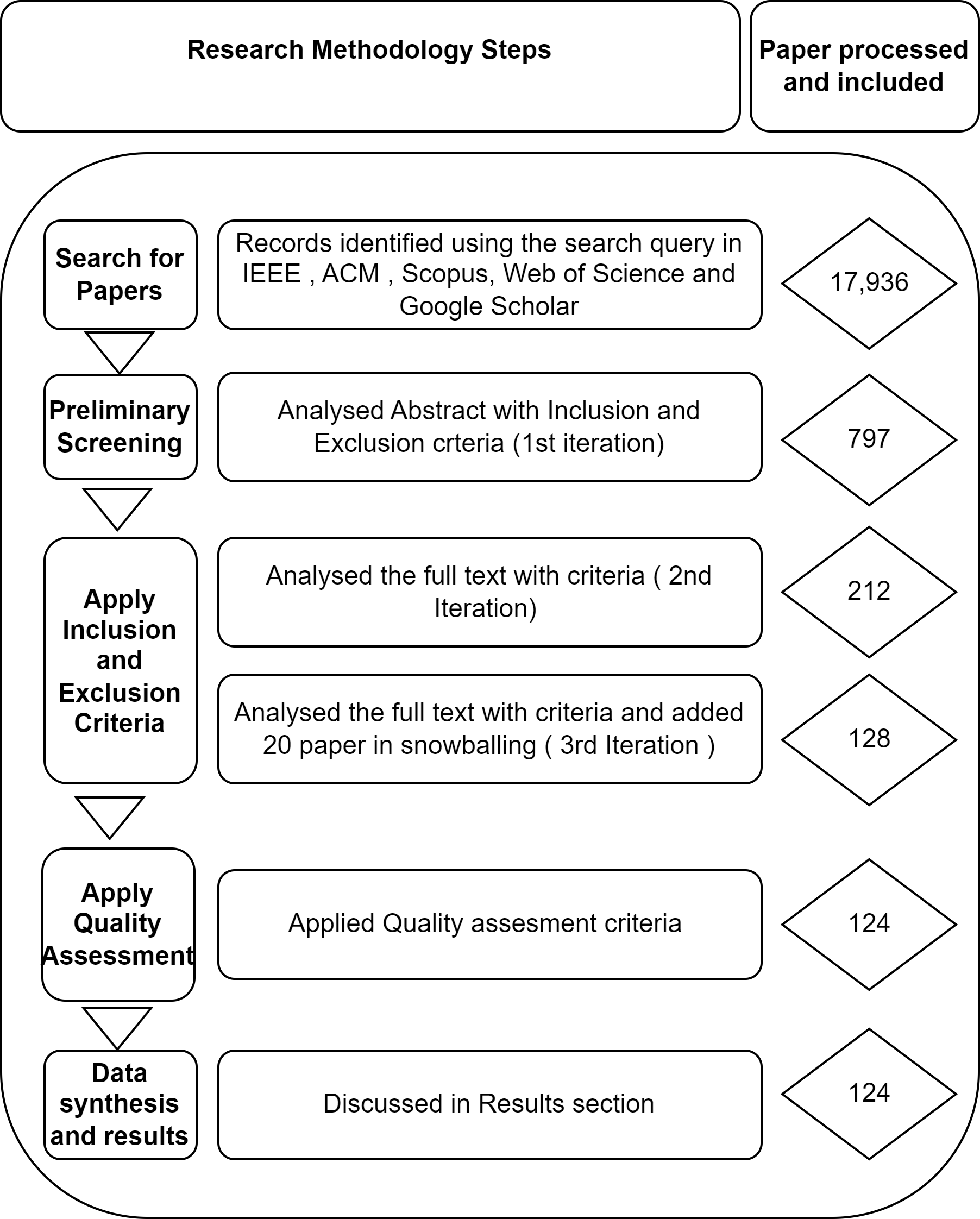}}
\caption{Systematic Literature Review Process}
\label{fig}
\end{figure}

\subsection{Step 4: Quality Assessment}
We performed a Quality Assessment (QA) as described in Table IV. 

Two independent reviewers evaluated around 128 papers and performed a Quality Assessment using a Table IV checklist adapted from Dyba and Dingsøyr  [22]. In addition to adapting the Quality Checklist, we further enhanced the quality assessment selection by performing this  Checklist [22] for each Inclusion criterion (IC1 - IC5).   We used a 5-point Likert scale for agreement for this assessment, where:  1 - Strongly disagree; 2 - Disagree; 3  - Neither agree nor disagree; 4 - Agree and 5 - Strongly agree for each given Inclusion criteria ( IC1 - IC5). We then normalised and converted this 5-point Likert Scale using a MinMax scaler to obtain a 2-Point dichotomous ("yes" or "no") scale adopted by Dyba and Dingsøyr  [22].For a paper to be included, two independent reviewers had to agree on at least one of the Inclusion criteria with a dichotomous response of "yes". In case of disagreements, the third reviewer evaluated the record independently, and an agreement based on the majority was reached [16, 20].  
We also employed Cohen Kappa inter-rater agreement statistics from this quality assessment response in a dichotomous ("yes" or "no") scale. We got the following agreement score for each inclusion criterion, as shown in Table III.  

\begin{table}[htbp]
\centering
\caption{Inter-Rater Agreement Scores for Inclusion Criteria}
\begin{tabular}{ll}
\hline
\textbf{Inclusion Criterion} & \textbf{Agreement Score} \\
\hline
IC1 & 0.7653 \\
IC2 & 0.7853 \\
IC3 & 0.8230 \\
IC4 & 0.7595 \\
IC5 & 0.8388 \\
\hline
\end{tabular}
\end{table}
The kappa statistic ranges from -1 to 1, with 1 indicating perfect agreement, 0 indicating random agreement, and -1 indicating perfect disagreement. Landis and Koch  [23] provide guidelines for interpreting kappa values: 0.0 to 0.2 suggests slight agreement, 0.21 to 0.40 suggests fair agreement, 0.41 to 0.60 suggests moderate agreement, 0.61 to 0.80 suggests substantial agreement, and 0.81 to 1.0 suggests almost perfect or perfect agreement. Krippendorff [24]  offers a more cautious interpretation, tentative conclusions can be made for values between 0.67 and 0.80, and definite conclusions can be made for values above 0.80. However, researchers often retain kappa coefficients below these cutoffs, as acceptable estimates of interrater reliability depend on study methods and the research question [25]. The agreement in our case is more the 0.75 for all the Inclusion criteria, which corresponds to the substantial agreement between the raters, as shown in Table III.

\begin{table}[htbp]
  \centering
  \begin{tabular}{p{8cm}}
    \toprule
   \textbf{Quality Assessment} \\
    \midrule
\\
\textit{Research Design}
\\

QA1. Is the research objective sufficiently articulated and adequately justified?
\\
QA2. Is the context in which the research was conducted clearly stated, including the industry, project setting, utilised products, participants, or observational units?
\\
\textit{Data Collection}
\\

QA3. Was the data collection carried out comprehensively? For instance, is there a discussion of the data collection procedures and how the research setting might have influenced the collected data?
\\
\textit{Data Analysis}
\\

QA4. Are the approach and formulation of the analysis effectively conveyed? For example, is there a justification and description of the used analysis method/tool/package?

QA5. Have alternative explanations and confounding factors been considered and discussed in the analysis?
\\
\textit{Findings and Conclusion}
\\

QA6. Do the findings and conclusions possess credibility? For instance, is the study's methodology sufficiently explained to instil confidence in the findings? Are the findings clearly stated, supported by the data, and consistent with existing knowledge and experience?

QA7. Have the limitations and credibility of the study been adequately discussed? \\

\bottomrule
  
  \end{tabular}
  \\
  \caption{Quality assessment criteria [22]}
  \label{tab:example1113}
\end{table}

\subsection{Step 5: Data Extraction and Analysis}

Once we selected relevant literature, we examined each paper thoroughly and extracted data per our RQ. After careful reading, we assigned one or more research questions to each paper based on its content. To systematically identify recurring concepts, themes, and patterns in the literature, we employed the open coding technique using NVIVO tools, with our research questions serving as the primary starting point. Kitchenham [16] also recommends using graphical user interface tools to code and visualise qualitative data analysis.

In qualitative research, a code refers to a word or phrase that succinctly summarises or captures the essence of a particular portion of data. Coding involves the analytical process of categorising data. In NVivo, coding entails gathering related material into a Node container. By opening a node, one can access all the references in the project that are coded to that specific node.

These nodes serve as codes representing themes or topics found within the data. Relationships, on the other hand, depict connections between various project items. Regarding coding approaches, there are two main methods: deductive and inductive. Deductive coding involves using a pre-established coding scheme, often derived from  RQs identified in a literature review. Inductive coding, conversely, entails generating codes while examining the collected data [30].

In our case, we started with a research question (RQ) as our foundation and employed both deductive and inductive coding approaches. This allowed us to extract emergent topics and relationships within the scope of our research question. The coding process played a crucial role in meaningfully organising and categorising the extracted data.

To illustrate how this coding step was carried out, let us consider an example quotation from a paper. The quotation states, "Data leakage: a splitting strategy may risk data leakage, i.e., leak information in the validation data that should not be available for model training into the training data, which may introduce bias and result in misleading evaluation results. This necessitates the need for an iterative trial and error approach to find the best splitting strategy" [LR6]. This quotation was coded to RQ1 under the sub-topic Challenges in "Data preprocessing."

Throughout the analysis, we paid explicit attention to the interdependence between Maintainability and Scalability challenges and solutions and their tradeoffs and potential strategies at different stages of the machine learning workflow and when building the machine learning system in the current ecosystem. The authors consistently engaged in discussions regarding emerging findings to ensure code consistency and maintain high levels of abstraction.

\subsection{Step 6: Data Synthesis}

In the data synthesis section, we aimed to analyse the extracted data and identify the interdependencies and influences among the different stages of the machine learning (ML) development process, specifically focusing on maintainability and scalability challenges.

To begin this synthesis step, we examined the data gathered from the literature review and coding process, which included information related to the challenges encountered at each stage of the ML workflow. By analysing the data, we aimed to understand how these stages are interconnected and how they impact the maintainability and scalability of the ML system throughout its development.

Using the identified relationships and dependencies, we created a mapping diagram to represent these connections visually. This diagram allowed us to visualise the interactions between the various stages of the ML workflow and understand how changes in one stage can affect the maintainability of the entire system.

Furthermore, we highlighted the tradeoffs and balance required between maintainability and scalability in the ML system or application during the synthesis process. By examining the data and considering the relationships between different stages, we identified instances where prioritising one aspect (e.g., maintainability) may result in tradeoffs with the other (e.g., scalability). Understanding tradeoffs and balance is crucial for making informed decisions during the ML development process.

Finally,   we presented the synthesised findings, insights, and implications for ML tool developers and researchers. This final step involved distilling the key takeaways from our analysis and offering recommendations or suggestions based on the identified challenges, interdependencies, and tradeoffs. These insights are valuable information for developing ML tools and guiding future research.

To ensure transparency and reproducibility, we will make all study artefacts publicly available at Zenodo Link, \href{https://zenodo.org/record/8024834}{Replication package}

Throughout this process, we adhered closely to the guidelines suggested for conducting SLRs by Kitchenham et al. [16 ]. We employed a similar approach to the Research design from H. Edison et al. [20] and Dyba and Dingsøyr  [22].

\subsection{Threats to Validity 
}
This section outlines four potential challenges to the validity of this systematic review and presents our strategies for addressing each of them. This section is adapted from H. Edison et al. [20]. 

\subsubsection{ Bias in Publication}
Publication bias arises when research with positive outcomes is more likely to be published than those with negative results [16], [17]. This may occur when method developers are vested in reporting the method or when organisations studied wish to downplay negative aspects. Reviews like ours are always vulnerable to such bias. In this study, we consider this risk to be significant. To minimise it, we performed a quality assessment and differentiated between robust empirical research and anecdotal reports. We opted not to include grey literature, such as works-in-progress, technical reports, blogs, and unpublished articles.

\subsubsection{Identifying Primary Studies}

In order to minimise the risk of overlooking or excluding relevant papers, we aimed to gather as many peer-reviewed articles as possible related to the maintainability and scalability of the ML system and application. We also employed a snowballing search during the full-text review by identifying relevant papers in the reference lists of each article [18]. It is impossible to entirely eliminate the risk of missing relevant articles. Inconsistent or varying terminology in ML research may have influenced identifying primary studies. We did not optimise the search string for maximum precision to be as inclusive as possible. Precision refers to the proportion of relevant items retrieved to all items retrieved [19]. 

\subsubsection{Consistency  in Selecting Primary Studies and Data Extraction
}

We formulated inclusion and exclusion criteria and assessed the level of agreement among reviewers. The first iteration focused on titles and abstracts. Reviewers' assessment results were compared, and any discrepancies were addressed during a meeting. A similar approach was followed during the second iteration of the full-text review. One ambiguity that emerged during the pilots related to defining Maintainability and Scalability in the ML context, as some studies did not explicitly report these contexts. Definitions were formally established and agreed upon. The iteration exercise was repeated, demonstrating that the issue was sufficiently resolved.

All the iteration was also conducted for the quality assessment checklist. Like the selection process, two independent reviewers evaluated the quality of primary studies. This quality assessment iteration involved around  128  papers to verify the agreement level among reviewers. Following the iteration, a meeting was held to assess the checklist and develop strategies for addressing vagueness or missing details in specific papers. Any differences in reviewers' assessments were also discussed and resolved with input from all reviewers.

\subsubsection{Reliability of Extracted Data
}

We discovered that some studies needed more precise information regarding a clear connection between SE and ML maintainability. In these instances, we had to rely on the researcher's judgment. For example, some papers could have specified clearly how different or unique ML maintainability and scalability are different from the SE perspective. In cases where the method was not explicitly stated, we used the study context and practices reported in related studies to classify the paper. Consequently, some of the extracted findings may be partially inaccurate. To mitigate this, each reviewer refined and verified the extracted data during the Quality Assessment stage.

\section{Results and synthesis}

\subsection{Publication Types, Sources, and Venues}
This section of the SLR provides essential insights into the rationale behind selecting the timeframe for this review and the distribution of selected papers.

Fig. 2 demonstrates a significant increase in the number of papers published each year from 2014 to 2023. The selection of the timeframe for the review, starting from 2014, is based on the observation that ML/DL tools and libraries started to merge around that time. Initially, fewer papers were available due to the immaturity and low adoption of ML libraries and tools. The prominence of deep learning only began after the breakthroughs in GPU technology and the AlexNet model in the ImageNet competition. As the field evolved and researchers validated the feasibility and effectiveness of new approaches, the interest and number of relevant papers increased, particularly in recent years.
\begin{figure}[H]
    \centering
    \includegraphics[width=1\linewidth]{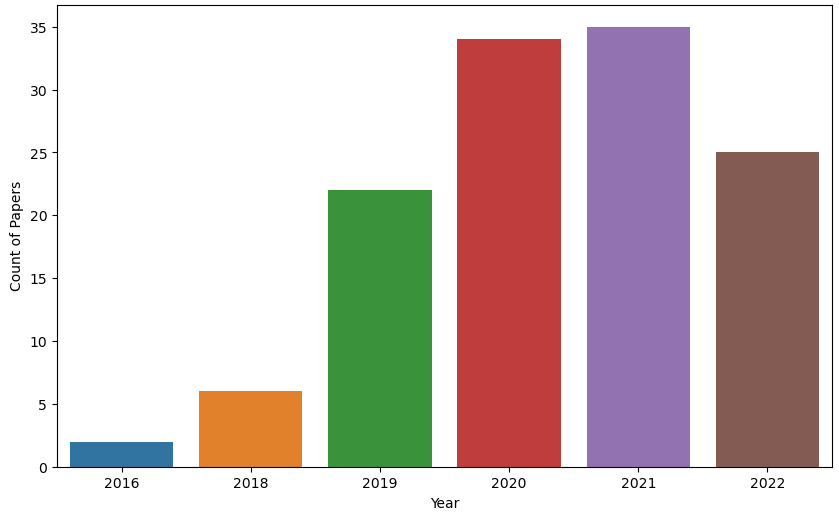}
    \caption{Number of Papers by Years}
    \label{fig:enter-label}
\end{figure}
Fig. 3 reveals the distribution of the selected paper; around 65 \% of the paper selected are from conferences and 30.9 \% of the paper are from Journal papers, and the remaining percentage is from the workshop and special papers. This distribution indicates a good balance of high-quality, peer-reviewed research articles.

\begin{figure}[H]
    \centering
    \includegraphics[width=1\linewidth]{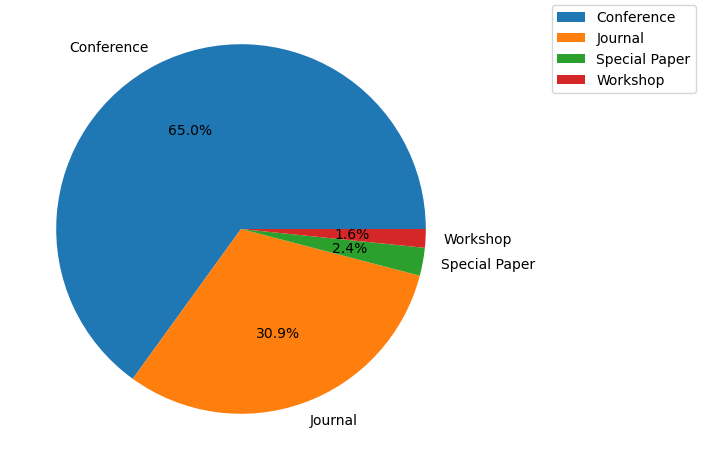}
    \caption{Distribution of Publication Type}
    \label{fig:enter-label}
\end{figure}

The top 15 publications are listed below in Table V, highlighting the key venues where the research has been published.

\begin{table}[H]
  \centering
 \caption{Top 15  Journal and Conference publication }
  \label{tab:metadata}
  \begin{tabular}{llr}
    \hline
    \textbf{Journal/Conference} & \textbf{Type} & \textbf{Count} \\
    \hline
    ICSE & Conference & 8 \\
    ACM Computing Surveys & Journal & 3 \\
    ASE & Conference & 3 \\
    SEAA & Conference & 3 \\
    SIGMOD & Conference & 2 \\
    ISSREW & Conference & 2 \\
    Future Generation Computer Systems & Journal & 2 \\
    Journal of Machine Learning Research & Journal & 2 \\
    KDD & Conference & 2 \\
    MEDES & Conference & 2 \\
    PROFES & Conference & 2 \\
    Parallel and Distributed Computing & Journal & 2 \\
    SANER & Conference & 2 \\
    ICST & Conference & 2 \\
    Big Data & Conference & 2 \\
    \hline
  \end{tabular}
\end{table}

\subsection{Authors' affiliations and demographics}

The authors' affiliations and demographics are also examined, revealing that multiple authors contribute to each paper, often affiliated with different academic institutions or industries, as shown in Fig 4. The geographic distribution of authors shows varying concentrations in different regions, indicating the global nature of research in this field as shown in Fig 5.

\begin{figure}[H]
    \centering
    \includegraphics[width=1\linewidth]{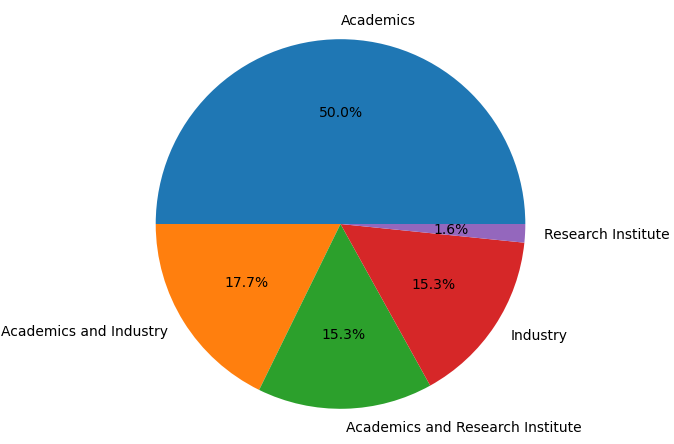}
    \caption{Distribution of Authors Affiliation}
    \label{fig:enter-label}
\end{figure}

\begin{figure}[H]
    \centering
    \includegraphics[width=1\linewidth]{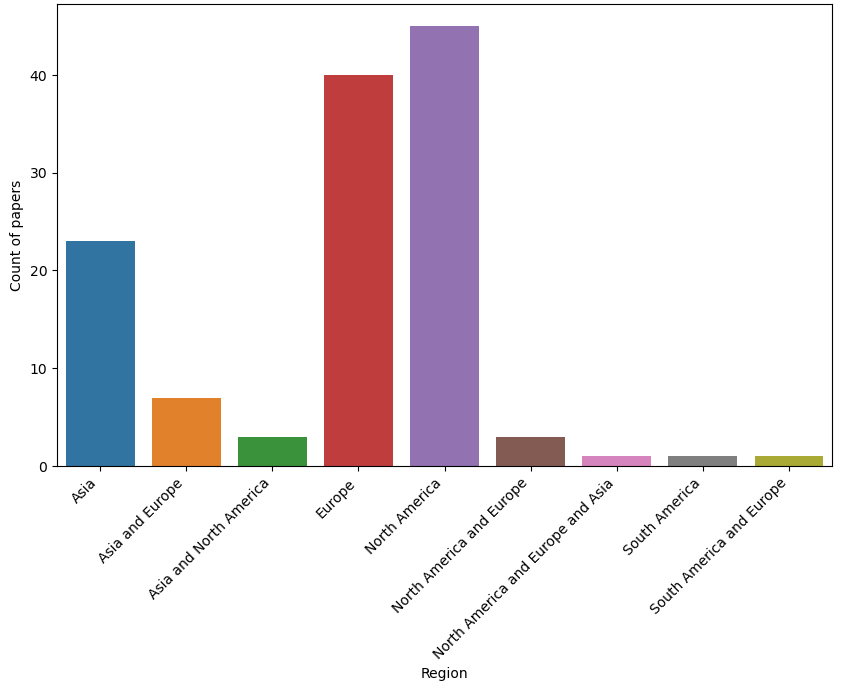}
    \caption{Authors Distribution by Region}
    \label{fig:enter-label}
\end{figure}

These visual representations and analyses provide
valuable insights into the research landscape,
showcasing the growth, trends, and dynamics within the field of study, and
contributing to the overall understanding of the selected literature.

\section{Results for  RQ's}

In this section, we summarise all the maintainability and scalability challenges and solutions, which we critically analysed from our literature review study and then coded into different ML workflow stages using NVIVO tools to answer all the Research questions. This section is divided into Six RQ, which we highlighted in Section 1. 
For the maintainability challenges and solutions we covered in RQ1 to RQ3, we coded the identified challenges and solutions for different ML stages in the ML workflow. We consolidated all the identified maintainability challenges in different stages of the ML workflow. Each challenge is indexed from CM01 up to CM41 and referenced in each  RQ1, RQ2 and RQ3 in Table VI, TableVIII and Table X, respectively. We have also similarly consolidated the corresponding maintainability solution indexed from SM01 to SM41 for every indexed maintainability challenge. These solutions are also referred to in  RQ1, RQ2 and RQ3 in Table VII, Table IX and Table XI, respectively.

Similarly, We have consolidated all the identified scalability challenges from RQ 4 to RQ 6, where CS01 up to CS13 in Table XII are the indexed scalability challenges. Each indexed challenge for these RQs has a corresponding solution or tools indexed from SS01 to SS13 in Table XIII.

\subsection{ RQ 1. Data Engineering  Maintainability  Challenges and Solutions}

\subsubsection{\textbf{Dataset Creation Challenges  }}

Collecting and preparing datasets for ML/DL research or applications can be challenging and time-consuming. The data collected from external or evolving sources may be susceptible to errors, missing data, outliers, susceptible to malicious attacks and may contain inherent bias ( \textbf{CM01, CM02} [LR1,2]). Creating labels for training data often requires the help of humans, but the repetitive nature of this task can quickly become tedious and demotivating for the annotator \textbf{(CM01} [LR9,15]). Furthermore, developing custom labelling tools for different applications is challenging due to the need for cross-disciplinary knowledge, and existing labelling tools are often difficult to customise (\textbf{CM01} [LR76]).

ML models can misclassify data due to adversarial attacks and variations in a production environment, making it challenging to obtain a mature model without natural variations (\textbf{CM02} [LR17,71] ). Existing data augmentation techniques to improve the model's robustness against corrupted data or adversarial attacks had significant tradeoffs in terms of accuracy. Probabilistic and Bayesian neural networks struggle under data shifts using these existing techniques. Another tradeoff is that using aggressive data augmentation schemes may improve clean accuracy at the cost of robustness against corruption or adversarial attack  (\textbf{CM02} [LR89]).

Additionally, datasets used in ML are often poorly documented, poorly maintained, and lack transparency in their creation process, contributing to errors and perpetuating bias (\textbf{CM03} [LR16]).

\textit{Summary: Maintaining datasets poses challenges, including errors, missing data, outliers, bias, adversarial attacks, variation in production environments, tedious label creation, custom tool development, and poor documentation, requiring efforts for long-term maintainability.}

\subsubsection{\textbf{Dataset Creation Solution}}

Creating high-quality labels for training data is essential for the performance of machine learning models. However, manual labelling is costly and time-consuming. Various approaches have been proposed to address this, such as using crowdsourcing platforms like Amazon Mechanical Turk, weak supervision and semi-supervised learning, and data programming ( \textbf{SM01} [LR3]). Data programming is a technique that uses human knowledge in the form of labelling functions and incorporates them into a generative model, which has succeeded in text and structured data applications. In addition to that, Weak supervision is also proven to semi-automatically generate labels of reasonable quality and quantity (\textbf{SM01} [LR2,3]). Shuffler is a tool that helps manage annotations for computer vision tasks like object detection and image classification. It uses a database to easily modify and store annotations in a human-readable format that can be manually edited. It does not care how images are stored and can load and modify annotations quickly ( \textbf{SM01} [LR3]). Inspector Gadget is another tool that combines crowdsourcing, data augmentation, and data programming to produce weak labels for image classification ( \textbf{SM01} [LR9]). 

Zhang et al. [LR76]  have created a modular design that allows developers to customise labelling tools for various applications. The design is implemented through a system called OneLabeler, which is designed to be flexible and reusable. OneLabeler has pre-built conceptual modules that developers can quickly implement to build customised labelling tools. The effectiveness of OneLabeler has been validated through an extensive case study and user study {\textbf{SM01}}. Additionally, gamification techniques have been shown to increase user engagement and satisfaction during the data annotation and label creation process {\textbf{SM01}} [LR15]. 

To improve the robustness of machine learning models, researchers have also focused on data augmentation techniques such as the "Worst ofk" method and Sensei, a search-based method using a genetic algorithm ({\textbf{SM02}}[LR17]). Augmix is a data augmentation technique that improves machine learning model robustness against unforeseen corruptions during deployment using stochastic augmentation operations and a consistency loss. Augmix significantly improves robustness and uncertainty measures on challenging image classification benchmarks and outperforms previous methods by a large margin in some cases ({\textbf{SM02}}[LR89]).

Wang et al. [LR40] present a defence strategy in an adversarial setting consisting of four main parts: (1) security evaluation mechanism, (2) defence strategy during the training phase, (3) defence strategy during the prediction/test phase, and (4) privacy protection method for the ML system. The security evaluation mechanism aims to evaluate the security and robustness of ML models against adversarial attacks. During the training phase, the defence strategy aims to enhance ML model robustness by sanitising the training data, improving the generalisation capability of the ML model, and using adversarial training. The defence strategy during the prediction/test phase follows two main directions: modifying the ML model and using various methods such as data compression, foveation-based, gradient masking, defensive distillation, and DeepCloak to enhance the robustness of the ML model. The privacy protection method aims to protect the privacy of the ML system. Different techniques, such as differential privacy (DP) and homomorphic encryption (HE), are used to ensure training data privacy ({\textbf{SM02}}[LR40]).

To ensure the responsible use of AI, transparency and accountability must be prioritised throughout the data development lifecycle. This includes treating datasets as guilty until proven innocent, identifying assumptions in writing, conducting independent peer reviews, evaluating datasets against experience and judgment, using visualisation tools, performing sensitivity studies, and understanding the limitations of datasets ({\textbf{SM03}} [LR16]).

\subsubsection{\textbf{Data Pre-processing challenges}}

Maintaining the quality of data during the preprocessing stage presents significant challenges in handling data errors, missing values, outliers, and bias ({\textbf{CM04}}[LR1]). Minor changes in data features can impact model accuracy as the model's and data features' performance are interdependent. Careful data handling in preprocessing and validation stages is required to ensure data quality and model fairness ({\textbf{CM04}}[LR1,6]). 

Determining the order and impact of steps such as data cleaning, sanitisation, and unfairness mitigation is unclear, adding to the challenge of maintaining data quality and fairness ({\textbf{CM05}}[LR1,2]). 

Addressing imbalanced data involves handling the unequal distribution of class labels, which can affect model performance. Data leakage refers to unintentional information leakage between training and validation sets, requiring careful handling to prevent biased results. Both challenges impact the maintainability of datasets during the preprocessing stage ({\textbf{CM06} }[LR5,6]).

Open data science faces challenges in dividing and combining tasks and source code among multiple contributors due to the need for a practical framework. Data analytics tools available today often need to be able to integrate easily with existing frameworks in multiple languages, leading to reduced productivity and efficiency. Despite the availability of Big Data frameworks like Apache Hadoop, Twister, Apache Spark, and Apache Flink, they are unable to meet the growing demand for faster and more integrable frameworks that can operate on both specialised and commodity hardware, especially with the increasing needs of ML ({\textbf{CM07}}[LR6,8]). While Big Data frameworks attempt to match user-friendly frameworks like NumPy, Python Pandas, or Dask by providing similar APIs, performance is sacrificed due to the overhead of switching between runtimes ({\textbf{CM07}}[LR8]).
   
\textit{ Summary: Maintaining data quality during preprocessing is challenging, involving handling errors, missing values, outliers, bias, determining step order, addressing imbalanced data, data leakage, and iterating and reevaluating the preprocessing steps, which can be a resource- and time-intensive with lots of trial-and-error processes to improve the model performance}.

\subsubsection{\textbf{Data pre-processing Solution}}

Missing values in a dataset can be handled by discarding examples or imputing values; some available imputation methods are the k-nearest neighbour (KNN) algorithm is a standard imputation method, including the KNN imputation (KNNI) and the weighted KNNI (WKNNI) variations. Another method is K-means Clustering Imputation (KMI) (\textbf{SM04}[LR5]).  
Karanikola et al. [LR5] proposed an improved Iterative Robust Model-based Imputation (IRMI) by combining boosting and decision tree theory. IRMI  uses the feature variables as predictors and the missing value as the target to estimate the missing value using a multivariant model. This new version outperforms other IRMI methods used in machine learning software (\textbf{SM04}  [LR5]).

The order of different data processing steps, such as data cleaning, sanitisation, and unfairness mitigation, and their impact on the model's accuracy and fairness need to be clarified. Tae et al. [LR1] compared and analysed different orders of data preprocessing steps and found that data sanitisation and cleaning should be done together, followed by unfairness mitigation, resulting in improved accuracy and fairness in the model. The authors developed a data processing framework called MLClean that provides better performance in terms of accuracy, fairness, and runtime using the proposed order (\textbf{SM05} [LR1]).

Imbalanced data can negatively impact modelling, and various data rebalancing techniques such as over-sampling, under-sampling, SMOTE, and ROSE can be applied to balance the distribution of classes (\textbf{SM06} [LR5]). Data leakage, a common issue in ML, can be mitigated by using alternative splitting methods such as Random splitting, Time-based splitting, and Baseline splitting ( \textbf{SM06} [LR6]). 

Ballet offers a lightweight solution for integrating features in open data science projects and improves model development tasks and feature selection techniques collaboratively. Commercial data science platforms like Domino Data Lab and Dataiku support all aspects of data science but lack explicit collaboration and incremental model development features (\textbf{SM07} [LR7]). 

Cylon is a data engineering toolkit integrating ML and other data processing systems. It aims to provide a more robust and efficient data engineering pipeline and increases user productivity and efficiency through seamless integration with various existing frameworks in multiple languages (\textbf{SM07} [LR8]).

\subsubsection{\textbf{Data validation challenges}}

Ensuring a reliable ML validation pipeline is challenging because of schema issues and data changes. External data is also vulnerable to drift, attacks, and errors, which makes proper error handling essential, especially with missing data. Additionally, data poisoning by malicious actors is also a big concern. Data validation tools, such as TensorFlow validation, generate a database schema from previous datasets and validate future datasets with the schema (\textbf{CM08} [LR2]). However, data drift and potentially unreliable sources make it difficult to maintain accurate models. ML models can also be challenging to understand and interpret, making it hard to assess their effectiveness without testing on representative cases. Ongoing validation challenges remain as the models and the problem itself may change, making it difficult to determine if the learned models accurately reflect current problem situations or if they are based on outdated training data sets  (\textbf{CM08} [LR12]).  

Integrating data validation tools into the development process of ML-enabled systems is recommended to handle data errors. However, this integration requires significant engineering resources for development and maintenance. There is a need to establish guidelines for ensuring overall data quality. The stochastic nature of ML models, data drifts and dependencies between features also pose challenges in validating and monitoring their effectiveness in solving intended use cases; as a result, many engineering teams tend to ignore data validation in their workflow if it is not mandatory (\textbf{CM09} [LR13]).

\textit{Summary: Maintaining a reliable data validation pipeline and infrastructure in ML systems is challenging due to schema issues, data changes, drift, attacks, errors, missing data, malicious actors, model interpretability, evolving problems, and the need for engineering resources and guidelines. Neglecting data validation without mandates exacerbates maintainability concerns.
}

\subsubsection{\textbf{Data validation Solution}}

Machine learning platforms, such as TensorFlow Extended (TFX), have practical data validation tools that can find and resolve errors in data using methods such as data visualisation and schema generation. For instance, Facets in TFX offer vital data statistics for machine learning, while tools like SeeDB and CUDE apply hypothesis testing to verify the significance of results. This data validation process is crucial for spotting problems in the data that could affect the machine learning pipeline. Also, TensorFlow Data Validation simplifies the process of creating database schemas from previous datasets and validating future datasets with the schema, improving the quality and dependability of training data (\textbf{SM08, SM09} [LR2]).  

Additionally, tools like Data Linter, Deequ, and TensorFlow Data Validation are used to inspect and suggest ways to improve training data representation automatically. Best practices for adopting a data validation process and tool for machine learning projects include defining data quality tests, providing actionable feedback, and treating data errors with the same severity as code. This approach can minimise manual effort, identify data errors early, and provide a testing approach for machine learning-enabled software systems  (\textbf{ SM08, SM09 } [LR13]).

\subsubsection{\textbf{Data management challenges}}

Managing large, evolving datasets in machine learning projects can be challenging due to the requirements for significant infrastructure to collect, process, store, and distribute the data. Efficient tools for managing large-volume live datasets need to be improved. Data management involves a range of tasks, including acquiring and integrating data from multiple sources, manipulating different data modalities, modifying annotations, serialising objects, and storing data in multiple formats, which often comes with challenges and antipatterns like glue code and pipeline jungles making version and maintaining a data pipeline complex  (\textbf{CM10} [LR3]). As datasets frequently change in exploratory phases, every alteration requires manual adjustments in the data pipeline, leading to making versioning and maintainability highly complex and error-prone all this makes it even more challenging to transition from exploratory data analysis to a regularised, reproducible production workflow (\textbf{CM10} [LR4]). 

Managing the machine learning process can be challenging, as automating pipelines for feature extraction, data acquisition, labelling, serialisation, and saving in different formats and distribution. This is not easy using traditional software engineering tools. For instance, converting models to different formats and ensuring compatibility across various platforms can lead to errors or loss of model performance, especially when using traditional software tools not specifically designed for machine learning workflows.
Tracking the data used at each workflow stage is crucial to ensure reproducibility and data provenance. However, traditional version control systems cannot efficiently handle the large amount of data machine learning applications use. Specialised tools are necessary to manage complex, non-linear workflow and address machine learning experiment concerns such as reproducibility, traceability, and auditability. Moreover, to ensure reproducibility and reuse in the workflow, data provenance, transformation steps, indexing, and intermediary result storage need to be tracked, which requires complex engineering and DevOps solutions (\textbf{CM10, CM11} [LR14,15, 68]). 

To improve the model development process, a more systematic approach is necessary for managing machine learning assets. This includes the need for better collaboration tools to address collaboration challenges among multiple developers and practical tools for comparing, analysing, and interpreting results to gain insights from machine learning experiments (\textbf{CM10} [LR68]).

\textit{Summary: Managing large, evolving datasets in ML projects is challenging due to infrastructure requirements, lack of efficient tools, diverse data tasks, constant dataset changes, and complexities in creating production pipelines and maintaining data provenance. The lack of collaborative teams, and data management tools, also exacerbates the maintainability challenges.}

\subsubsection{\textbf{Data Management Solution}}

There is a growing need for publicly available tools for data management that can facilitate tasks such as data acquisition, labelling, feature engineering, and serialisation in different formats and distributions. One example of such a tool is Shuffler, a toolbox for annotating and labelling images. It utilises a relational database and SQL query for storing and manipulating annotations. It allows for chaining multiple basic manipulation operations and adding new custom functions or operations. Shuffler supports loading, modifying, and storing images in agnostic formats and also supports object detection, semantic segmentation, and object matching tasks (\textbf{SM10} [LR3]). Another tool helpful in data management is JUNEAU, which aids in indexing, searching, and reusing tabular data like CSV and relational datasets. This is achieved by replacing the backends and extending the user interface of Jupyter Notebooks. JUNEAU acts like a backend data lake management subsystem that integrates relational and key-value stores to capture and index any external files loaded by the notebook and also captures intermediate data computational steps within the cells and notebooks (\textbf{SM10, SM11} [LR4]).

Machine learning experiments management tools like MLFlow, NeptuneML, and WandB aim to improve the efficiency of ML pipelines by reducing manual intervention and addressing challenges in managing machine-learning-specific assets such as reproducibility and traceability. Additionally, versioning tools such as DVC, Pachyderm, ModelDB, and Quilt Data are becoming increasingly popular in machine learning projects. These tools are used to specify data and model pipelines and manage model experiments, similar to infrastructure-as-code. They aim to reduce manual intervention and improve the efficiency of ML pipelines in open-source projects {\textbf{SM11}}[LR14]. Furthermore, Experiment management tools fall under machine-learning lifecycle management tools. Examples include MLFlow, Neptune, KubeFlow, and ModelDB provide functionalities to store, track, and version assets from different experiment runs. Some experiment management tools specialise in supporting deep learning  (\textbf{SM10,11} [LR68]).

\begin{table*}[htbp]
  \centering
  \begin{tabular}{p{1.5cm} p{7cm} p{2.5cm} p{3cm}}
    \toprule
    Challenge no  & Description of the challenges & ML  stage & Paper \\
    \midrule

CM01 &Creating and labeling datasets &Dataset Creation &LR1,2,9,15,76 \\
CM02 &Data susceptibility to error, drift, bias, and attacks&Dataset Creation &LR1,2,17,71,89 \\
CM03 &Poor documentation and transparency &Dataset Creation &LR16  \\
&  \\
CM04 &Missing data &Data Preprocessing &LR1,6 \\
CM05 &Order of data processing steps &Data preprocessing &LR1,2 \\
CM06 &Data imbalance and leakage &Data preprocessing &LR5,6 \\
CM07 &Collaboration difficulties due to limited tools and integration &Data preprocessing &LR6,8 \\
&  \\
CM08 &Maintaining reliable validation pipeline&Data validation &LR2,12 \\
CM09 &Maintaining data quality and validation infrastructure &Data validation &LR13 \\
&  \\
CM10 &Managing evolving ML datasets and metadata assets. &Data Management &LR 3, 4,14,15, 68 \\
CM11 &Missing tools for reproducibility, traceability, and auditability &Data Management &LR14,15,68 \\
&  \\

  \bottomrule
  \end{tabular}
  \caption{Maintainability challenges found from the LR study and the paper discussing these challenges 
 for a given ML stage in the workflow}
  \label{tab:example141}
\end{table*}

\begin{table*}[htbp]
  \centering
  \begin{tabular}{p{1.5cm} p{1.5cm} p{10cm}}
    \toprule
    Index & Solutions no & Solution/Tools/frameworks/methods \\
    \midrule

CM01 &SM01 &Crowdsourcing, AWS Mechanical Turk, Weak Supervision [LR2,3], Semi-Supervised Learning and Data programming techniques, Shuffler [LR3], Inspector gadget [LR9], Gamification [LR15], OneLabeler[LR76] \\
CM02 &SM02 &Worst ofK, Sensei [LR17], AugMIX [LR89, DeepCloak methods [LR40] \\
CM03 &SM03 & Guidelines from [LR16] \\
& & \\
CM04 &SM04 &KNNI, WkNNI, KMI, IRMI [LR5] {for Missing data ( imputation)}. \\
CM05 &SM05 &MLClean [LR1] \\
CM06 &SM06 &SMOTE , ROSE [LR5]. {data imbalance }, Random , Baseline and time based splitting { for data leakage }[LR6] \\
CM07 &SM07 &Ballet [LR7], Cylon [LR8] , Domino Data lab , Dataiku [LR7]. \\
& & \\
CM08 &SM08 &Tensorflow validation, Facets in TFX, SEEDB, CUDE [LR2], Data Linter, Deequ [LR13] \\
CM09 &SM09 &Tensorflow validation in TFX [LR2,13] \\
& & \\
CM10 &SM10 &Shuffler[LR3], JUNEAU [LR4] \\
CM11 &SM11 &JUNEAU[LR4], DVC, MLFlow, Pachyderm, MOdelDB, Quilt Data[LR14], MLFLow, NeptuneML, KubeFlow, WandB, ModelDB [LR68] \\
& & \\

  \bottomrule
  \end{tabular}
  \\
  \caption{ Maintainability solution for the identified Challenges that we indexed for each ML stage}
  \label{tab:example1113}
\end{table*}

\subsection{RQ2. Model Engineering Maintainability  Challenges and Solutions}

\subsubsection{\textbf{Model and Concept  Drift challenges}}

Maintainability challenges in ML can arise from concept drift, which refers to the changes in data distribution and relationships between variables over time. Rapid data and societal behaviour changes cause machine learning algorithms to lose accuracy. This can lead to the obsolescence of models trained on historical data, and it can be difficult to predict when these changes will occur in real-world applications (\textbf{CM12} [LR6,18, 20]). Researchers have created various methods, such as supervised, semi-supervised, unsupervised, statistical, and evolutionary algorithms, to handle concept drift, but none is suitable for all drift types. Machine learning ensembles for drift detection have been proposed but require retraining base learners and a strategy for selecting the best estimator, potentially affecting detection speed (\textbf{CM13} [LR19]).
 A dataset shift can occur when there is a change in the distribution of data points between the training and testing environments, leading to a decrease in accuracy. There are three data shifts: covariate shift, prior probability shift, and concept shift. The sources of data shift can cause a drop in system utility (\textbf{CM12} [LR77]). Drift detection in data distribution-based approaches can track changes in the location, density, and range of the data and can process both labelled and unlabeled data but can only detect certain types of drift and only address drift acting on a single model or a single pair of source and target streams (\textbf{CM12} [LR79]). Despite being a known problem for decades, concept drift remains a critical issue for ML applications today. Detecting model drift is a challenging and expensive task, as it requires expertise in drift detection algorithms, integration into existing pipelines and continuous maintenance to detect new drifts, as it is impossible for algorithms to identify all drifts. These can lead to model obsolescence and performance degradation (\textbf{CM14} [LR19,20]).

 \textit{Summary: Maintaining ML models faces challenges from concept drift due to changing data distributions and societal behaviour, causing accuracy loss. Various drift detection methods exist, but none are universally effective. Altered data distributions between training and testing environments, including covariate, prior probability, and concept shifts, can reduce accuracy; detecting and addressing these shifts are complex, expensive, and ongoing tasks, contributing to model obsolescence and performance decline.}

\subsubsection{\textbf{Model Drift Solution}}

Concept drift is a change in data distribution over time that reduces the accuracy of machine learning models. There are various techniques to mitigate the impact of concept drift, such as statistical testing, online learning, and the Streaming Ensemble Algorithm. Nishida et al.[LR6]  proposed that models can be updated incrementally rather than retraining from scratch when concept drift occurs. Other techniques include the Streaming Ensemble Algorithm and Kappa Updated Ensemble, which use majority-voting and dynamic weighing, respectively (\textbf{SM12} [LR6]). Drift detection is measured by the number of drifts in the data stream using the ADWIN  drift detector. This widely-used drift detector uses a Relative Standard Deviation (RSD) metric. A larger mRSD often indicates that the overall extent of concept drifts is more significant (\textbf{SM14, SM12 } [LR18]). The MAPE-K  (Monitor-Analyse-Plan-Execute over a shared Knowledge) is a software pattern implemented in the Driftage framework, which is used to detect drifts and make decisions in multi-agent systems. it includes agents for monitoring, analysis, planning, and execution. The framework is implemented in Python and communicates through an XMPP server  (\textbf{SM14, SM12} [LR19]).

Various tactics can be used to mitigate the impact of data shift on machine learning components, including component replacement, human-based labelling, transfer learning, unlearning, and retraining/hyperparameter optimisation. Each tactic has its costs and benefits, and the appropriateness of each tactic depends on the specific context and problem being addressed  (\textbf{SM12} [LR77]). 
The QuaD method aims to improve concept drift detection on dynamic data streams across multiple machine-learning models. This is achieved using feature stores, SHAP values, and Collaborative Filtering. The approach is the first to examine the collective behaviour of concept drift across multiple models and combines classical drift detectors to use labelled and unlabeled data. It relies on the data abstraction provided by emerging feature stores (\textbf{SM12, SM14} [LR79]).

Yang et al. [LR90] discuss the different approaches to adapting machine learning models to concept drifts, categorised into passive and active approaches. Passive approaches do not adapt to concept drifts, and instead, the models are periodically retrained on new data to maintain their accuracy. On the other hand, active approaches continuously monitor data streams and adapt the models as needed.   The author also proposes a new concept drift detection method that can handle classification and regression problems and determine the amount of data necessary for model updating. The method involves continuously updating online sequential extreme learning machines (OS-ELMs) and quantifying how much newly collected data modify the updated models. The proposed method outperforms alternative state-of-the-art concept drift detection methods, and updating the prediction model when concept drift is detected improves overall accuracy while minimising the number of model updates  (\textbf{SM13} [LR90]).

\subsubsection{\textbf{HPO challenges}}

Maintainability challenges in ML models include difficulties tuning models and hyperparameters, which can be time-consuming and require much computational effort. Hyperparameter Optimization (HPO) is a critical aspect of ML and can be challenging due to the high dimensionality of the search space and the need to balance exploration and exploitation. Various approaches such as Random Search, Grid Search, Bayesian Optimization, and Gradient-based methods have been proposed to solve the HPO problem, but each of them has its own limitations (\textbf{CM15, CM16} [LR22, 24]). Automated solutions for hyperparameter tuning are available but come with challenges, such as the lack of a standard API among current open-source solutions. Different HPO algorithms have different strengths, such as working well for low-dimensional numerical problems or high-dimensional structured model spaces (\textbf{ CM16} [LR21]). Without expert knowledge, it is often done through trial and error, which is very time-consuming and resource intensive. HPO impacts models' performance, efficiency, and convergence rate, and wrong choices can affect learnability and training progress (\textbf{CM15} [LR23, 24]). Moreover, it requires setting up and maintaining an orchestration pipeline to run the optimisation and keep track of the parameters and results for reproducibility (\textbf{CM15 } [LR21, 22]).

\textit{Summary: Keeping machine learning models working well is complex due to the difficulty in tuning model settings and hyperparameters, which can be time-consuming and computationally demanding. Finding the best hyperparameters is critical but complex, involving various methods and strategies, each with its limitations. Automated solutions exist, but they also have challenges. All these factors  impact model performance, efficiency, and training progress and require careful orchestration for reproducibility.}

\subsubsection{\textbf{HPO Solution}}

Researchers have proposed various methods for tuning machine learning models, such as Bayesian approaches, evolutionary algorithms, multi-armed bandits, and Neural architecture search by learning, to speed up the process. Efforts have been made to automate the process, such as Google Vizier, Amazon SageMaker, and SigOpt. However, these do not allow for customisation or easy integration with other resources. Open-source projects like Optunity and Tune offer user-friendly APIs but also have limitations in terms of customisation and integration (\textbf{SM15, SM16} [LR21]). Auptimizer is an open-source automated optimiser framework that addresses the challenges of using HPO in practice by reducing the effort to use and switch HPO algorithms, providing scalability for cloud/on-premise resources, simplifying the process to integrate new HPO algorithms and new resource schedulers, and tracking results for reproducibility (\textbf{SM15} [LR21]). HyperNOMAD package is a tool that uses the NOMAD algorithm to tune the hyperparameters of deep neural networks (DNNs). Cloud-based services like AWS Sagemaker, Azure Machine Learning, and Google Cloud AI platform offer hyperparameter tuning. However, users can only choose from the algorithms provided and have limited options for customisation (\textbf{SM15 } [LR22]). Open-source frameworks like Hyperopt, Auptimizer, Katib, and Tune offer more flexibility but may not be designed for reliable and efficient use of resources (\textbf{SM15} [LR23,24]). Ultron-AutoML is a service that can be installed on a Kubernetes cluster or a cloud platform like AWS, Google Cloud, or Azure ML. The framework is designed with reliability, efficiency, and scalability in mind to ensure the completion of a user's HPO job in a timely and cost-effective manner. Additionally, it allows for ease of use, flexibility, and customisation for the user  (\textbf{SM15}  [LR24]).

\subsubsection{\textbf{Model monitoring }}

Maintaining and improving machine learning (ML) models in production is a high cost, and engineers face challenges such as monitoring fine-grained quality and diagnosing errors in  ML  models, applications, and data pipelines. There is limited tooling for model lifecycle support, and engineers need tools to improve and maintain quality in the face of changes to input distribution and new production features (\textbf{CM17} [LR25]). Additionally, using system logs to monitor the behaviour of deployed models is a crucial issue in maintaining ML systems, as the community is still in the early stages of understanding the key metrics and methods for monitoring data and models (\textbf{CM18} [LR26]). Each model requires a unique approach, and metrics teams use self-developed or highly customised dashboard platforms to monitor the models. However, standardisation is lacking, which may lead to a bug-prone, ad-hoc and non-reusable measurement system. Automation is needed for model monitoring; therefore, teams have created tools that are not easily shared across the organisation and are difficult to maintain (\textbf{CM17} [LR20,27]). 

No standard approach or metrics exist for monitoring each model. Furthermore, a change in the distribution of data points between the training and testing environments can decrease accuracy and ultimately reduce system utility (\textbf{CM18} [LR77]) which may be difficult to notice without proper monitoring tools.  
Monitoring machine learning models is a complex task that requires an ongoing understanding of metrics and methods. There are challenges, such as feedback loops, where models can influence their behaviour over time. Automation and standardisation are needed to improve monitoring, as well as address issues like prediction bias ( \textbf{CM18} [LR 27,28]).

\textit{Summary: The challenges of monitoring and maintaining machine learning models in production include the absence of standardised tools, metrics, and automation, leading to ad-hoc monitoring systems and potential degradation of system utility due to changing metrics and model drift}

\subsubsection{\textbf{Model Monitoring Solution}}

Monitoring the behaviour of a deployed machine learning model is necessary to ensure it operates as expected, which involves testing the model using real data while it is online; this is called Online testing  (\textbf{SM18} [LR27]). As machine learning models are deployed in production systems, they may face degradation due to non-stationary data distribution, hardware degradation, and system updates. To address this,  a two-stage process of monitoring the model's staleness and updating it to improve its quality was proposed. The monitoring phase involves evaluating the model's performance by comparing it with defined success criteria and deciding whether to update the model based on a cost-benefit analysis (\textbf{SM18} [LR85]). Various tools and methods are available to practitioners to monitor the model's behaviour and evaluate its performance with real data. For instance, semi-automated model monitoring can be achieved using MLOps, where triggers can be set for performance degradation, diagnostic tools used, and model drift addressed. Fully-automated model monitoring can also be achieved by deploying and monitoring the models with triggers to acknowledge performance degradation. Utilising MLOps to undergo a transition towards fully automated monitoring of models requires CI/CD integration, CT pipeline, certification of models, governance and security controls, model explainability, auditing, reproducible workflow and models, and mechanisms to perform end-to-end QA test and performance checks  (\textbf{SM18}  [LR28]).

Furthermore, generating logs for ML and managing the evolution of log data in a DevOps environment optimised for ML requires effort and attention to introducing new processes and activities rather than technical frameworks. Adopting this approach can help investigate detected anomalies by linking the machine-understandable entry with the human-readable log ( \textbf{SM17 } [LR26]). 

Machine-learning management systems such as Overton streamline the model creation, deployment, and monitoring processes, allowing developers to focus on higher-level tasks and build deep-learning-based applications without writing any code (\textbf{SM17} [LR25]). 

Experiment management tools like MLFlow, NeptuneML, and WandB help manage machine-learning-specific assets like reproducibility and traceability and even monitoring to some extent during training and testing. However, to fully address the challenges in managing machine learning models, there is a need to assess the current tool landscape, develop new tools, and improve engineering processes in production (\textbf{ SM17} [LR68]).

\subsubsection{\textbf{Model Deployment challenges}}

Deploying Machine Learning models in real-world environments presents a series of complex challenges that are critical to both deployment and maintenance. These include managing dependencies, maintaining glue code, monitoring, logging, and handling unintended feedback loops  (\textbf{CM19} [LR11,20]). The difficulties are further pronounced in the migration and quantisation processes, where understanding their impacts on prediction accuracy and performance becomes vital. Some issues related to maintainability, such as keeping software and hardware up to date and maintaining reproducible results, often lead to increased engineering and human resource costs (\textbf{CM19} [LR28]).

The release and maintenance of ML models involve multiple intricate steps like packaging, validation, deployment to production, and continuous updating and training of the model with new data. Ensuring a secure and efficient process demands automated deployment and a dedicated CI/CD pipeline with DevOps integration, necessitating collaboration between technical experts and ML professionals (\textbf{CM21} [LR28]).

Deploying Deep Learning (DL) software on platforms like mobile devices or the cloud emphasises the challenge of compatibility and reliability. Challenges include data extraction, inference speed on various platforms, and environment setup, all of which impact the maintainability  (\textbf{CM20} [LR29, 30]).

A fintech company's experience highlights the challenge of varying hardware and platform parameters that can affect the model behaviour and result in errors. Moreover, the dynamic allocation of resources and the determination of when to scale down a system can be complex  (\textbf{CM20} [LR27]).

Due to model complexity, large training datasets, and data sensitivity, the increasing difficulty of reproducibility in ML compounds maintainability issues. Though sharing source code and data has improved reproducibility, retraining models and data sharing remain challenging. Using large-scale models necessitates extensive resources and expert knowledge in cloud, security, networking, and distributed processing. Platforms like Kubernetes can enable scalable inference but aren't specifically designed to enhance reproducibility, ML workflow, and lifecycle processes. Further efforts are required to address these intertwined deployment and maintainability challenges in ML (\textbf{ CM21} [LR72]).

\textit{Summary: Deploying and maintaining Machine Learning models in real-world environments entail complex challenges, including dependency management, reproducibility, compatibility across platforms, dynamic resource allocation, and the need for specialised knowledge and collaboration, all contributing to heightened engineering and human resource costs}
 
\subsubsection{\textbf{Model Deployment Solution}}

Deploying deep learning models on mobile devices is vital for making them widely accessible. Frameworks such as TF Lite and Core ML are used to convert trained models to formats that can run on mobile devices. Model quantisation, a process of reducing the precision of model weights, is commonly used to reduce memory and computation requirements. Chen et al. [LR29] found that the majority of deployment issues occur during the model conversion stage, with unsupported operations being the most common problem (\textbf{SM20} [LR29]). Chen et al. [LR30] also provide strategies for addressing different deployment issues. These include targeted learning of required skills and improving the usability of documentation provided by framework vendors [LR30].

DLHub is a learning system that aims to address the unique challenges in scientific ML, such as the need for model publication and sharing and the ability to run models on various computing resources. It allows for the publication of models with descriptive metadata, persistent identifiers, and flexible access control (\textbf{SM21} [LR33]).

 Integrating a deployed model requires building its infrastructure and making it consumable and supportable. Reusing data and models saves time and effort. It requires researchers to be involved in the development process with other engineers, using the same version control and participating in code reviews for better results (\textbf{SM19} [LR20]). Implementing MLOps environments enables continuous integration and delivery through a CI/CD pipeline and continuous retraining through a CT pipeline, leading to a safe and repeatable path for ML model development, deployment, and updating (\textbf{SM19, SM21} [LR28]). 
 
 Orchestration platforms like Kubernetes enable scalable inference in cloud computing. However, managing open-source ML inference platforms like KFserver, Polyaxon, and Kubeflow require expert knowledge in cloud, security, networking, and distributed processing. MLC is a cloud-based platform designed to improve ML conferences' reproducibility and review processes by sharing trained ML models. It is infrastructure agnostic, includes a Command Line Interface (CLI) for easy transfer of trained models, and provides researchers with an easy-to-use CLI to publish their findings (\textbf{SM19, SM21} [LR72]).

\subsubsection{\textbf{Model Governance challenges}}

Governing machine learning (ML) models present pronounced challenges due to the necessity for robustness, the ability to generalise to unseen inputs and compliance with specific functional requirements. Model verification, involving requirement encoding, formal verification, and test-based verification, is complex and must be precise to meet business and regulatory demands (\textbf{CM22 } [LR20]).

The development of ML models is intricate, requiring specialised systems to support various stages such as training, inference, and model serving. This complexity can hinder the governance of the model in production and scientific requirements like reproducibility and large-scale execution, causing maintainability challenges (\textbf{CM23} [LR33]). Current version control platforms like Git are still inadequate in tracking changes in data, hyperparameters, model artefacts, and dependencies. The complicated nature of managing AI components compounds the already existing difficulties in handling diverse teams and evolving ML models  (\textbf{CM23} [LR34]).

Operationally, ML systems face data uncertainty from unknown generation processes, posing significant risks to prediction reliability. This necessitates a well-defined method to counteract the risk of performance degradation, reflecting another dimension of maintainability challenges  (\textbf{CM23, CM24} [LR71]). With AI's growing role in safety and health, existing risk mitigation strategies are only partially effective, leading to fresh risk sources. Therefore, a revised risk management approach is essential for AI systems to address these unique challenges  (\textbf{CM24} [LR73]).

The opacity and unpredictability of ML systems add to the technical difficulties in ensuring accountability, underlining data governance as a crucial issue in AI governance. Existing governance structures must be better suited to manage the societal issues tied to AI, stemming from inadequate information to understand the technology and the regulatory lag. These aspects underscore the necessity for further research into the key actors and values in AI policies, emphasising the ongoing challenges of inclusivity and diversity in AI governance. 
 (\textbf{CM22, CM24} [LR88]).

\textit{ Summary: Governing machine learning (ML) models encompasses multifaceted challenges, including the need for robustness, generalisation, compliance, complexity in verification, inadequacy in version control, unpredictability, risk management, and ongoing challenges in inclusivity and diversity, all of which contribute to the broader difficulties in maintaining and governing AI systems.}

\subsubsection{\textbf{Model Governance Solution} }

The growing importance of machine learning in research and industry demands standardised infrastructure for model development, publication, sharing, and evaluation. A key challenge involves creating standardised model packages, metadata schema, and repositories for reproducing results. DLHub addresses these challenges by enabling model publication with detailed metadata and flexible access control ( \textbf{SM23} [LR33]).

Model risk assessment is crucial during machine learning model development and deployment, requiring a specialised team to review documentation for compliance with regulations and standards [LR27]. In addition, to avoid the risk associated with the model, there is a need for the process of model verification involving requirement encoding, formal verification, and test-based verification to ensure the model meets business and regulatory needs ( \textbf{SM24} [LR20]). 

Mining Software Repository4ML (MSR4ML) enhances productivity and developer awareness in machine learning projects by automatically identifying and tracing artefacts in Git-based ML projects, improving the traceability of connections between data, code, and ML models (\textbf{SM23} [LR34]).

Experiment management tools like MLFlow, NeptuneML, and WandB address challenges in managing ML-specific assets, such as reproducibility and traceability. However, these tools are still evolving, and refining engineering processes are necessary (\textbf{SM23} [LR68]).

Steimers et al. propose using the three components of trustworthy AI from the European Commission's High-Level Expert Group on Artificial Intelligence (AI HLEG) as a guideline to classify sources of risk in a taxonomy. These components are lawful, ethical, and robust, ensuring compliance with laws, ethical principles, and robustness from technical and social perspectives. Sources of risk can be categorised into two blocks: ethical aspects such as fairness, privacy, automation, and control, and various aspects impacting the AI system's reliability and robustness, including its ability to maintain performance under varying conditions. AI systems can become more trustworthy by addressing ethical aspects and factors affecting reliability and robustness, promoting adoption and societal benefits (  \textbf{SM22, SM24} [LR73]).

Ensuring the reproducibility of machine learning models is critical for scientific rigour and robust applications. However, the stochastic and non-convex nature of ML training procedures and randomised data splits can make reproducing ML models difficult. To mitigate the risk of non-reproducibility, relevant model information such as the ML source code, data sets, and computation environment should be tracked and stored, and tool-based approaches such as version control and metadata handling can also help ensure reproducibility. Also,  a need to maintain experimental documentation for monitoring and understanding model changes  (  \textbf{SM23} [LR85]).

Governing AI systems through code-enforced rules holds potential, but implementation challenges exist. Concrete specifications for AI governance frameworks and identifying responsible government parties are needed  (\textbf{SM22} [LR88]).

\subsubsection{\textbf{Model Training challenges}}

Training a model can be challenging due to its computational demands and difficulty validating and replicating results. Additionally, deep learning may not be the best approach for all types of problems, and it can be computationally expensive and challenging to fine-tune  (\textbf{CM25} [LR35]).

Retrained and incremental modelling are two standard methods for training ML models, but evidence to justify the choice between these methods is often lacking. Choosing between retrained and incremental modelling does not affect model interpretation, but it does affect learning assumptions and algorithm variants, leading to different accuracy and training time tradeoffs (\textbf{CM26} [LR18]).

Training models can also be costly, particularly in the deep neural network, and concerns have been raised about the environmental impact of training ML models. Additionally, training performance at runtime with evolving data streams is an even more significant maintainability challenge because the need for constant retraining with new data adds to the complexity of managing, integrating, and deploying the training pipeline to other systems and applications  (\textbf{CM27} [LR18, 20]). Furthermore, Model training also poses maintainability challenges, including setting up the infrastructure to automate the pipeline and monitor performance (\textbf{CM25} [LR32]). 

When training in on-premises environments, practitioners are often limited by the restricted usage of libraries and software packages and only allow the use of audited tools and platforms within the organisation for sensitive data. All dependencies must be approved before use and are stored in a private repository with only allowed packages that have been internally audited. The challenges in modelling include 1) limitations in using the latest ML technologies, 2) the importance of baseline models, 3) tracking all experiments, which often require a custom spreadsheet, and 4) defining problem-specific performance metrics, hindering the definition of standards  (\textbf{CM27} [LR27]).

\textit{Summary: Training and maintaining machine learning models present multifaceted challenges, including computational demands, validation difficulties, the choice between retrained and incremental modelling, environmental concerns, complexity in managing evolving data streams, infrastructure setup for automation, and limitations in on-premises environments such as restricted usage of software and constraints in using the latest technologies, hindering standardisation.}

\subsubsection{\textbf{Model Training Solution}}

There are many ways to train models to improve performance for some Deep learning models besides incremental and retraining, using a method called transfer learning, which involves reusing an existing model instead of training a new one from scratch. This can help to improve model performance by using high-quality training data (\textbf{SM26 } [LR2]). There are two categories of transfer learning algorithms based on domain similarity: homogeneous transfer learning and heterogeneous transfer learning. Homogeneous transfer learning algorithms assume that the source and target domains share the same feature space but have different data distributions. In contrast, heterogeneous transfer learning algorithms assume that the source and target domains have different feature spaces (\textbf{SM26 } [LR69]).

DL model may not be the best solution for all scenarios. Another viable alternative is using simpler models, such as shallow network architectures,  decision trees, and random forests. These models can be used as a starting point to accelerate the deployment of a machine learning solution, provide valuable feedback, and prevent over-complicated designs. Additionally, they can be used as a way to test the concept of a proposed machine-learning solution and establish the end-to-end setup(\textbf{SM25 } [LR20]). Furthermore, Majumder et al. also show using simpler alternatives. They proposed the use of local learning, which involves training separate models for specific regions of the data, as it has been shown to improve performance (\textbf{SM25 } [LR35]).

Chen et al. [LR18] discuss two predominant modelling methods: Retrained modelling and incremental modelling are two methods for training machine learning models. Retrained modelling trains a new model from scratch each time new data is available, while incremental modelling updates the existing model as new data becomes available. The choice between these two methods depends mainly on the learning algorithm's characteristics and the data's nature. Overall, the study found that the incremental modelling has consistently better accuracy on Bagging for ensemble learning algorithms while the retrained one shows less error on Boosting. The study also found that the training time of incremental modelling is more robust, while that of the retrained one varies depending on the subject adaptable software. It is concluded that the decision to use incremental or retrained modelling can be a tradeoff, and it is vital to consider the tradeoff between accuracy and training time when choosing between these two methods  (\textbf{SM26 } [LR18]). 

Machine-learning experiment management tools like MLFlow, NeptuneML, and WandB address challenges in managing machine-learning-specific assets, such as reproducibility and traceability. However, these tools still need to be fully matured, and their development is limited by factors such as a lack of interoperability and tight coupling with specific libraries. To address these limitations, there is a need to assess the current tool landscape, develop new tools, and improve engineering processes. Experiment management tools, including specialised deep learning support, fall under machine-learning lifecycle management tools such as MLFlow, Neptune, KubeFlow, and ModelDB, which provide functionalities to store, track, and version assets from different experiment runs  (\textbf{SM25, SM27 } [LR68]).

\subsubsection{\textbf{ML Testing challenges}}

Machine learning (ML) systems present distinct challenges in software testing and maintenance due to their evolving nature and emergent behaviour based on learning (\textbf{CM28} [32]). Traditional testing methods, which depend on expected outputs and coverage, often do not apply to ML systems because of their vast input spaces and unpredictable behaviour. Deep neural networks (DNNs), for instance, do not offer statistical guarantees for out-of-distribution data, meaning new inputs that differ from the training data distribution (\textbf{CM30} [38]). The deployment of an ML system includes generating a prototype model, conducting offline testing before deployment, and making predictions afterwards. Online testing complements offline testing by identifying bugs once the model is in a natural environment [39]. As ML systems continually improve through learning, conventional test oracles become outdated (\textbf{CM29, CM31} [39]). 

Testing ML systems is complicated as they derive from training data rather than predefined rules. The absence of specifications or source code makes understanding their behaviour challenging. Issues like data pipelines and redundant experimental code paths also pose testing challenges (\textbf{CM28} [32, 40 ]. There is also a need for testing for valid behaviours; many ML system classes may have multiple valid behaviours. Additionally, individual test failures in ML systems may not always indicate a bug, making debugging a challenging task  (\textbf{CM28} [36].

ML systems (MLSs) face other hurdles, including messy datasets, lengthy model training, and the challenge of interpreting and predicting ML model behaviour. The vast input space of MLSs makes it hard to reveal data-sensitive faults. Testing MLSs demands new ML-specific criteria, such as mutation adequacy. MLSs testing occurs at various levels, from input testing to system testing. Data-box testing, which uses training and test data to analyse model behaviour, is essential due to the ample input space of MLSs. Other challenges include generating realistic input data, scalability assessment, and addressing the oracle problem (\textbf{CM28, CM29, CM30, CM31} [86]).

There are also challenges in emulating faults in data to gauge their impact on ML systems. While frameworks exist for the flexible use of fault models, handling data faults remains difficult (\textbf{CM28} [41,43]). The repercussions of faults or attacks in the ML pipeline can be severe, especially in critical systems like autonomous vehicles. Techniques like mutation testing and fault injection have been employed to evaluate ML models. However, there is no comprehensive tool for evaluating ML applications for input data faults, which can arise from adversarial attacks or noisy data (\textbf{CM28, CM31} [71]).

Other challenges include ensuring data quality, especially for high-dimensional data like natural language, and the labour-intensive nature of test data collection (\textbf{CM31} [83]). The entangled  ML components in complex systems make efficient test execution difficult. The randomness inherent in many ML algorithms can also lead to test flakiness, making it hard to differentiate between genuine test failures and noisy executions (\textbf{CM30} [78]).

\textit{Summary: Testing Machine Learning (ML) systems poses unique challenges due to their evolving and unpredictable nature, vast input spaces, and lack of traditional testing methods that can be supported for ML. These challenges encompass issues like understanding behaviour without predefined rules, handling messy datasets, lengthy model training, emulating data faults, ensuring data quality, and identifying and maintaining good test oracles, which can lead to test flakiness and make debugging a complex task.}

\subsubsection{\textbf{ML Testing Solution}}

The evaluation of unit test generation techniques for machine learning (ML) libraries was conducted through an empirical analysis of five popular ML libraries using tools like EVOSUITE and Randoop  (\textbf{SM28} [LR36]). It was found that most ML libraries have inadequate unit test suites, as evidenced by quality metrics such as code coverage and mutation score. However, tools like EVOSUITE and Randoop only offer limited improvements in these metrics. DeepXplore and Themis have been developed to identify bugs and discrimination in ML systems (\textbf{SM28} [LR39]).

Zhu et al. introduced a novel approach for functional regression testing of ML systems, applied to an ML-based spelling checker/corrector (Speller). This method, which leverages production data for test case updates, perturbs data for coverage and uses clustering for identifying failure patterns, proved more effective than manual testing in generating test cases, detecting failures, and improving coverage (\textbf{SM28} [LR37]).

In a comprehensive examination of ML systems, Berend et al. [LR38] assessed the impact of deep learning (DL) systems in production by evaluating adversarial retraining with data distribution awareness, finding that this awareness during testing and enhancement phases improved retraining by up to 21.5\%, a result that aligns with other studies pinpointing challenges and gaps in current ML testing practices and proposing future research directions, including error detection in data, learning programs, and frameworks, as well as reliable oracle acquisition [LR32,39]; this broader context is further enriched by Pastor et al. [LR32], who provided a thorough review of current testing practices, encompassing black-box and white-box approaches for ML model error detection and numerical-based techniques for ML code implementation error detection, thereby collectively highlighting the importance of multifaceted approaches to error detection and improvement in ML systems.

The dpEmu framework, still a prototype, enables easy modelling of data faults in ML systems and serves various purposes, such as evaluating the tradeoff between accuracy and robustness (\textbf{SM28} [LR41]). Li et al. introduced TensorFI, a flexible and easy-to-use framework for evaluating the resilience of TensorFlow-based ML applications (\textbf{SM28} [LR42]).

CALLISTO, a tool that addresses training data issues, improves ML system quality by using entropy to quantify prediction uncertainty and generating tests to identify mislabeled or low-quality data (\textbf{SM28} [LR44]). Panichella et al. [LR45] discussed mutation testing in ML, particularly DL and suggested further research.

Dutta et al. conducted the first large-scale study on seed usage in ML project testing. The study, which employed a tool called XSEED, recommended alternative strategies to use seeds and provided valuable insights into seed usage in ML project testing (\textbf{SM30} [LR78]).

Testing ML systems (MLS) presents significant challenges, such as their stochastic nature and faults introduced during training. Potential solutions include developing automated and dependable oracles, creating common frameworks and benchmarks, and utilising techniques like input mutation and search-based approaches for generating diverse test data (\textbf{SM29} [LR86]).

Fault injection techniques, used to evaluate application and tool reliability, have led to the development of several tools like Ares, TensorFI, and PyTorchFI for different ML frameworks. To fill the gap of an end-to-end evaluation tool for ML applications, TensorFlow Data Mutator (TF-DM) was developed to target various data faults for TensorFlow 2-based ML programs. TF-DM helps thoroughly assess TensorFlow 2-based ML models, improving their robustness and resilience, which is vital for safety-critical systems (\textbf{SM28} [LR71]).

\begin{table*}[htbp]
  \centering
  \begin{tabular}{p{1.5cm} p{7cm} p{2.5cm} p{3cm}}
    \toprule
    Challenge no  & Description of the challenges & ML  stage & Paper \\
    \midrule

CM12 &Models becoming inaccurate due to data distribution changes &Model / Concept drift & LR6,18,20,77,79 \\
CM13 &Choosing the right drift detection algorithm. &Model/ Concept drift & LR19 \\
CM14 &Maintaining drift detection pipeline. &Model / Concept drift & LR19, 20 \\
& \\
CM15 &Time-consuming and computationally intensive &HPO & LR22,23,24 \\
CM16 &Determining the right HPO approach. &HPO & LR21,22,24 \\
&  \\
CM17 & Monitoring ML models in production. &Model Monitoring & LR25,20,27 \\
CM18 & Challenges in standardisation, metrics selection, feedback loops, and model drift. &Model Monitoring & LR26,27,28,77 \\
&  \\
CM19 &Deploying ML models in real-world settings &Model Deployment & LR11,20,28 \\
CM20 &Many different types of hardware and software platforms. &Model Deployment & LR29,30,27 \\
CM21 & Maintaining reproducibility and updating software/hardware. &Model Deployment & LR72,28 \\
&  \\
CM22 &Ensuring ethics, compliance, robustness, and generalization &Model Governance & LR20,88, 73 \\
CM23 & Reproducibility and tracking dependencies and artefacts. &Model Governance & LR33,34,71 \\
CM24 &Managing risks associated with the model. &Model Governance & LR73,88,71 \\
&  \\
CM25 &Maintaining and replicating results of model training. &Model training & LR32, 35 \\
CM26 &Choosing between retraining and incremental modelling. &Model training & LR18 \\
CM27 &Managing and integrating training pipelines. &Model training & LR18,20,27 \\
& \\
CM28 &Challenges in model testing and maintenance. &Model Testing & LR32,36,40,41,43,71,86 \\
 CM29 &Oracle problem in ML testing&Model Testing & LR39,86 \\
 CM30 &Testing ML algorithms due to their stochastic nature. &Model Testing & LR 38, 78, 86 \\
 CM31 &Collecting and ensuring test data quality. &Model Testing & LR39,86,83,71 \\
&  \\

  \bottomrule
  \end{tabular}
  \caption{Maintainability challenges found from the LR study and the paper discussing these challenges 
 for a given ML stage in the workflow}
  \label{tab:example141}
\end{table*}

\begin{table*}[htbp]
  \centering
  \begin{tabular}{p{1.5cm} p{1.5cm} p{10cm}}
    \toprule
    Index & Solutions no & Solution/Tools/frameworks/methods \\
    \midrule

CM12 &SM12 &Statistical testing, online learning, and the Streaming Ensemble Algorithm. Incremental learning rather than retraining [LR6]. ADWIN drift detector [LR18], Driftage [LR19], Quad [LR79] \\
CM13 &SM13 & No clear options or winner, some claimed methods Online sequential extreme learning machines (OS-ELMs) [LR90] \\
CM14 &SM14 &ADWIN drift detector[LR18], Driftage framework [LR19], QUAD[LR79]. \\
& & \\
CM15 &SM15 &Google Vizier, Amazon SageMaker, and SigOpt.Optunity and Tune , Auptimizer [LR21] . HyperNOMAD [LR22]. Hyperopt, Auptimizer, Katib, and Tune [LR23,24]. Ultron-AutoML [LR24] \\
CM16 &SM16 &Bayesian approaches, evolutionary algorithms, multi-armed bandits, Reinforcement Learning, and Neural architecture search[LR21] \\
& & \\
CM17 &SM17 &Overton [LR25], WandDB, MLflow, NeptuneML, MOdelDB, KubeFLow[LR68] \\
CM18 &SM18 &  Online Testing [LR27] , Embracing MLOps  [LR28,85]\\
& & \\
CM19 &SM19 &  Infrastructure maintenance and collaboration between team [LR20] , Automation and MLOps [LR28 , 72]  \\ 
CM20 &SM20 &TF Lite and Core ML[LR29] , \\
CM21 &SM21 &DLHub[LR33], MLC[LR72], KFserver, Polyaxon, and Kubeflow [LR72]. \\
& & \\
CM22 &SM22 &  EU AI HLEG trustworthy AI components[LR73] , Governing AI system through Code-enforced rules [LR88]  \\
CM23 &SM23 &DLHUB [LR33], MSR4ML framework [LR34], MLFlow, NeptuneML, and WandB, KubeFlow, and ModelDB [LR68], Track and store relevant model information and metadata[LR85]  \\
CM24 &SM24 &  EU AI HLEG[LR73],  Model verification  involving  requirement encoding, formal verification, and test-based verification [LR20]
 \\
CM25 &SM25 &Use Simpler model [LR20], Local learning[LR35] MLFlow, NeptuneML, and WandB, KubeFlow, and ModelDB [LR68] \\
CM26 &SM26 &  Homogeneous and heterogeneous Transfer Learning [LR69] \\
CM27 &SM27 & Neptune, KubeFlow, and ModelDB [LR68] \\
 \\
CM28 &SM28 &EVOSUITE and Randoop [LR36], Speller [LR37], DeepXplore, Themis [LR39] , dpEmu [LR41], TensorFI [LR42] , CALLISTO [LR44] , Ares, TensorFI, and PyTorchFI [LR71] , TensorFlow Data Mutator (TF-DM) [LR71] \\
CM29 &SM29 & Potential solution and recommendation [LR86]\\
CM30 &SM30 & XSEED [LR78] \\
CM31 &SM31 & No good solution found in this SLR\\
& & \\

  \bottomrule
  \end{tabular}
  \\
  \caption{ Maintainability solution for the identified Challenges that we indexed for each ML stage}
  \label{tab:example1113}
\end{table*}

\subsection{RQ 3. Current Ecosystem  Engineering   Challenges and Solutions in Building an ML System or Applications}

\subsubsection{\textbf{AUTOML challenges}}

Automated Machine Learning (AutoML) is an emerging field that searches for the best-performing model for a given task, dataset, and evaluation metric, reducing the human resources and expertise required to develop machine learning models and decreasing the time-to-market for these models [LR80]. It uses heuristics and iteration to find the most accurate model for a task, and different types of AutoML software are available. However, it still requires understanding the underlying concepts [LR55]. In addition to that, AutoML also faces several challenges, including limitations of current AutoML libraries and systems, focused on specific use cases and being proprietary, time-consuming trial-and-error testing of ML algorithms and configurations, difficulties with network morphism and Bayesian optimisation in Neural Architecture Search (NAS) (\textbf{CM32 } [LR51, 60, 61 ]). Additionally, the AutoML libraries and systems, such as ATM, Vizier, Rafiki, Google AutoML, DataRobot, and Azure Machine Learning Studio, have limitations, such as focusing on a specific subset of ML use cases and being designed as proprietary applications that do not support community-driven integration of innovations (\textbf{CM32} [LR51,62, 64]). 

Bajaj et al. evaluate existing commercial AutoML services and found some gaps, such as the lack of customizability, transparency, and unique enhancements, as well as the use of only full-precision models (\textbf{CM33}  [LR80]). Auto ML minimises expert intervention by autonomously selecting a model and algorithm, but data preprocessing automation is often overlooked (\textbf{CM33} [LR63]).  

\textit{Summary: AutoML faces challenges including limitations in current libraries and often proprietary for specific use cases,  time-consuming trial-and-error testing, difficulties with network morphism and Bayesian optimisation in NAS, and issues with commercial services such as lack of customizability and transparency, while also often overlooking data preprocessing automation.}

\subsubsection{\textbf{AutoML Solution}}

 Auto-WEKA was the first AutoML tool, but its scope was limited. It is an  AutoML framework focusing only on classification tasks and cannot handle missing data. Katib is another recent AutoML platform that requires user hints and may be challenging for non-experts (\textbf{SM33} [LR80]).   The Machine Learning Bazaar is a new framework that simplifies the development of machine learning and automated machine learning software systems with ML primitives, a unified API, and a specification for data processing and ML components demonstrated through 5 real-world use cases and 2 case studies with an open-source variant called AutoBazaar (\textbf{SM32, SM33 } [LR51]). Various meta-learning approaches have been developed as frameworks, such as Auto-WEKA, Auto-Sklearn, and SmartML, with a proposed framework that is implemented as a microservice architecture for increased scalability and maintainability (\textbf{SM32 }[LR60]). A  novel framework has been presented for efficient neural architecture search using network morphism and Bayesian optimisation, with an open-source AutoML system called Auto-Keras  (\textbf{SM32} [LR61]); 

Several automated machine-learning platforms have been proposed for various applications. One such platform, designed for Smart City applications, is an autonomic machine learning platform that automates machine learning processes based on autonomic levels [LR62]. Another platform, called AutoTrain, is an automatic training system for small-scale image classification problems. It includes features like sample equalisation, Bayesian optimisation, dynamic adjustment model, and model selection module for automation. Additionally, Google has proposed the AutoAugment method for automatically searching for suitable data enhancement strategies, while DeepAugment focuses on data augmentation using a Bayesian algorithm (\textbf{SM33} [LR63]).

A new tool called ALOHA has been proposed for task-specific applications to address the need for a more targeted approach to automating the deep learning pipeline for specific use cases [LR64].

Walmart has developed an enterprise-scale AutoML framework called WALTS to democratise machine learning within their organisation. WALTS is designed to explore models from a pool of candidates and is made scalable and robust through the technology stack used. Ultron is an AutoML framework designed by a different Walmart team limited to only hyperparameter optimisation. Commercial tools like Google Cloud AutoML, H2O AutoML, and Dataiku AutoML are also compared to WALTS  (\textbf{SM33} [LR80]).

\subsubsection{\textbf{Machine Learning System Quality Challenges}}

Maintaining AI/ML systems poses significant challenges, including software testing, model development, data management, project management, infrastructure, and requirements engineering. These challenges are further categorised into subcategories such as defining AI ethics requirements, interpretability, and scalability, which include issues like the lack of methods and tools to implement AI ethics, difficulties in consistently labelling large datasets, and integrating ML models into scalable software systems while assessing data quality (\textbf{CM34} [LR10, 57 ]). DL software also faces similar challenges in terms of quality and reliability and communication across teams during feature engineering, model selection, and hyperparameter tuning (\textbf{CM35} [LR38, 54]). There is a lack of knowledge and frameworks for introducing fault tolerance in ML systems and ensuring quality assurance for AI-based systems (\textbf{CM34} [LR56]). The slow adoption of AI in business applications is due to concerns about the reliability and maintainability of AI applications in business-critical systems, leading to the need for a rigorous quality management framework [LR58]. Existing software architecture practices need to be adapted to account for the data-dependent behaviour of ML components, and new architecture patterns and tactics are necessary to address quality attributes such as explainability, data centricity, verifiability, monitorability, observability, fault tolerance, security, and privacy (\textbf{CM34} [LR59]).

The availability of high computing power, large amounts of data, and open-source ideas have facilitated significant progress in artificial intelligence. However, a human-centred approach and compliance with basic safety principles are essential to developing such systems. The complexity of tasks and operational environments in machine learning-based AI systems can result in incomplete specification and operational environment, which is the source of risk (\textbf{CM34} [LR73]). To ensure human safety and health, developing safe and trustworthy AI systems is necessary. While established risk reduction measures in software development are limited in mitigating these risks, some research proposes assurance cases to support quality assurance and certification of AI applications. Assurance cases have been proposed to support quality assurance and certification of AI applications, but a detailed list of concrete criteria for these cases is still lacking. The development of safe AI systems requires a good understanding of the components of trustworthy artificial intelligence, and risk management for systems that use AI must be carefully adapted to the new problems associated with this technology \textbf{CM34} [LR73]). The successful collaboration between software engineering and data science teams is crucial for developing ML-enabled systems, which requires coordination and integration from multiple experts or teams. This collaboration poses additional challenges due to the exploratory nature of ML development, the need for continuous monitoring and evolution, the difficulty of testing ML systems, and non-traditional quality requirements such as explainability and fairness. While technical aspects of developing ML components have received significant research attention, human factors such as coordination and documentation of responsibilities and interfaces and planning for system operation and evolution have been neglected (\textbf{CM35}  [LR75]).

The demand for machine learning is rapidly expanding, but surveys reveal that practical ML projects must meet their sponsors' or clients' expectations. Key challenges in the ML life cycle include data and software quality and a need for more guidance through standards and development process models specific to ML applications. To address these needs, a Japanese industry consortium called QA4AI was founded (\textbf{CM34} [LR85]). It is essential to understand the impact of different platforms on deep learning software, as they have varying capabilities in supporting deep learning software. Practical guidelines are provided for developers and researchers to better understand the characteristics of deep learning frameworks and platforms and develop and deploy high-quality deep learning systems effectively.

Developing and deploying models on different platforms and devices introduces more system quality challenges. PyTorch provides more stable training and validation processes than TensorFlow, CNTK, and MXNET, and computing differences across frameworks can lead to the misclassification of deep learning models. It is important to note that existing model conversion between frameworks is unreliable and requires special attention and inspection. Additionally, deep learning models from different frameworks exhibit different robustness against adversarial attacks  (\textbf{CM34}  [LR87]).

\textit{Summary:  Practical challenges in the ML life cycle, such as data and software quality, are being addressed by industry consortiums, but the development and deployment of models across different platforms introduce additional system quality challenges. These challenges encompass a lack of methods and tools, difficulties in labelling datasets, integrating models into scalable systems, ensuring quality assurance, and adapting existing architecture practices. The complexity and risk associated with ML-based AI systems necessitate a human-centred approach, compliance with safety principles, and collaboration between software engineering and data science teams.}

\subsubsection{\textbf{ML System Quality  Solution}}

With machine learning-based AI systems, the problem of incomplete specification is a consequence of the complexity of the task and operational environment, which can be regarded as the source of risk. In order to achieve a comprehensive understanding of individual sources of risk, a thorough taxonomy of risk sources for AI systems has been proposed, which lays the groundwork for examining individual risk sources, understanding their origins, and devising appropriate risk mitigation measures (\textbf{SM34}   [LR73]). 

To enhance collaboration, formalised practices are advised, such as creating a mutual understanding of project goals and scope, setting up effective communication channels, and documenting decisions and outcomes. The CRISP-ML(Q) process model is introduced to develop ML applications with quality assurance methodology, encompassing six phases from defining scope to maintaining the deployed application. This model allows for iterative adjustments, project postponement, or cancellation and aligns with risk-based thinking. Future research aims to evaluate the proposed model across various industries and applications to establish a universal standard process model for ML application development (\textbf{SM35}  [LR85]).

While input/output checkers help detect statistical deviations over time, other quality aspects such as robustness, security, and data privacy should be taken into account for AI systems (\textbf{SM34}   [LR56, 57]).

Santhanam et al. [LR58] advocate for traditional methods like manual inspection and testing to identify defects in ML systems and maintain stable performance, a perspective that complements the findings of Masuda et al. [LR67], who enumerate seven leading software quality techniques for ML applications, including Deep Learning, Fault Localization, Prediction, MLaaS, Multi-Agent, Search-Based, and Model Checking. These approaches are further supported by research showing that incorporating data distribution awareness during testing and enhancement can improve outcomes by up to 21.5 \%  [LR38]. Additionally, the interconnected challenges of conveying ML model quality in large teams have been explored [LR54]. Tools like MLCask, a version control system, have been introduced to track changes in ML pipelines and facilitate collaboration (\textbf{SM34} [LR50]), reflecting a comprehensive effort to enhance quality control and collaboration in ML development.

To tackle collaboration issues between data scientists and software engineers, investment in interdisciplinary teams, clear documentation of responsibilities and interfaces, recognising engineering work as crucial, and prioritising process and planning are recommended (\textbf{SM35}  [LR75]).

\subsubsection{\textbf{MLOPS challenges}}

Machine Learning Operations (MLOps) is a field that applies the DevOps principles of collaboration, automation, and continuous improvement to machine learning workflows. However, MLOps faces numerous challenges, such as high costs, lack of data scientists, privacy concerns, difficulty in determining the final model, and issues with deployment, integration, and communication with end-users  (\textbf{ CM36 } [LR28]).

Additionally, there are challenges in tracking and comparing experiments, a need for more version control, and difficulties in deploying models. Continuous Delivery for Machine Learning (CD4ML) is an approach to automate the lifecycle of machine learning applications. However, it still faces challenges in implementing data operations and relying on datasets from multiple organisations (\textbf{ CM36 } [LR52]). Incorporating MLOps in a multi-organisational setting also presents unique challenges related to integration, data privacy, security, and regulatory compliance. Addressing these challenges requires custom solutions to maintain adherence to governance, auditing, and regulations (\textbf{CM36 } [LR52]).

Research has been conducted on creating ML platforms with DevOps capability and running ML pipelines, identifying potential performance bottlenecks, such as GPU utilisation. Automation in the ML process, from data preprocessing to application runtime monitoring, can help users concentrate on application development  (\textbf{ CM36 } [LR53]). 

Although several tools like MLFlow, Amazon SageMaker, Spack, and EasyBuild have been proposed to improve the workflow of ML project development, rebuilding of ML models, shareable models, and continuous integration and data management, they are still in the early stages of development, and their workflows are not easy to put into practice. Therefore, there is a need for more research to address the challenges of MLOps and improve its adoption in the industry  (\textbf{ CM37 } [LR53,74]).

\textit{Summary: MLOps, the application of DevOps principles to machine learning, faces challenges including high costs, lack of expertise, privacy issues, deployment difficulties, lack of version control, and complexities in multi-organisational settings, necessitating further research and custom solutions for improved industry adoption.}

\subsubsection{\textbf{MLOPs Solution}}

John et al. conduct a systematic literature review and a grey literature review to derive a framework for the adoption of MLOps (Machine Learning Operations) in companies, present a maturity model outlining four stages, validate it by studying three case companies and aim to provide standard guidelines on how to effectively integrate MLOps into existing software development practices ( \textbf{SM37} [LR28]). Zhou et al. [LR53] also discuss various machine learning platforms that provide solutions for orchestrating ML pipelines and managing data storage, access control, pipeline execution, job scheduling, and performance monitoring. These platforms are designed to provide a generic solution for different ML tasks in hybrid environments, allowing developers to describe ML workflows through pipeline configuration or software development kits (SDKs).

Granlund et al. [LR52] propose the use of integration mechanisms for ML/AI to create multi-organisational AI/ML systems, focus on the integration and scaling of systems that include ML components, use the example of Continuous Delivery for Machine Learning (CD4ML) ( \textbf{SM37} )  as a concrete example of MLOps pipelines, and discuss various machine learning platforms that have been developed to address the end-to-end lifecycle management of ML applications such as TensorFlow Extended (TFX), ModelOps, and Kubeflow that provide solutions for orchestrating ML pipelines and management of data storage, access control, pipeline execution, job scheduling, and performance monitoring, and can be used in hybrid environments to provide a generic solution for different ML tasks ( \textbf{SM36} [LR52]).

Rzig et al. [LR74] also highlight the benefits of adopting DevOps tools within ML projects, such as improved code sharing, integration speed, and issue resolution. It also highlighted the need for support for automatically synchronising DevOps configuration files to reduce maintenance overhead. The analysis identified bug-fixing commits as the primary event that triggers DevOps file changes. This indicates the need for co-evolution analysis of functional code and DevOps configuration files to facilitate early bug detection. The analysis also showed that adopting DevOps tools positively impacts code sharing, integration speed, and issue resolution. The benefits of DevOps outweigh the associated costs, and it is recommended that data scientists and ML developers adopt DevOps tools within their projects ( \textbf{SM37}  [LR74]).

\subsubsection{\textbf{Building ML system}  challenges}

Machine learning (ML) models hold the potential to benefit real-world applications greatly; however, their success hinges on proper implementation and management. Ensuring reproducibility can be challenging due to limited access to training data, insufficient documentation, or inconsistent software and platform use. Developing ML models involves addressing a range of concerns, such as handling significant inputs, ethical considerations, model complexity, accuracy, data limitations, uncertainty, documentation, missing values, and data integration from multiple sources ( \textbf{CM38} [LR10, 11]). Data dependency, model staleness, and drift are crucial issues that can lead to unstable models, necessitating continuous monitoring.

Acquiring sufficient data, particularly labelled data for supervised learning, can be arduous. Developing AI/ML differs from traditional software development and demands a comprehensive understanding of the interconnected steps involved, such as data preprocessing and model deployment ( \textbf{CM39} [LR11, 20]). The actual challenges of ML extend beyond algorithms, emphasising the need to focus on the entire lifecycle, including data collection, documentation, model monitoring, and risk assessment. Effective ML model development calls for parallel experimentation, hyperparameter optimisation, code and model repositories, and collaboration among data scientists for version control and scalability ( \textbf{CM39} [LR14, 27, 28]).

The growing popularity of ML has given rise to versioning tools, but these tools must be synchronised with other software artefacts [LR14]. Deploying ML models in real-world situations can be time-consuming and fraught with challenges, including the risk of failure due to a lack of expertise, data bias, and high costs. Maintaining and enhancing deployed models is essential for cost-effectiveness, but there is a scarcity of tools to support the model lifecycle ( \textbf{CM40} [LR27]). 

ML models can be valuable business resources with potential benefits, but their efficacy relies on reproducibility, accessibility, and management. A framework is necessary for constructing and deploying ML-based software analytics and business intelligence solutions, which often become stuck in a prototypical stage due to complex setup and maintenance ( \textbf{CM40, CM41} [LR31]). DNNs introduce challenges in AI system engineering, including data quality assurance, expert knowledge integration, and testing and debugging ( \textbf{CM39} [LR47]). A unified environment is needed for prototyping, deploying, and maintaining ML models at scale, addressing advanced feature engineering, real-time data handling, and support for multiple frameworks ( \textbf{CM40} [LR48]). Although cloud-based AI platforms have simplified AI adoption, they often lack the complexity and flexibility required by practitioners ( \textbf{CM38} [LR49]). Collaborative ML environments also encounter coordination and maintenance issues with pipeline changes and updates, necessitating version control semantics and efficient storage ( \textbf{CM40} [LR50]). Data scientists frequently struggle to find complete systems for specific tasks, leading to intricate and fragile pipeline jungles ( \textbf{CM40} [LR51]).

Offering intelligent ML strategies to product engineers is challenging for several reasons. First, the product's long-term objectives might need to align with ML's loss functions. Second, data collected from the product may deviate from the training data. Third, correlations identified by the ML system do not always imply a causal link to improve product metrics. Fourth, assessing an ML system's performance may necessitate A/B testing due to the absence of clean data. Fifth, traditional A/B testing often overlooks personalised treatments. Sixth, an ML platform is essential to train, host, and monitor numerous models to achieve economies of scale. Seventh, real-time feature extraction and inference are required, despite more efficient asynchronous batch processing. Lastly, product engineers, especially those who are new to ML, need a simple, standardised, and future-proof way to incorporate innovative strategies into products ( \textbf{CM38, CM41} [LR70]).

\textit{Summary: Building and managing ML systems present multifaceted challenges, encompassing issues like data acquisition, reproducibility, model complexity, deployment, monitoring, and coordination. These challenges extend beyond algorithms to the entire lifecycle, requiring comprehensive understanding, aligning with the business objectives, collaboration, and tools for version control, scalability, real-time handling, and integration into products, often complicated by factors such as data quality and scarcity, high costs, and the need for specialised expertise.}

\subsubsection{\textbf{Building ML System Solution }}

A McKinsey global survey highlights a 25\% YoY growth in machine learning, with streaming-based architecture replacing Data-Oriented Architecture (DOA) to address micro-service data flow issues. Tools like DVC, MLFlow, Pachyderm, ModelDB, and Quilt Data aid data engineers and scientists in versioning data and models, while the Technology Readiness Levels for ML (TRL4ML) framework helps prioritise data in ML projects ( \textbf{SM38} [LR14]). Frameworks have been proposed for constructing data pipelines using self-supervised, semi-supervised, or representation learning algorithms [LR24]. Overton is a software architecture enabling deep learning application development, maintenance, and monitoring through data file manipulation  ( \textbf{SM39} [LR25]).

Building ML models in fintech applications necessitates a comprehensive understanding of governance, risk management, compliance [27], and to address this complexity, a framework has been proposed that supports the entire lifecycle of ML models, from data collection to retraining deployed models [46], while ongoing research is actively tackling challenges such as Automated and Continuous Data Quality Assurance, Domain Adaptation, Hybrid Model Design, Interpretability, Software Quality, ALOHA Toolchain, and Human AI Teaming, all of which are integral components in constructing a robust and efficient ML system [47].

Flame, a web application, utilises biologically annotated chemical structures for model creation, offering advanced management tools and supporting the entire development cycle ( \textbf{SM39} [LR31]). ML platforms such as TFX, KubeFlow, MLflow, H2O, Skymind Intelligence Layer, Uber's Michelangelo, Facebook's FBLearner Flow, Data Robot, Polyaxon, Comet, Atalaya, Amazon's SageMaker, and Microsoft Azure Machine Learning are available [LR48]. ThunderML is a Python toolkit for industrial AI model creation and deployment ( \textbf{SM39} [LR49]). MLCask supports versioning for ML pipelines as an end-to-end analytics system ( \textbf{SM39} [LR50]).

Selecting the suitable algorithm for specific use cases can take time for non-experts [LR60]. KOI is a software architecture for managing the lifecycle of iteratively trained ML systems ( \textbf{SM39} [LR65]). The Looper ML platform offers simple APIs and architecture covering the entire ML lifecycle, allowing for the widespread adoption of ML models in software systems. Companies like Google, Meta, LinkedIn, and Netflix are adopting data-centric, vertical ML platforms that provide end-to-end solutions without code through high-level, declarative abstractions ( \textbf{SM39} [LR70]).

Flow-Based Programming (FBP) and Model-Driven Software Development (MDSD) simplify ML application development using FBP and MDSD principles. By modelling Spark ML's Java APIs as composable components, non-programmers can create ML programs graphically with a higher level of abstraction ( \textbf{SM38} [LR81]). Products such as WEKA, Azure Machine Learning Studio, KNIME, Orange, BigML, mljar, RapidMiner, Streamanalytix, Lemonade, and Streamsets support high-level ML programming ( \textbf{SM38} [LR81]).

Deploying ML-based projects on GitHub is challenging due to data issues, computationally intensive training, and real-time deployment requirements. Openja et al. [LR82] studied the use of Docker in deploying 406 open-source ML projects, finding Docker's standard functionalities include data management, interactive development, task scheduling, and model management, with applications in logging, monitoring, cloud-based development, model management, and CI/CD ( \textbf{SM38} [LR82]).

CompareML offers a unique way to evaluate and compare ML libraries and services, enabling practitioners to select the best algorithm and provider for their data without requiring deep ML knowledge ( \textbf{SM38} [LR84]).

Cirrus is an ML framework that automates data centre resource management for ML workflows using serverless infrastructures like AWS Lambdas and S3. It is designed for serverless computation and iterative ML training. It provides an easy-to-use interface to perform scalable ML training leveraging the high scalability of serverless computation environments and cloud storage. It overcomes the resource constraints of serverless computing by using an ultra-lightweight worker runtime, streaming training mini-batches, and a stateless worker architecture. Cirrus outperforms both serverless and specialised ML frameworks in terms of time-to-accuracy ( \textbf{SM39} [LR91]).

\begin{table*}[htbp]
  \centering
  \begin{tabular}{p{1.5cm} p{7cm} p{2.5cm} p{3cm}}
    \toprule
    Challenge no  & Description of the challenges & ML  stage & Paper \\
    \midrule

 CM32 &Proprietary AutoML libraries and systems. &AutoML & LR51, 60, 61, 62, 64 \\
 CM33 &Gaps in AutoML services. &AutoML & LR63,80 \\
&  \\
 CM34 &Achieving AI software quality &ML System quality & LR10,57,56, 59, 73, 85,87 \\
CM35 & impact of human factors on ML model quality. &ML System quality & LR38,54, 75 \\
&  \\
 CM36 &Model deployment and integration challenges &MLOPS & LR28, 52,53 \\
 CM37 &Challenges in implementing MLOPS tools and workflows. &MLOPS & LR53, 74 \\
&  \\
 CM38 &Developing ML models. &Building ML system & LR10,11,49, 70 \\
 CM39 &Deploying ML models in real-world scenarios. &Building ML system & LR11,20, 27,47, 14,28 \\
CM40 &Maintaining deployed models. &Building ML system & LR27,51,50,31,48 \\
CM41 &Aligning ML models with long-term product goals. &Building ML system & LR31,70 \\
  \bottomrule
  \end{tabular}
  \caption{Maintainability challenges found from the LR study and the paper discussing these challenges 
 for a given ML stage in the workflow}
  \label{tab:example141}
\end{table*}

\begin{table*}[htbp]
  \centering
  \begin{tabular}{p{1.5cm} p{1.5cm} p{10cm}}
    \toprule
    Index & Solutions no & Solution/Tools/frameworks/methods \\
    \midrule

CM32 &SM32 &AutoBazaar[LR51]. Meta-Learning approaches, Auto-Weka , Auto-Sklearn and [LR60] , Auto-Keras [LR61], \\
CM33 &SM33 &AutoBazaar[LR51], Katib, Dataiku AutoML, H2O AutoML, Google cloud AutoML, Ultron,   Properteriay WALTS [LR80] , Auto-Keras [LR61] \\
& & \\
CM34 &SM34 &MLCask [LR50] , CRISP-ML(Q)[LR85], QA4AI [LR85] \\
CM35 &SM35 & Recommendation from [LR75], CRISP-ML(Q) [LR85]
 \\
CM36 &SM36 &TensorFlow Extended (TFX), ModelOps, and Kubeflow [LR52] \\
CM37 &SM37 &CD4ML[LR52], MLFlow, Amazon SageMaker, Spack, and EasyBuild, [LR74] \\
& & \\
CM38 &SM38 &DVC, MLFlow, Pachyderm, ModelDB, and Quilt Data [LR14], high-level ML programmings, such as WEKA, Azure Machine Learning Studio, KNIME, Orange, BigML, mljar, RapidMiner, Streamanalytix, Lemonade, and Streamsets [LR81], Technology Readiness Levels for ML [LR20], Weka, Orange, RapidMiner, and Knime, PowerBI, PyCaret, and Cloud AutoML [LR84] \\
CM39 &SM39 &Overton[LR25], Flame [LR31], ThunderML [LR49], MLCask[LR50],KOI [LR65], Looper [LR70], Cirrus[LR91] \\
CM40 &SM40 &TFX, KubeFlow, MLflow, H2O, Skymind Intelligence Layer, Uber's Michelangelo, Facebook's FBLearner Flow, Data Robot, Polyaxon, Comet, Atalaya, Amazon's SageMaker, and Microsoft Azure Machine Learning [LR48] \\
CM41 &SM41 & No good solution found in this SLR \\
  \bottomrule
  \end{tabular}
  \\
  \caption{ Maintainability solution for the identified Challenges that we indexed for each ML stage}
  \label{tab:example1113}
\end{table*}

\subsection{RQ 4. Data Engineering  Scalability  Challenges and Solutions}

\subsubsection{\textbf{Data Engineering Scalability Challenges}
}

Scalability in data engineering is paramount for managing large-scale data analysis, machine learning, and real-time applications. To ensure scalability, current solutions utilise High-Performance Computing (HPC) and cloud infrastructures; however, exascale systems will soon be essential for handling extreme-scale data analysis, requiring scalable data mining solutions, advanced analysis tools, and apps that operate on architectures with dependable storage and high-performance processors (\textbf{CS01} [LR92]). Parallel data access can enhance data speeds, but fault resilience concerns arise with a scale which is still an open challenge (\textbf{CS01} [LR92]). 

Effective model training demands vast amounts of production data located in close proximity for optimal performance. A scheduling system is crucial for maintaining a consistent machine learning platform, while processes for managing new training data and monitoring frameworks to detect and address failures are necessary (\textbf{CS07} [LR101]).In addition, developing applications for large and diverse core systems requires addressing challenges such as energy consumption, multitasking, scheduling, reproducibility, and resilience. Programming constructs and runtime systems will be vital for enabling data analysis programming models, runtime models, and hardware platforms to tackle these challenges and support genuine large-scale data analysis applications (\textbf{CS13} [LR92]).

Companies like Facebook leverage both GPUs and CPUs to cater to the intense computational demands posed by various machine learning models, emphasising the need for novel parallel computing strategies for real-time analysis from diverse data sources (\textbf{CS01, CS02} [LR95]). Furthermore, Large-scale online services face cold start and sparse data challenges due to new content and short item lifespans in consumer-to-consumer platforms (\textbf{CS03 } [LR101]). Inadequate data structures and the lack of a taxonomy for metadata related to user content exacerbate these difficulties. Obtaining feature values in real-time is also challenging, as data is often dispersed, synthesised, and must be aggregated before use (\textbf{CS02} [LR101]).

Lastly, the quality and quantity of training data are crucial to achieve optimal performance in machine learning. Nonetheless, collecting and processing large-scale real data is expensive, challenging, and raises privacy and legal concerns (\textbf{CS04}  [LR113]). Synthetic data presents a promising solution to these challenges, as it is cost-effective, offers rich annotations, and provides complete control over data, which can alleviate privacy, bias, and licensing issues. However, the tools for effective synthetic data generation are less mature than those for architecture design and training, resulting in fragmented generation efforts (\textbf{CS04} [LR109]).

\textit{Summary: Addressing the challenges of scalability in data engineering involves managing large-scale data analysis, developing scalable solutions for exascale data systems, ensuring optimal machine learning model training, and addressing issues of energy consumption, multitasking, and resilience across diverse core systems, with companies like Facebook requiring novel parallel computing approaches and synthetic data emerging as a cost-effective alternative despite its current limitations.}

\subsubsection{\textbf{Data Engineering Scalability Solutions}
}

The surge in extensive data analysis has led to increased adoption of cloud-based systems for scalable analysis. This has given rise to a new Data Analysis as a Service (DAaaS) model, which offers data analysis solutions through a service-oriented approach. DAaaS consists of three models: DAIaaS, DAPaaS, and DASaaS, which collectively support the structured development of extensive data analysis systems and applications, such as the Data Mining Cloud Framework (DMCF) (\textbf{SS01} [LR92]).

Delight is an end-to-end framework for performing deep learning tasks in resource-constrained settings. It adaptively learns and customises the hybrid data structure to improve memory footprint, energy consumption, and runtime and provides a user-friendly API for rapid prototyping  (\textbf{SS01} [LR100]). Another solution is Cylon, which is a high-performance, distributed memory data parallel library for structured data processing that outperforms other big data frameworks (\textbf{SS01} [LR8]). 

Innovative parallel computing approaches are essential for real-time analysis and prediction using vast and diverse data sources. Technologies like Kafka, Kafka-ML, Amazon Kinesis, Kafka Streams, and Faust have emerged to handle data ingestion, processing, and integration with ML/AI  (\textbf{SS02}  [LR97, LR121]).

In large-scale online services, cold start and sparse data issues arise due to new content and short item lifespans on consumer-to-consumer platforms. The challenges are compounded by poor data structures and the lack of metadata taxonomy linked to user content. A scheduling system is required to ensure consistency and quality in the machine learning platform, with processes in place to manage new training data and a monitoring framework to detect and handle failures. Incremental training or warm starting, which leverages previous models, should be utilised to avoid starting training from scratch [LR101]. Using data from a surrogate task is also critical to mitigate cold start and sparse data issues. Warm starting initialises the new model with the neural network and weights from a previous model  (\textbf{SS03} [LR101]).
Furthermore, Transfer learning by using pre-trained models can provide solutions for data-hungry deep neural networks [LR118]. Self-supervised learning is a branch of unsupervised learning that leverages input data itself as supervision to extract knowledge from large-scale unlabeled data. Self-supervised pre-training has become a focus of current AI research, especially for NLP tasks where annotating a textual dataset as large as ImageNet is nearly impossible. Pre-trained models from one task can be used to perform well on another task, reducing the need for large amounts of annotated data. Efficiency in large-scale pre-trained models can be improved through system-level optimisation, efficient learning algorithms, and model compression strategies  (\textbf{SS03}  [LR118]). 

The quality and quantity of training data are paramount to achieving good machine learning performance. However, collecting and processing data at a large scale is expensive, challenging and raises privacy and legal concerns [LR113]. Furthermore, techniques like Federated Learning (FL) and Split Learning (SL) are two methods that enable individual data owners to train a shared model over their joint data without exchanging it  (\textbf{SS04} [LR113]). Solutions to address these challenges include using synthetic data, transfer learning, and proper validation and verification processes. Synthetic data can be generated by software programs to mimic real-world data, reducing the need for large amounts of real-world data. Kubric is an open-source Python framework that generates photo-realistic scenes with rich annotations and seamlessly scales to large jobs distributed over thousands of machines. Kubric supports reuse, replication, and shared assets and has a standard export data format for porting data into training pipelines  (\textbf{SS04} [109]). Transfer learning involves using pre-trained models from one task to perform well on another task, which can reduce the amount of data required to train a machine learning model  (\textbf{SS04} [118]). Validation and verification processes are crucial for ensuring the quality of big data applications. The CMA system is highlighted as an example of a framework for rigorously validating image data and verifying software tools and machine learning algorithms. By implementing these solutions, accurate results can be obtained, and risks associated with using machine learning in sensitive areas such as healthcare and biology can be reduced  (\textbf{SS04} [102]).

\subsection{RQ 5. Model Engineering Scalability  Challenges and Solutions}

\subsubsection{\textbf{Model Engineering Scalability Challenges}
}

Scalability in Model Engineering is a multifaceted challenge encompassing data scaling, model size, and count. These complexities give rise to challenges such as cold start, sparse data, distributed ML model trade-offs, efficiency-effectiveness balance, continuous retraining, and ML serving infrastructure design [LR101]. The scalability of architectures and application frameworks are often use-case-specific and lack standardised deployment methodologies [LR95,96]. Configuring the vast array of hyperparameters for peak performance is cumbersome. Moreover, understanding and deciphering vast datasets, discerning patterns, and extracting insights for non-technical users have become daunting. The intricacies of the extensive data infrastructure further escalate the difficulty in configuring and conducting these processes (\textbf{CS05} [LR99]). 

Distributed model training splits the neural network graph or dataset into multiple sub-graphs or parts to be trained simultaneously on multiple machines, requiring a high-performance resource and job orchestration system to reduce communication and memory costs  ( \textbf{CS09} [LR101].

Training deep neural networks (DNNs) is time-consuming, and deploying tasks on multiple nodes can lead to communication overhead, becoming a bottleneck in distributed training. High-performance hardware accelerators can reduce computation overhead, but communication overhead remains an issue (\textbf{CS06} [LR116]). Furthermore, communication overhead increases exponentially with cluster scale growth, undermining parallel training benefits. Additionally, ML systems need end-to-end software support for various components affecting data provenance, model training, and inference (\textbf{CS06} [LR123]).

Finding optimal hardware and software configurations for ML workloads on HPC clusters poses a challenge. This is evident from a study focused on the TensorFlow workflow for image recognition, which emphasised the importance of optimised arithmetic libraries tailored for specific architectures and the integration of Arm Performance Libraries into TensorFlow. When evaluated on an HPC cluster with Marvell ThunderX2 CPUs, it was evident that the performance during the training phase largely depended on these factors ( \textbf{CS11} [LR107]).

Similarly, on-device learning has its own set of challenges. The most prominent are insufficient user data, back-propagation difficulties during model updates, and peak processing speed limitations. Addressing these makes the real-world implementation of on-device learning frameworks a formidable task, thus necessitating further studies (\textbf{CS09 } [LR106]).

Distributed ML training can generate significant data traffic, which, if not appropriately managed, can result in network saturation and disruption. To mitigate this, techniques like compression and scheduling algorithms have been introduced. It is worth noting that during off-peak hours, an abundance of idle servers exists, providing a rich source of computational resources for distributed ML training. However, using these resources comes with its challenges. Specifically, Schedulers face challenges in balancing loads across heterogeneous hardware while considering network topology and synchronisation costs, which can impact training speed and quality. ML applications require the separation of data and training workloads due to differing characteristics (\textbf{CS07} [LR95,96,101]).

There has been a noticeable shift in executing ML workloads on modern hardware like GPUs, veering away from MapReduce-based systems [LR94]. However, deploying research-generated ML models can be challenging due to the need to modify the training pipeline and longer deployment cycles caused by differences between frameworks, making regular retraining and incremental training or warm starting essential (\textbf{CS08} [94, 101]). 

The long-term viability of distributed machine learning depends on addressing many open challenges, including performance, fault tolerance, privacy, and portability. While federated machine learning is making strides in especially managing input privacy issues, there is a scarcity of open frameworks that effectively address distributed computing challenges like scalability and resilience (\textbf{CS09} [LR111]).

Edge computing environments in distributed machine learning bring forth limited capacities, energy conservation, data privacy, and communication bandwidth issues. In particular, the communication overhead is a significant obstacle (\textbf{CS09}  [LR112]).

Modern statistical ML experiences a tug-of-war between accuracy and efficiency. There is always a balance to strike between correctness and performance. This balance becomes incredibly delicate in scenarios like real-time distributed ML systems in autonomous vehicles, where there is a constant play between consistency, latency, and accuracy. Additionally, devices constrained in terms of resources grapple with limitations in local computational abilities. Offloading tasks to remote systems carries its speed costs [LR122]. Such challenges are even more pronounced in ML systems for innovative services due to the limitations of IoT devices. Achieving scale in distributed machine learning operations for edge intelligence is essential but complex due to the IoT's heterogeneous and geographically dispersed nature, requiring careful consideration of energy efficiency, privacy, and real-time inference requirements (\textbf{CS09}  [LR122,123]).

\textit{Summary: Model Engineering scalability faces challenges in data scaling, model size, and count. Distributed training, though essential, brings about overheads and complexities, especially in configuring and optimising systems. On-device learning and edge intelligence present unique data insufficiency and energy efficiency hurdles. Effective hardware and software configurations for ML workloads remain elusive, with a noticeable shift towards modern hardware. Addressing these multifaceted challenges is critical for the sustainable evolution of distributed machine learning systems.}

\subsubsection{\textbf{Model Engineering Scalability Solution}}

Shahoud et al. [LR60] developed a new framework that applies meta-learning to automate model selection and parameter tuning in Big Data environments. The framework is crafted to be modular, scalable, and generic, employing container and microservice technologies to support various model selections, emphasising time series model selection. Ultron-AutoML is an open-source HPO (hyperparameter optimisation) framework designed to prioritise user-friendly customisation and ease of use. It automates multiple processes, such as containerisation, script execution, model checkpointing, and parallelisation, enabling users to efficiently design and execute HPO tasks while retaining flexibility (\textbf{SS05} [LR60, LR110 ]).

In the HPC ecosystem, several parallel/distributed software frameworks, including Apache Mahout [LR94] and MLlib [LR112], provide scalable machine learning algorithms for diverse applications like recommendation mining, clustering, classification, and frequent itemset mining (\textbf{SS08} [LR112, LR94 ]). 

Bighead [LR48] is another solution, offering a cost-effective and scalable end-to-end ML platform that supports major frameworks and manages the entire model lifecycle (\textbf{SS08}).

The ASML framework aims to solve the open research question of deploying end-to-end ML pipelines in a generic and standardised manner. It presents technical details and application scenarios to serve as examples for implementing other ML pipelines, along with a performance evaluation to showcase its efficiency (\textbf{SS08} [LR96]). A comparative study of TensorFlow Serving, SageMaker, Clipper, and DLHub determined that TensorFlow Serving was superior in terms of invocation time and request time, making it a more efficient option for specific use cases (\textbf{SS08} [LR119]). 

Transfer learning and self-supervised pre-training techniques help in addressing the challenges of creating large annotated datasets for deep neural networks and can also improve efficiency of large-scale pre-trained models through system-level optimisation, which includes single-device and multi-device optimisation methods; efficient learning algorithms, which involve techniques like replaced token detection, warmup strategies, and sharing self-attention patterns; and model compression methods, such as parameter sharing, model pruning, knowledge distillation, and model quantisation, which can help in eliminating the need to deploy the new model and avoid some cost-effective retraining (\textbf{SS09 } [LR118]).

Deep learning models have become increasingly complex and resource-intensive, necessitating various techniques and tools to overcome these challenges. Hardware components, such as GPUs, FPGAs, TPUs, and neuromorphic hardware, provide acceleration for training and inference, while frameworks like cuDNN and NCCL enable efficient parallelisation strategies  (\textbf{SS06} [LR103]). Parallelisation techniques, including data, model, pipeline, and hybrid parallelism, allow the training of large deep learning models on distributed infrastructure {(\textbf{SS06} [LR103]). Popular deep learning frameworks, such as CNTK, Deeplearning4j, Keras, MXNet, PyTorch, SINGA, and TensorFlow, support distributed and parallel training, offering customisation options for layers, loss functions, and operators (\textbf{SS06} [LR119]). The Switch Transformer model is a technique that optimises the Mixture of Expert (MoE) routing algorithms, reducing computational costs and simplifying routing implementation for improved performance (\textbf{SS06} [LR115]). The ZeRO optimisation technique minimises memory redundancies in data- and model-parallel training, enhancing memory efficiency and computational granularity (\textbf{SS06} [LR116]).DeepSpeedMoE is an end-to-end MoE training and inference solution that employs novel architecture designs and model compression techniques to reduce the model size and improve inference latency and cost. This solution enables the serving of massive MoE models with faster and cheaper inference compared to quality-equivalent dense models (\textbf{SS06} [LR117]).

 Cloud providers, such as Google, Microsoft, Amazon, and IBM, offer machine learning as a service, making it more accessible to a broader audience without the need for costly infrastructure or specialised staff (\textbf{SS06} [LR120]). This accessibility fuels the development of new systems and approaches contributing to the machine-learning ecosystem, making it easier for small businesses and individuals to adopt machine-learning technology.   

Profound learning scheduling challenges can be addressed with various solutions, such as heuristic algorithms that help determine the mapping of processes to resources and when to execute them and elastic ML frameworks that dynamically allocate resources to optimise utilisation and reduce completion times. Decentralised data-parallel systems are used for improving scalability in large-scale distributed DL systems, while specialised resource schedulers focus on DL-specific optimisation goals  (\textbf{SS07} [LR119]). Cloud providers simplify the process for developers by offering machine learning as a service and pre-packaged ML software like TensorFlow, MXNet, and PyTorch to abstract away all the complexities related to balancing and scheduling in a distributed cluster (\textbf{SS07} [LR120]).

Kafka-ML is an open-source framework that provides a comprehensive solution for managing data streams and integrating them with ML/AI pipelines. It offers a web UI for training and inference. It supports popular ML frameworks like TensorFlow, making it an attractive choice for businesses and individuals seeking to leverage machine learning and artificial intelligence technologies (\textbf{SS08} [LR121]).

Various factors can enhance distributed deep learning, including architectural approaches, algorithm-level optimisations, gradient compression, and low-level network infrastructure optimisations (\textbf{SS06, SS09} [LR114]).
To ensure efficient distributed machine learning, several techniques are employed to balance workloads and reduce computation time, including Bulk Synchronous Parallel (BSP), Stale Synchronous Parallel (SSP), Approximate Synchronous Parallel (ASP), Barrierless Asynchronous Parallel (BAP), Total Asynchronous Parallel (TAP), and Hybrid Communication (HybComm) (\textbf{SS07, SS09} [LR120]). 

An automated software framework for distributed deep learning called Blind Learning (BL) allows for optimal model training from distributed datasets over multiple clients, involving parallel training of all clients, server model updating once per training round, and synchronisation of client-side models after each training round. This approach addresses the challenges of centralised sharing of healthcare data in deep learning  (\textbf{SS09} [LR113]).

FEDn, a hierarchical federated learning framework, is designed for real-world testing and industrial applications [LR111]. This scalable framework supports cross-device and cross-silo use cases, allowing privacy-enhancing technologies like differential privacy and multiparty computation to be integrated. This flexibility enables FEDn to be utilised in various applications, from production-grade tools to research purposes, exploring the performance of novel federated algorithms (\textbf{SS09} [LR111]). 
To tackle challenges in executing machine learning algorithms in distributed IoT environments, a four-layer edge-cloud continuum architecture is proposed, consisting of device, edge, fog, and cloud layers [LR112]. This architecture minimises data movement, ensures load balance and fault tolerance, respects application and system requirements, and employs simulation tools for testing IoT systems and networks. A practical implementation is a bagging-based ensemble learning algorithm for image classification, which trains weak learners on edge devices and aggregates them at higher levels to generate a more accurate global model (\textbf{SS09} [LR112]). Furthermore, When implementing decentralised algorithms on IoT systems and networks, it is essential to consider various aspects such as data location, geographical distribution, algorithm features and configurations, task scheduling, data persistence, application constraints, and coding and testing. 

\subsection{RQ 6. Current Ecosystem  Scalability Challenges and Solutions}

\subsubsection{\textbf{Current Ecosystem Engineering  Challenges}}

Amidst the ever-evolving technology landscape, the challenges of ensuring scalability in the current ecosystem are multifaceted. Integrating efficient support for ML workloads on HPC clusters demands optimal hardware and software configurations, as emphasised by a study involving TensorFlow's image recognition workflow, where training phase performance hinged on tailored arithmetic libraries for specific architectures and the incorporation of Arm Performance Libraries into TensorFlow (\textbf{CS12} [LR107]). Although distributed machine learning is becoming increasingly popular, many open challenges are still crucial to its long-term success. These challenges include performance, fault tolerance, privacy, and portability  (\textbf{CS12} [LR120]).

Traditional HPC systems using  Hadoop and MapReduce paradigms need to be better-suited for machine learning tasks due to their limitations in handling iterative algorithms, frequent data access and modification, and limited ability to exploit the potential of modern hardware like GPUs fully. Therefore, there has been a shift towards new platforms and frameworks such  as TensorFlow and PyTorch, which are designed specifically for machine learning and can run on a variety of hardware platforms, including GPUs, to provide better support for modern machine learning applications., Apache Mahout is a library for machine learning that can be used in distributed dataflow systems. However, the Hadoop platform has been redesigned to support other parallel processing paradigms   (\textbf{CS12} [LR94]). In addition, Real-time analysis and prediction of big data requires a novel parallel computing approach. Popular options are Apache Kafka and Flume for ingestion and Spark Streaming and Storm for stream processing. However, their use in ML is limited  (\textbf{CS12} [LR97]). 

Furthermore, many popular libraries for data science are written in Python, an interpreted programming language with high overheads, resulting in performance trade-offs. To address this, highly performant code written in C or C++ is added to frameworks to create domain-specific libraries that offer both productivity and high performance [LR110]. Past works in this area have focused on either using HPC frameworks underneath existing data science and machine learning frameworks or building new frameworks with HPC technologies. However, some efforts may miss optimisations that are only possible when higher-level semantics are exposed (\textbf{CS13} [LR110]). 

Ensuring cost efficiency in large-scale machine learning/deep learning (ML/DL) workflows can be complex due to hardware costs, time and resource constraints, algorithm complexity, and data costs (\textbf{CS11} [LR106]). There is still a  mismatch between tensor-based and non-tensor-based applications in implementing ML on dataflow systems,  dynamic neural networks has the potential to reduce its impact, but it is still in the research phase and not yet suitable for production and mobile deployment, but transfer to production can be facilitated through the ONNX toolchain [LR95]. 

Cloud computing for edge intelligence has some concerns due to concerns like privacy and resource costs. On-device learning is a solution but still needs more user data, backward propagation blocking, and limited processing speed. Cloud computing may not be suitable for edge-intelligent features due to privacy and latency concerns [LR106]. Overall implementing on-device learning systems can be challenging and highlight issues with data insufficiency, model overfitting, and hardware limitations  (\textbf{CS12 } [LR106]).

To manage large datasets, many large-scale machine learning (ML) implementations use multiple servers or virtual machines (VMs) that process local data and synchronise periodically. Managing this infrastructure is complex, so many companies use external cloud providers to run their ML jobs. However, cloud resources can be costly, especially for long-running large ML jobs ( \textbf{CS11} [LR108]).

Lastly, ensuring reproducibility in large-scale ML systems is challenging due to data volume and distribution, complex training pipelines, platform and framework dependencies, and resource availability (\textbf{CS10} [LR98]), and the quest for an effective debugger system is hindered by a limited understanding of ML diagnosis and diverse sources of errors, resulting in ad-hoc and non-reusable ML serving infrastructure solutions  (\textbf{CS10} [LR104]).

\textit{Summary: Optimising ML on HPC clusters demands specialised configurations, but distributed ML has unresolved challenges. Popular ML tools often prioritise portability over performance. Large-scale ML setups are complex, pushing many towards costly cloud resources. On-device learning addresses cloud limitations but faces its challenges. Reproducibility and effective debugging in ML remains problematic due to multiple dependencies and diagnosis complexities.}

\subsubsection{\textbf{Current ecosystem  Engineering  Solutions}}

The ONNX (Open Neural Network Exchange) offers interoperability between various machine learning frameworks and vendor-optimised libraries  (\textbf{SS13 } [LR95]). Container technologies like Docker enable replication of runtime environments with all dependencies in container images [LR95]. Curious Containers project enhances reproducibility by providing applications as container images, loading data directly into containers, and storing information in a single file  (\textbf{SS10, SS11} [LR98]). Other frameworks for efficient training and managing ML models using big data software and microservice architecture with user-friendly interfaces have been discussed b Shahoud et al.  (\textbf{SS11} [LR99]).

Model management is crucial, involving tracking, storing, and indexing trained models for sharing, querying, and analysing. ONNX facilitates interoperability between different DL frameworks, while systems like ModelDB and ModelHub provide automatic tracking, indexing, and querying of ML models  (\textbf{SS10, SS11} [LR119]) . DistRDF2ML is a framework for creating explainable ML pipelines for knowledge graphs using Apache Spark, Jena, and SANSA Stack, with reduced processing time by increasing resources and memory  (\textbf{SS11} [LR104]).

Lwakatare et al. suggest using data from surrogate tasks to address cold start and sparse data issues, highlighting TensorFlow's ease of use in heterogeneous distributed systems. However, the black-box nature of function calls in deep learning graphs complicates the debugger design, so a TensorFlow debugger was developed  (\textbf{SS10} [LR101]).

ADCME incorporates MPI for parallelising numerical solvers in distributed memory HPC architectures, providing small overhead and custom parallel algorithm flexibility. JVM-based solutions like Apache Spark are popular in ML production systems  (\textbf{SS12} [LR93]) .

Mayer et al. [LR119] discuss deep learning (DL) infrastructure, including hardware components (GPUs, FPGAs, TPUs, and neuromorphic hardware), large-scale infrastructure, and low-level libraries like cuDNN and NCCL. Performance is measured by throughput, latency, and energy consumption, with evaluations conducted on GPU interconnects and FPGAs vs. GPUs. Tools like IBM Fabric for Deep Learning (FfDL), Djinn, Tonic, Ray, and Ease.ml are also mentioned  (\textbf{SS12} [LR119]).

Amazon SageMaker provides scalable, elastic model training on large data streams, using a streaming model for linear update time, pause/resume capabilities, and elasticity. Despite mathematical complexity, the streaming model offers significant advantages, including cost-effective processing, fixed memory footprint, and near-linear scalability for runtime and compute costs (\textbf{SS11, SS12} [LR124]). It reduces training time and cost for customers and uses the MXNet library to operate seamlessly across CPUs and GPUs. The system's streaming model enables beneficial techniques for hyperparameter optimisation workloads, such as adding more data to the model every day and stopping models that do not perform well early on (\textbf{SS11, SS12} [LR124]). A study comparing TensorFlow Serving, SageMaker, Clipper, and DLHub in terms of performance found that TensorFlow Serving had the best performance in terms of invocation time and request time (\textbf{SS11} [LR119]).

Tasgetiren et al. proposed a hybrid distributed computing architecture for the banking industry, employing the facade design pattern to reduce complexities and including components for data preprocessing, model training, and reporting. The prototype demonstrated negligible processing load and compatibility with multiple subsystems without code changes (\textbf{ SS12} [LR105]).

Programming constructs and runtime systems are essential for addressing challenges in big data analysis applications [LR92]. Efforts have been made to apply high-performance computing (HPC) technologies to data science frameworks, creating domain-specific libraries that balance productivity and performance [LR110]. Various approaches exist for high-performance data science and machine learning frameworks. Some prioritise programmability and portability over performance, while others utilise HPC frameworks under existing data science frameworks or build new software stacks using HPC technologies [LR110]. Examples include Legate Numpy, which provides a numpy-like interface on top of the Legion programming model, and Arkouda, built on Chapel [LR110]. However, these still need to consider the existing Python data science ecosystem fully. SHMEM-ML is a domain-specific library for distributed array computations and machine learning, leveraging HPC software stack performance to accelerate workflows. It targets the complete machine learning workflow, offering a general-purpose nd-array abstraction for efficient distribution, transformation, and manipulation. OpenSHMEM serves as the communication layer, enabling high-performance networking across distributed processes. SHMEM-ML interoperates with the Python machine learning ecosystem, using Apache Arrow, and achieves up to 38x speedup in distributed training performance compared to the Horovod framework without model metric regression (\textbf{SS13} [LR110]).

\begin{table*}[htbp]
  \centering
  \begin{tabular}{p{1.5cm} p{7cm} p{2.5cm} p{3cm}}
    \toprule
    Challenge no  & Description of the challenges & ML  stage & Paper \\
    \midrule

CS01 &Managing  Data pipeline and its infrastructure for large-scale ML is complex &DE / ME & LR92,95 \\
& & & \\
CS02 &Real-time analysis from large, diverse data pools requires new computing approaches. &DE  & LR95,101 \\
& & & \\
CS03 &Sparse data and lack of metadata taxonomy create cold start challenges in large-scale online services. & DE / ME & LR101 \\
& & & \\
CS04 &Collecting and processing data on a large scale raises privacy and legal concerns. &DE & LR109,113 \\
& & & \\
CS05 &Configuring optimal hyperparameters is time-consuming. &ME/ CEC & LR99 \\
& & & \\
CS06 &Distributed model training and its drawbacks which diminishes parallel training advantages. &ME & LR116,123 \\
& & & \\
CS07 &Balancing and scheduling load across heterogeneous hardware can affect training speed and quality. &ME / CEC  & LR95,96,101 \\
& & & \\
CS08 &Deploying and serving DL models requires modifying training pipelines and regular retraining. &ME & LR94,101 \\
& & & \\
CS09 & training in constrained environments pose challenges, particularly with privacy-sensitive data. &ME /  CEC  & LR111,112.122,123, \\
& & & \\
CS10 &Understanding ML diagnosis, reproducibility, and diverse error sources are challenging. & ME / CEC & LR98.104 \\
& & & \\
CS11 &Managing infrastructure for large-scale ML workflows is complex and costly. &CEC / ME  &LR106,107 \\
& & & \\
CS12 &Efficiently supporting ML workloads on clusters requires optimal hardware and software configurations. & CEC & LR94,97,107,120 \\
& & & \\
CS13 & programmability and portability prioritisation in ML models and frameworks can lead to performance trade-offs.& CEC & LR110 , 92  \\

  \bottomrule
  \end{tabular}
  \caption{Scalability  challenges found from the SLR study and the paper discussing these challenges }
  \label{tab:example1412}
\end{table*}

\begin{table*}[htbp]
  \centering
  \begin{tabular}{p{1.5cm} p{1.5cm} p{10cm}}
    \toprule
    Index & Solutions no & Solution/Tools/frameworks/methods \\
    \midrule

CS01	&SS01	&DeLight [LR100], Data Mining Cloud Framework (DMCF), Data Analysis as a Service (DAaaS)  [LR92]. Cylon [LR8]  \\
		
CS02	&SS02	&Kafka [LR97] , Kafka-ML [LR121], Amazon Kinesis [LR121], Kafka Streams , Faust and Spark [LR121]  Apache SAMOA, Apache Flink, Apache Spark, Spark Streaming and  TensorFlow Serving  [LR121]\\

CS03 & SS03&Warm starting initialises the new model with the neural network and weights from a previous model [LR101], Transfer Learning, and Self-supervised pre-training [LR118] 
\\
		
CS04	& SS04	&Kubric [LR109], CMA [LR102], Transfer Learning [LR118], Federated Learning (FL) and Split Learning (SL) [LR113] \\
		
CS05	& SS05	&Ultron-AutoML [LR110], Meta-learning to automate model selection and HPO [LR60] \\

CS06	& SS06	& Switch Transformer[LR115], ZeRO optimisation technique[LR116],  DeepSpeed2 [LR116],  and DeepSpeedMoE[LR117]. Cloud providers[LR120], KafkaML, Kubeflow, Sagemaker, Tensorflow Serving [LR121] \\ 
CS07 & SS07	& Elastic ML[LR119], Cloud providers offer ML as a service to abstract away the complexities and other synchronous and asynchronous parallelisation techniques  [LR120]  \\
    
CS08	&SS08	&Apache Mahout [LR94] and MLlib in HPC [LR112], Bighead [LR48], ASML [LR96], DLHub [LR119], TensorFlow Serving, SageMaker, Clipper [LR119], Kafka-ML [LR121],  \\

CS09  &SS09	&  Four layered edge-cloud architecture  and recommendation from [LR112], Recommendation and  Techniques from [LR114], Balancing  workloads and reducing computation time using synchronous and asynchronous parallelisation techniques  [LR120], Blind learning [LR113], FEDn [LR111], Transfer learning and self-supervised pre-training techniques [LR118].\\

CS10  &SS10	& Curious Container [LR98 ],  ModelDB and ModelHub [ LR119] , TensorFlow debugger [LR101] \\
CS11 & SS11	& Curious Container [LR98], A framework for microservice architecture for training and managing model [LR99],  DistRDF2ML [LR104], ModelDB and ModelHub [ LR119], Amazon SageMaker [LR124], TensorFlow Serving, SageMaker, Clipper, and DLHub [LR119]\\

CS12 & SS12	&ADCME[LR93],  IBM Fabric for Deep Learning (FfDL), Djinn, Tonic, Ray, and Ease.ml [LR119], Amazon SageMaker [LR124], Hybrid distributed architecture showcasing compatibility with multiple systems without code changes  [LR105]  \\
CS13 & SS13	&OpenSHMEM-ML [LR110]  \\

  \bottomrule
  \end{tabular}
  \caption{Scalability solution for the identified Challenges that we indexed in Table V}
  \label{tab:example142331}
\end{table*}

\section{Synthesis}

\subsection{RQ 7. Interdependence of maintainability challenges }

Our synthesis from SLR  data provides insights into the interdependencies within various stages of the ML workflow and their influence on maintainability. This interconnectedness is visually represented by arrows in Fig 6 and detailed in Table XIV. An arrow originating from one stage and pointing to another signifies how the starting stage influences the maintainability of the destination stage. Arrows pointing both ways denote a mutual impact between the two stages. Each arrow in Fig 6 corresponds to a specific row in Table XIV, where we further explain the interdependence from our  SLR. This framework aims to assist practitioners in understanding and evaluating the interdependencies and associated maintenance challenges across an ML system's lifecycle.

Additionally, our findings highlight a problematic trend we have termed "Repetitive Maintenance" (illustrated as a blue oval with arrows in Figure 6). Here, challenges like inadequate model accuracy, performance issues, and other quality concerns identified during the Model testing, governance, and monitoring stages may lead to recurring and costly adjustments in the stages of dataset creation, data preprocessing, data management, data validation, HPO, and model training.

\begin{figure}[H]
\centerline{\includegraphics[scale=0.75]{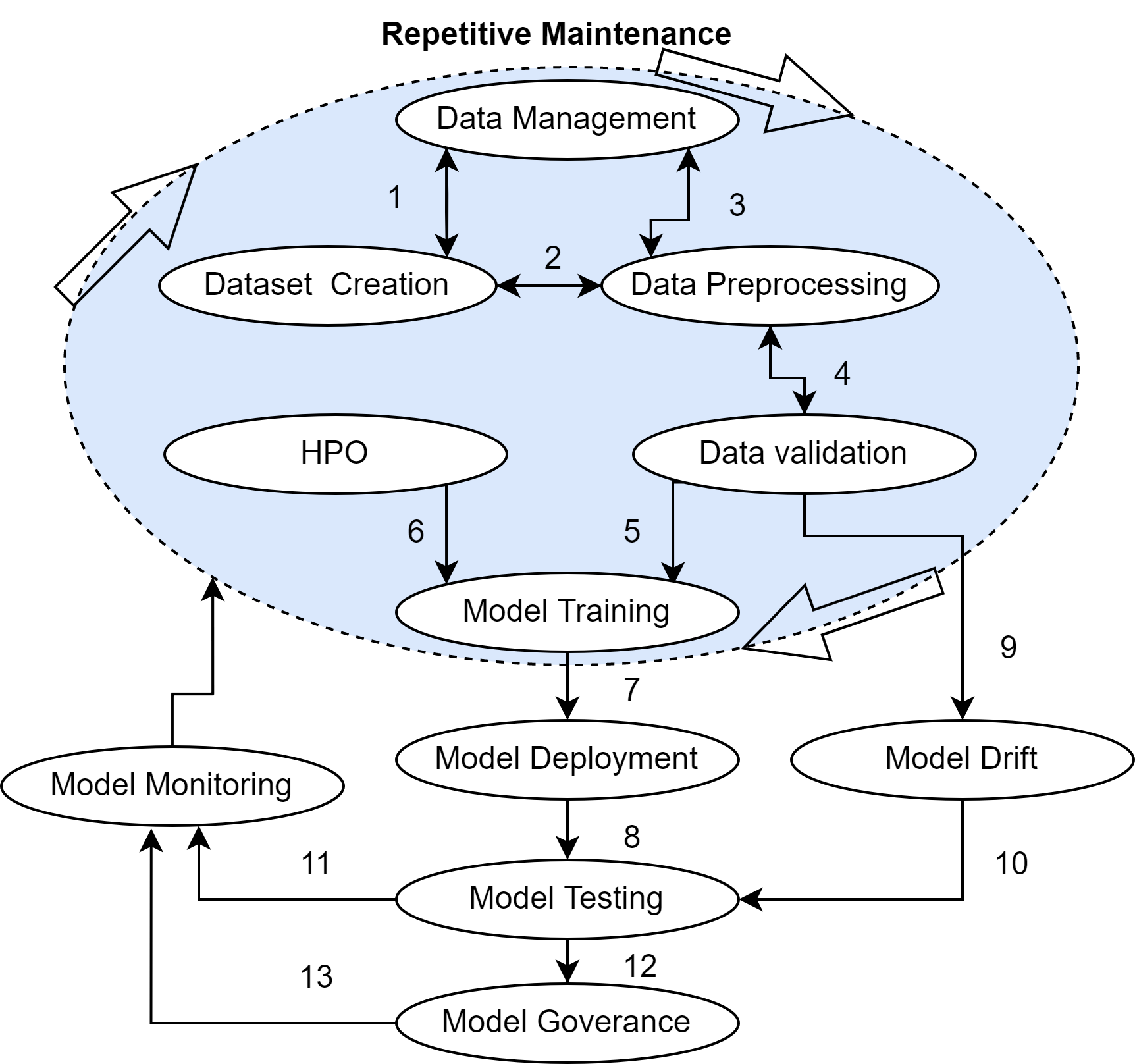}}
\caption{Mapping interdependence of Maintainability challenges in different ML lifecycle and workflow stages.}
\label{fig}
\end{figure}
\subsubsection{Dataset creation and Data Management  interdependent maintainability challenges}

Dealing with evolving datasets involves several stages of the data engineering pipeline, including data acquisition, manipulation, serialisation, and storage [LR1, 2, 3, 4]. Creating labels for training data is another challenge in dataset creation and management. This task typically requires human input, which can be time-consuming and demotivating for the annotator, especially when the task becomes repetitive [LR9,15]. These stages are interconnected because they require ongoing maintenance and attention to ensure the data is appropriately processed and stored for future use in the pipeline.

\subsubsection{Dataset Creation and   Data Preprocessing interdependent maintainability challenges }

Datasets can be riddled with errors and quality issues, ranging from missing data and outliers to unfairness and even adversarial or poisoned data. To address these issues, a preprocessing pipeline is required to clean and transform the dataset before it can be used in training. However, even minor changes to the dataset can significantly impact the model's performance, as the relationship between dataset features and model performance is complex. Furthermore, external or evolving data sources may be vulnerable to errors, malicious attacks, and inherent bias [LR1,2]. As a result, preparing data for machine learning (ML) and deep learning (DL) tasks can be challenging, requiring significant time and effort to handle dirty, biased, missing, and poisoned data during preprocessing and validation to ensure data quality and model fairness. Maintaining data quality is an ongoing challenge, and imbalanced data can pose a significant obstacle to modelling, requiring data re-balancing techniques to address. Traditional methods of randomly splitting data into training and validation sets can also pose a risk of data leakage [LR5, 6]. Additionally, integrating data analytics tools with existing frameworks in multiple languages is often challenging [LR6,8]. All these factors can lead to a repetitive maintenance anti-pattern.

\subsubsection{Data Preprocessing and Data Management   interdependent maintainability challenges }

To ensure reproducibility and reuse in the highly experimental data preprocessing steps, it is crucial to track the provenance of transformation steps and intermediary results [LR3]. However, due to the data-dependent behaviour and entanglement between data features and model performance, maintaining this pipeline can be challenging and may result in correction cascades. Moreover, transitioning from exploratory data analysis to a regularised, reproducible production workflow can be complicated when datasets constantly change in new tasks or applications [LR4]. Data scientists typically use pipelines to extract features, experiment with hyperparameters, and deploy models, but automating this process can be challenging with traditional software engineering tools. Additionally, tracking and managing  the data used at each step of the workflow to ensure reproducibility and data provenance is challenging due to the large amount of data used by ML applications that traditional version control systems cannot efficiently handle [LR14,15, 68].

\subsubsection{Data Preprocessing and  Data Validation interdependent maintainability challenges }

Data processing poses several challenges, such as the possibility of encountering dirty data, evolving data, and errors caused by bugs in the data source. The stochastic nature of these challenges makes it tough to eliminate all data errors during the preprocessing stage [LR1, 2]. Hence, it is necessary to have a validation pipeline that consistently monitors and scrutinises these stages to ensure that the training data is quality. Proper handling of errors, including missing data, is crucial for optimal algorithm performance. Additionally, data poisoning by malicious actors remains a concern, mainly when dealing with external data prone to errors, attacks, and drift [LR1,2]. Ongoing data validation challenges persist, particularly as models and problems change over time, making it difficult to determine if the learned models accurately reflect current situations or are based on outdated training data sets [LR12]. Therefore, Data preprocessing can be challenging due to dirty and error-prone data, necessitating a validation pipeline to monitor the quality of training data consistently.

\subsubsection{Data Validation and Model Training  interdependent maintainability challenges}

Validating machine learning models is a complex task, as most models are black boxes that require constant monitoring and assessment of data to identify issues and evaluate data quality. Moreover, maintaining the model can be problematic due to its continuous updates using new data in an online learning system [LR12]. Without proper data validation strategies, errors or undesired behaviour in the data could result in degraded training performance and training-serving skew, resulting in the anti-pattern of Repetitive Maintenance. The models may change with new data, making it difficult to determine whether the learned models accurately reflect current situations or are based on outdated training data [LR 6, 18, 20]. Thus, ongoing monitoring and maintenance of data quality are critical to prevent errors and ensure the model's accuracy.

\subsubsection{HPO and Model training interdependent maintainability challenges }

HPO is a laborious and tedious task that requires manual configuration of the appropriate hyperparameter settings. This process is often carried out through trial and error without the guidance of expert knowledge [LR22, 24]. The models' effectiveness, efficiency, and convergence speed are directly linked to the hyperparameter choices. An incorrect selection of these parameters can significantly impact the model's learnability and coverage rate during the training phase, ultimately leading to retraining the model with new parameters, resulting in the wastage of time and resources [LR 21, 23, 24]. Furthermore, to ensure reproducibility, an orchestration pipeline needs to be set up and maintained to execute the optimisation and keep track of the parameters and results [LR 21, 22].

\subsubsection{Model training and Model Deployment interdependent maintainability challenges}

The choice of training techniques affects the performance and deployment of a model, with incremental learning being more accurate for highly fluctuated systems and retraining for stable systems [LR18]. Challenges in deployment arise from managing dependencies, maintaining glue code, unintended feedback loops, and setting up monitoring and logging capabilities [LR 11, 20]. Furthermore, deploying DL software on mobile devices faces issues with compatibility and reliability [LR29, 30]. The deployment process involves packaging, validating, and serving the model in production [LR28]. It is crucial to consider all these factors when deciding on a modelling or training approach and what deployment strategy is best for an intended use case.

\subsubsection{Model Deployment and Model Testing interdependent maintainability challenges 
}
Deploying trained models and integrating and testing them with other applications can be challenging due to various constraints, such as different OS and hardware environments, power and memory limitations, and vendor-specific optimisation packages and libraries [LR11,20, 30]. To ensure an efficient  process of creating, training, deploying, and updating ML models, it is crucial to have an automated deployment system, CI/CD pipeline, and other  DevOps integration, which demands collaboration between ML professionals and other technical experts [LR28]. Testing machine learning systems is complicated due to the ever-evolving nature of ML systems and the learning-based emergent functional behaviour of these systems [LR32, 36]. In addition,  the deployment life cycle involves generating a prototype model, offline testing, and online testing [LR 36], consequently leading to repetitive maintenance.  

\subsubsection{Data validation and model drift interdependent maintainability challenges.}

Model drift occurs due to several reasons, such as changes in data, data seasonality, and fluctuations in data collection, which can lead to model obsolescence and performance degradation [LR 2, 12]. When there are fluctuations in data collection or seasonality, it changes data that a single data validation algorithm cannot detect [LR 6,18, 20]. Although researchers have various methods to deal with these drifts, there are challenges [LR19]. Implementing them can be costly, requiring drift detection algorithms, engineering solutions into existing pipelines, and ongoing maintenance to detect new drifts [LR19,20]. Furthermore, the problem and models may change, making it difficult to determine whether the learned models accurately reflect the current problem or are based on outdated training datasets [LR19,20]. 
 
\subsubsection{Model Drift and Model testing  interdependent maintainability challenges}

Machine learning systems are subject to drift, which can cause rapid obsolescence of test cases and present new challenges for detecting and creating unit tests[LR38, 39, 78, 86]. ML systems differ fundamentally from traditional software, which poses unique challenges for maintaining unit and regression tests and can result in the anti-pattern of Repetitive Maintenance [LR 36]. The stochastic and data-dependent nature of ML systems means that societal and data changes can cause a loss of accuracy, rendering historical models obsolete. Traditional functional testing methods are unsuitable for ML systems due to their large input spaces and unpredictable functional behaviour [LR 32, 38 ]. Moreover, the continual learning of ML systems can render test oracles irrelevant [ LR 39]. Deep neural networks (DNNs) do not provide statistical guarantees for out-of-distribution data, and their impact on errors still needs to be fully understood, despite progress in testing [LR 38].

\subsubsection{Model testing and Model Monitoring  interdependent maintainability challenges }

ML systems can influence their behaviour over time, leading to hidden feedback loops where the input to the model is indirectly adjusted to influence its behaviour [LR 27,28]. Monitoring and testing for these feedback loops, data errors, performance metrics, and drifts in an evolving data-dependent system is a complex challenge that requires an understanding of the data and model quality attributes to monitor [LR 26]. Traditional testing approaches are not directly applicable to complex data-dependent models, and online testing is necessary to detect bugs after deployment [LR 38, 39]. Overall, it is crucial to comprehensively test and monitor machine learning systems to ensure their reliability and effectiveness,

\subsubsection{Model testing and Model Governance interdependent maintainability challenges }

High-risk ML applications require cross-disciplinary efforts to define quality metrics and requirements specifications and rigorous formal verification tests to ensure regulatory and ethical compliance [LR 73, 88]. Governance in ML  is also challenging due to the need for robustness, generalisation to unseen inputs, and meeting functional requirements [LR20]. Traditional functional testing methods need to apply to ML systems adequately, and the constant improvement of ML systems through learning from new data renders traditional test oracles obsolete. ML testing also faces challenges due to the stochastic nature and data dependency of ML, which impacts Model Governance and leads to Repetitive Maintenance.

\subsubsection{Model Governance and Model Monitoring  interdependent maintainability challenges }

Maintaining ML models can be challenging due to the need for robustness, generalisation to unseen inputs, and meeting functional requirements. To ensure that the model meets business and regulatory needs, the model verification process involves requirement encoding, formal verification, and test-based verification [LR 20]. Version control platforms like GitHub lack support for ML projects' specific needs, such as data versioning and links between ML and software artefacts, making it challenging to trace and evolve ML models [LR33, 34]. While specialised ML platforms exist, they still require manual specifications of ML components and connections. In addition, engineers are tasked with building custom solutions to effectively monitor the ML application, providing visual tools, and implementing access privileges for team members. This monumental task requires ongoing maintenance to ensure high-quality model governance [LR 20,27].

\begin{table*}[htbp]
\caption{Maintainability challenges and their interdependence}
\label{my-label}
\begin{tabularx}{\textwidth}{@{} p{0.7cm} |p{4 cm}  | X @{}}
\toprule
\textbf{No} & \textbf{Interdependent Stages} & \textbf{Impact on Maintainability and its challenges} \\
\midrule
1 & Dataset Creation $\longleftrightarrow$ Data Management & When dealing with evolving datasets, these stages are interconnected because there is constant back-and-forth maintenance between them, with data acquisition from different sources, facilitation and manipulation of various types of data, serialisation, and storage of multi-format data for reuse further down the pipeline. \\
\midrule
2 & Dataset Creation $\longleftrightarrow$ Data Preprocessing & Datasets can have many errors and quality issues like missing data, outliers, and unfairness, and could also be adversarial and poisoned. Consequently, the dataset goes through a preprocessing pipeline, cleaned and transformed before being used in the training pipeline. The relationship between dataset features and model performance is complex, so even minor changes in the dataset can significantly impact the model's performance, resulting in the Repetitive Maintenance anti-pattern. \\
\midrule
3 & Data Preprocessing $\longleftrightarrow$ Data Management & The highly experimental data preprocessing steps demands provenance tracking, querying, and storing transformation steps and intermediary results to ensure reproducibility and reuse. However, due to the data-dependent behaviour and entanglement between the data features and model performance, maintaining this pipeline can be challenging and may lead to correction cascades. \\
\midrule
4 & Data Preprocessing $\longleftrightarrow$ Data Validation & Typical data processing challenges include dirty data, evolving data or data drift, and errors caused by bugs in the data source. Because of the stochastic nature of these challenges, it is challenging to eliminate all data errors during the preprocessing stage. Therefore, a validation pipeline is needed to continuously check and reexamine these stages to ensure good quality training data. \\
\midrule
5 & Data Validation $\longrightarrow$ Model Training & Since most machine learning models are complex black boxes, validating them involves constantly assessing and monitoring data to pinpoint issues and evaluate their quality. Maintaining the model can be problematic due to its continuous updates using new data in an online learning system. Without proper data validation strategies, errors or undesired behaviour in the data could cause degradation in training performance and even training-serving skew, which may lead to the Repetitive Maintenance anti-pattern. \\
\midrule
6 & HPO $\longrightarrow$ Model Training & Configuring the right sets of hyperparameters is a prolonged and manual process often done through trial and error without expert knowledge. The models' effectiveness, efficiency, and convergence speed are directly linked to the hyperparameter choices made. Incorrect selection of these parameters can significantly impact the model's learnability and coverage rate during the training phase, ultimately leading to the need for retraining the model with new parameters, resulting in the wastage of both time and resources. \\
\midrule
7 & Model Training $\longrightarrow$ Model Deployment & The choice of modelling and training techniques affects how a model performs and where it is deployed. For example, incremental learning is more accurate for highly fluctuated and adapted systems, while retraining is better for stable systems. Challenges in deployment also arise when transitioning from a prototype to production, where glue code, ad hoc, and brittle pipelines have to be managed and require setting up monitoring and logging capabilities. \\
\midrule
8 & Model Deployment $\longrightarrow$ Model Testing & Deploying trained models and integrating and testing them with other models or applications can be challenging due to various constraints such as different OS and hardware environments, power and memory limitations, and vendor-specific optimisation packages and libraries. To ensure an efficient process of creating, training, deploying, and updating ML models, it is crucial to have an automated deployment system, CI/CD pipeline, and other DevOps integrations, which demand collaboration between ML professionals and other technical experts. Testing machine learning systems is complicated due to the ever-evolving nature of ML systems and the learning-based emergent functional behaviour of these systems. \\
\midrule
9 & Data Validation $\longrightarrow$ Model Drift & Model drift occurs due to changes in data, data seasonality, and fluctuations in data collection, which can lead to model obsolescence and performance degradation. When there are fluctuations in data collection or seasonality, it changes data that a single data validation algorithm cannot detect. Although researchers have various methods to deal with these drifts, none is perfect. Implementing them can be costly, requiring drift detection algorithms, engineering solutions into existing pipelines, and ongoing maintenance to detect new drifts. Furthermore, the problem and models may change, making it difficult to determine whether the learned models accurately reflect the current problem or are based on outdated training datasets. \\
\midrule
10 & Model Drift $\longrightarrow$ Model Testing & Machine learning systems are subject to drift, which can cause rapid obsolescence of test cases and present new challenges for detecting test cases and creating unit tests. ML systems differ fundamentally from traditional software, which poses unique challenges for maintaining unit and regression tests and can result in the anti-pattern of Repetitive Maintenance. The stochastic and data-dependent nature of ML systems means that societal and data changes can cause a loss of accuracy, rendering historical models obsolete. Traditional functional testing methods are unsuitable for ML systems due to their large input spaces and unpredictable functional behaviour. Moreover, the continual learning of ML systems can render test oracles irrelevant. Deep neural networks (DNNs) do not provide statistical guarantees for out-of-distribution data, and their impact on errors is not yet fully understood, despite progress in testing. \\
\midrule
11 & model Testing $\longrightarrow$ Model Monitoring & ML systems can influence their behaviour over time, leading to hidden feedback loops where the input to the model is indirectly adjusted to influence its behaviour. Monitoring and testing for these feedback loops, data errors, performance metrics, and drifts in an evolving data-dependent system is a complex challenge that requires an understanding of the data and model quality attributes to monitor. Traditional testing approaches are not directly applicable to complex data-dependent models, and online testing is necessary to detect bugs after deployment. Overall, it is crucial to comprehensively test and monitor machine learning systems to ensure their reliability and effectiveness, preventing the Repetitive Maintenance anti-pattern. \\
\midrule
12 & model Testing $\longrightarrow$ Model Governance & High-risk ML applications require cross-disciplinary efforts to define quality metrics and requirements specifications and rigorous formal verification tests to ensure regulatory and ethical compliance. Governance in ML is also challenging due to the need for robustness, generalisation to unseen inputs, and meeting functional requirements. Traditional functional testing methods do not adequately apply to ML systems, and the constant improvement of ML systems through learning from new data renders traditional test oracles obsolete. ML testing also faces challenges due to the stochastic nature and data dependency of ML, which impacts Model Governance and leads to Repetitive Maintenance. \\
\midrule
13 & Model Governance $\longrightarrow$ Model Monitoring & Maintaining ML models can be challenging due to the need for robustness, generalisation to unseen inputs, and meeting functional requirements. To ensure that the model meets business and regulatory needs, the model verification process involves requirement encoding, formal verification, and test-based verification. Version control platforms like GitHub lack support for ML projects' specific needs, such as data versioning and links between ML and software artefacts, making it challenging to trace and evolve ML models. While specialised ML platforms exist, they still require manual specifications of ML components and connections. In addition, engineers are tasked with building custom solutions to effectively monitor the ML application, providing visual tools, and implementing access privileges for team members. This monumental task requires ongoing maintenance to ensure that model governance is high quality. \\
\bottomrule
\end{tabularx}
\end{table*}

 \section{RQ 8 . Interdependence of Scalability challenges }

Data engineering, model engineering, and  Current ecosystem scalability challenges are interdependent stages in developing a Scalable ML system. The interdependence of different stages in an ML workflow implies that changes in any stage can impact other stages' scalability. Here is a breakdown of how these stages are interconnected, followed by Table XV with summaries  :

\subsubsection{Data Engineering and Model Engineering interdependent Scalability challenges }

Scaling machine learning workflows, particularly data engineering and model engineering presents complex challenges that are closely interlinked. The availability and quality of training data are crucial for the optimal performance of ML models hinges on the quality and availability of training data. However, large-scale data collection and processing are resource-intensive and bring forth concerns related to privacy, legality, and cost [LR113,109,102].

Efficient management of vast data volumes mandates scalable ingestion and preparation pipelines. These pipelines must be in sync with distributed computing frameworks such as Spark, allowing for parallel processing and horizontal scaling as data volumes expand [LR120]. Key considerations include data aggregation, in-memory parallel operations, and locality-based data selection to reduce communication overhead in distributed systems [LR92]. Furthermore, these pipelines must optimise data for compatibility with ML models, ensuring reduced training time and enhanced accuracy [LR 95, 96, 120].

Large-scale online services face challenges like cold start and sparse data due to new content and short item lifespans on consumer-to-consumer platforms [LR101,118]. Real-time feature value acquisition is complex as data is often scattered, synthesised, and requires aggregation before use. Poor data structures and a lack of taxonomy for metadata associated with user content further complicate matters [LR101,118].

Scaling ML systems requires addressing dataset size,  model size, and model count. it is recommended to consider challenges like cold start and sparse data, tradeoffs in distributed ML models, balancing efficiency and effectiveness in ML workflows, continuous ML model retraining, and designing ML serving infrastructure [LR101]. Although data and model parallelism are active research areas, scalable system architectures and application frameworks are often complex and specific to individual use cases, needing a general and standardised deployment approach [LR95,96].

Distributed machine learning offers benefits like time and energy savings, increased reliability, and eliminating the need for large data volumes on a single machine. However, implementing distributed ML algorithms and frameworks on edge devices can be challenging due to limited computing and storage capacities, data privacy, energy savings, and restricted bandwidth for communication[LR112, 122, 123].

Regarding hardware considerations,  Facebook deploys both GPUs and CPUs, emphasising the need for innovative parallel computing methodologies to cater to the rising computational demands of varying configurations [LR95]. Model engineering echoes this sentiment, revealing challenges associated with optimising hardware and software configurations for machine learning workloads, especially on high-performance computing clusters [LR107]. This ties back to the industry trend of moving ML workloads to modern hardware solutions like GPUs [LR94].

Data engineering and model engineering are intertwined, especially in scalability challenges [LR92][LR99]. Both domains face significant challenges with resource management. For instance, data engineering emphasises the significance of having a robust scheduling system for consistent machine learning, focusing on the management of new training data and monitoring frameworks [LR101]. Model engineering extends this concern, pointing out the necessity of high-performance resources and job orchestration for efficient distributed training. This orchestration is critical for minimising communication and memory costs, especially when training models on distributed systems [LR101].

In conclusion, the multifaceted challenges in data engineering and model engineering are intrinsically connected. Addressing the escalating complexities in both data and models demands a holistic and collaborative approach that harnesses the strengths and addresses vulnerabilities.

\subsubsection{Data Engineering and  Current Ecosystem Interdependent scalability challenges }

Scalability challenges in data engineering are multifaceted and often interconnected. The looming need for exascale systems to handle extreme-scale data analysis and high-performance processors is intertwined with the concern of fault resilience in parallel data access [LR92]. As we demand more from these systems, the challenge of energy consumption becomes an inherent part of the solution [LR92]. 

Further complicating the landscape is the connection between model training and data management. For models to be trained effectively, they require proximity to vast amounts of data and robust scheduling systems  [LR101]. However, obtaining this large-scale data for training presents its own set of hurdles, including cost considerations and pressing privacy concerns  [LR113]. The intersection of real-time data analysis and streaming must be noticed. As the emphasis grows on real-time data processing from diverse sources, the limited applicability of popular streaming tools in machine learning landscapes further accentuates the scalability challenge [LR95] [LR97]. 

The challenge extends to hardware and computational paradigms. As companies like Facebook underscore the imperative of parallel computing strategies to cater to growing computational demands, it brings to the forefront the limitations of traditional HPC systems like Hadoop and Map Reduce. This necessitates transitioning towards frameworks like Spark and TensorFlow Extended and PyTorch, designed especially for machine learning  [LR95] [LR94]. Nevertheless, the quest for optimal performance has its challenges, as performance tradeoffs emerge with the adoption of  some popular data science libraries in interpreted languages like Python [LR110].

Infrastructure and cost-efficiency present another intertwined challenge. While large-scale ML necessitates a complex infrastructure, often distributed across multiple servers or virtual machines, mitigating this complexity through cloud resources introduces new cost challenges, particularly for extended ML workflow [LR108], [LR106]. 

Lastly, the intertwined issues of ensuring reproducibility in large-scale ML systems, developing efficient debugging systems, and the complexities of data handling and platform dependencies. These challenges are intricate and relate to data volume, distribution, and reliance on various platforms and frameworks [LR92], [LR98], [LR104].

These interrelations manifest that scalability challenges in data engineering cannot be viewed in isolation. Addressing them demands a holistic approach that considers the intricate web of interdependencies.

\subsubsection{Model Engineering and  Current Ecosystem Interdependent scalability challenges }

Scalability is a critical aspect of machine learning (ML) workflows, from data ingestion to deployment. In particular, model engineering and deploying ML applications and infrastructure are closely linked and require scalable systems to handle training and deployment effectively [LR120, LR95].

To achieve scalability, both model engineering and infrastructure must be optimised to handle complex models and accommodate growth in data volume and model complexity. This optimisation can be achieved by using distributed computing frameworks and horizontally scalable systems [LR95]. However, managing highly scalable infrastructure and more complex models can introduce additional complexity and maintenance overheads, which require extra resources or expertise to manage effectively [LR95, 96]. Furthermore, training deep neural networks (DNNs) is time-consuming, and communication overheads can become a bottleneck in distributed training. High-performance hardware accelerators can reduce computation overhead, but communication overhead remains an issue [LR116].

ML systems require end-to-end software support for various components affecting data provenance, model training, and inference. Balancing loads across heterogeneous hardware while considering network topology and synchronisation costs can impact training speed and quality [LR123]. ML workflow also requires separating data and training workloads due to their differing characteristics [LR95, 96, 101].

To efficiently support ML workloads on high-performance computing (HPC) clusters, it is essential to identify ideal hardware and software configurations. However, the popularity of Python-based data science libraries can lead to performance tradeoffs due to the language's high overheads. This issue can be addressed by integrating HPC technologies with these frameworks to develop domain-specific libraries that balance productivity and performance [LR110].

Cost efficiency in large-scale ML/DL workflows is challenging due to hardware costs, time and resource constraints, algorithm complexity, and data costs. Additionally, ensuring reproducibility in large-scale ML systems is complex due to data volume and distribution, complex training pipelines, platform and framework dependencies, and resource availability [LR98]. Effective debugger systems are necessary but challenging to develop due to limited ML diagnosis comprehension and diverse error sources, leading to ad-hoc and non-reusable ML serving infrastructure solutions [LR104].

Various techniques and tools can overcome these challenges. Hardware components such as GPUs, FPGAs, TPUs, and neuromorphic hardware accelerate training and inference, while frameworks such as cuDNN and NCCL enable efficient parallelisation strategies [LR103]. Parallelisation techniques, including data, model, pipeline, and hybrid parallelism, allow the training of large deep-learning models on distributed infrastructure [LR103]. Architectural approaches, algorithm-level optimisations, gradient compression, and low-level network infrastructure optimisations can enhance distributed deep learning [LR114]. Cloud providers such as Google, Microsoft, Amazon, and IBM offer machine learning as a service, making it more accessible to a broader audience without costly infrastructure or specialised staff [ LR120].

Lastly, hyperparameter configuration can be overwhelming, especially for non-technical users. The addition of big data infrastructure compounds this issue and requires addressing scalability and resilience to manage and orchestrate different experimental HPO pipelines in  distributed systems [LR60, LR110, LR101, LR118].

Overall, achieving scalability in ML workflows requires considering multiple factors and adopting various techniques and tools to optimise model engineering, infrastructure, and support systems.

\begin{table*}[htbp]
 \caption{Scalability challenges and their interdependence}
\label{my-label}
\begin{tabularx}{\textwidth}{p{.2cm} | p{3cm} | p{13.8cm} }
\toprule
\textbf{No} & \textbf{Interdependent Stages} & \textbf{Impact on Scalability and its challenges} \\ 
\midrule
1 & Data Engineering $\longleftrightarrow$ Model Engineering & Scaling data engineering and model engineering workflows in machine learning is a complex challenge that requires addressing data availability, quality, and privacy concerns. Efficient management of large volumes of data and compatibility with distributed computing frameworks like Spark is necessary for scalable data ingestion and preparation pipelines. Data preparation pipelines should be optimised for handling vast data quantities and seamless compatibility with ML algorithms. Additionally, addressing concerns like energy consumption, multitasking, scheduling, reproducibility, and resilience is crucial for developing applications for large core systems. Poor data structures and a lack of taxonomy further complicate scaling in intricate systems \\

\\

\midrule
2 & Data Engineering $\longleftrightarrow$ Current Ecosystem & Scalability in data engineering and deploying an ML application or infrastructure are closely linked. To effectively handle large datasets, both data engineering and infrastructure must be optimised for scalability. This includes the use of distributed computing frameworks and horizontally scalable systems. Advanced parallel computing is pivotal for real-time data analysis, as demonstrated by giants like Facebook. Complexities in data management, privacy, and associated costs overlap in both domains. Performance optimisation across diverse hardware remains a shared hurdle. The intertwined challenges in reproducibility and platform dependencies highlight the need for holistic solutions.  \\
\\

\midrule
3 & Model Engineering $\longleftrightarrow$ Current Ecosystem  & Scalability in ML workflows, particularly in model engineering and deployment, necessitates optimised infrastructure for increasing data and model complexity. Distributed computing and scalable systems can offer solutions, but they bring their complexities, with communication overheads posing significant challenges, especially in DNNs. ML workflows need end-to-end software support, balancing heterogeneous hardware loads and data-workload separation. While Python-based libraries are popular in data science, they may lead to performance tradeoffs. Cost, reproducibility, debugger systems, and hyperparameter configuration in big data contexts present additional challenges, though advancements in hardware, parallelisation techniques, and cloud services offer promising solutions.
 \\

\\

\bottomrule
\end{tabularx}
\end{table*}

\section{RQ 9 . Tradeoffs in achieving Scalability and Maintainability }

 To achieve scalability and maintainability, the tradeoffs and competing priorities must be carefully weighed, requiring a thorough understanding of the system components and business requirements as summaries of different consideration for these tradeoffs is listed in Table XVI . 
 
The different tradeoffs in achieving Scalability and Maintainability are as follows :

\subsubsection{Complexity vs Simplicity}

When building machine learning (ML) systems, finding the right balance between complexity and simplicity is crucial. As ML systems become more complex, their accuracy and power may increase, but maintaining and updating them can become challenging due to many moving parts and dependencies. This is because complex systems often have many interconnections between components, making them difficult to keep track of, especially as the system scales [LR33][LR34][LR35]. Additionally, industries require models that can handle large datasets and perform real-time processing, making scalability and low latency important factors [LR33].

However, managing the complexity of AI components can add to existing challenges in managing diverse teams and ML workflows and tracing and evolving ML models. Existing version control platforms like Git cannot still track changes in data, hyperparameters, model artefacts, and dependencies, making managing ML components more complex [LR34].

On the other hand, simpler ML models may be easier to maintain and update but may have lower performance than complex models. Deep learning, commonly used in complex models, may only sometimes be the best approach for all problems and can be computationally expensive and challenging to fine-tune [LR35]. Simpler models, such as shallow network architectures, decision trees, and random forests, can be used as a starting point to accelerate the deployment of ML solutions and prevent over-complicated designs [LR20]. However, simplifying an ML system can result in slower performance or lower accuracy, impacting the system's overall effectiveness [LR35].

When choosing between retrained and incremental modelling, there is a tradeoff between learning assumptions and algorithm variants, leading to different accuracy and training time tradeoffs [CM26][LR18]. Training models can also be costly, and concerns have been raised about the environmental impact of training ML models [LR18][LR20]. Deploying distributed ML systems in power and resource-limited environments presents challenges in balancing accuracy [LR112]. In modern statistical ML, there is a tradeoff between correctness and performance, with resource-constrained devices facing limitations in local computation capabilities [LR122]. Real-time distributed ML systems face unique challenges in balancing consistency, latency, and accuracy [LR122].

Therefore, finding the right balance between complexity and simplicity when building ML systems requires considering the application's specific needs, available resources, and the expertise of the team building and maintaining the system. This includes managing the complexity of AI components, choosing appropriate models, considering the tradeoffs between retrained and incremental modelling, and being aware of training ML models' costs and environmental impacts.

\subsubsection{Speed vs Accuracy:
}
Regarding machine learning scalability, there is a crucial consideration of the tradeoff between speed and accuracy. Generally, the faster a model can be trained, the less accurate it may be, while a more accurate model may require more time to train. Simpler machine learning models may be easier to maintain and update but may not perform as well as more complex models [LR112]. This is because simple models may capture a different level of complexity in the data, leading to lower accuracy or slower performance.

Deep learning may not be the best approach for all problems, as it can be computationally expensive and challenging to fine-tune [LR35]. A potential solution is to use simpler models such as shallow network architectures, basic PCA-based approaches, decision trees, or random forests as a starting point to accelerate the deployment of a machine learning solution, provide valuable feedback, and prevent over-complicated designs. These models can also be used to test the concept of a proposed machine learning solution and establish the end-to-end setup [LR20].

One solution to mitigate the tradeoff between speed and accuracy is to increase computational resources by using more powerful hardware or parallelising the training process. However, this can increase costs and create other challenges, such as communication overhead in distributed systems [LR116][LR123]. Furthermore, increasing model complexity can improve performance but may make the model more prone to overfitting, reducing its ability to generalise to new data [LR112].

To address these tradeoffs, techniques such as hyperparameter optimisation and distributed machine learning can be used. Hyperparameter optimisation involves finding the best values for the various parameters that affect the model's performance [LR60, LR110]. However, this can be daunting, particularly for non-technical users [LR122,123]. On the other hand, distributed machine learning involves training the model on multiple machines or devices, which can improve efficiency and reduce training time. However, it also requires addressing challenges such as communication overhead and the need for end-to-end software support [LR120].

In conclusion, addressing the tradeoff between speed and accuracy is crucial to achieving scalable and accurate machine learning models.

\subsubsection{Scalability vs Performance:
}
Scalability and performance are often at odds with each other, and improving scalability may lead to decreased system performance [LR60]. For example, distributing the workload across multiple machines or using larger batch sizes can enhance scalability but decrease the speed of individual computations. Therefore, it is crucial to balance these tradeoffs to meet the required performance standards and ensure scalability for future growth [LR110].

Various parallel and distributed software frameworks, such as Apache Mahout, MLlib [LR112], Bighead [LR48], and ASML [LR96], provide scalable machine-learning algorithms for diverse applications. Transfer learning and self-supervised pre-training techniques are utilised to address the challenges of creating large annotated datasets for deep neural networks. Researchers also focus on improving the efficiency of large-scale pre-trained models through system-level optimisation, including single-device and multi-device optimisation methods, efficient learning algorithms, and model compression methods [LR118].

The scalability components of large and complex systems address issues such as data scaling, model size, and the number of models, with challenges like cold start and sparse data, tradeoffs in distributed ML models, balancing efficiency and effectiveness in ML workflows, continuous ML model retraining, and difficulties in designing ML serving infrastructure [LR101]. Active research areas include data and model parallelism. However, scalable system architectures and application frameworks tend to be complicated and specific to particular use cases, needing a general and standard deployment approach [LR95,96].

In implementing decentralised algorithms on IoT systems and networks, it is essential to consider factors such as data location, geographical distribution, algorithm features and configurations, task scheduling, data persistence, application constraints, and coding and testing [LR112]. With deep learning models becoming increasingly complex and resource-intensive, hardware components (GPUs, FPGAs, TPUs, and neuromorphic hardware), frameworks supporting distributed and parallel training, and parallelisation techniques are necessary [LR103]. Cloud providers like Google, Microsoft, Amazon, and IBM offer machine learning as a service, making it more accessible to a broader audience without costly infrastructure or specialised staff [LR120].

Training deep neural networks can be time-consuming due to the computation required [LR114]. Deploying DNN training tasks on multiple nodes can result in frequent communication requirements and communications overhead, which can become a critical bottleneck in distributed training. While high-performance hardware accelerators like GPUs and TPUs can decrease the computation overhead, communications overhead remains unchanged. This creates a tradeoff between functionality, usability, memory, and compute/communication efficiency [LR116]. As cluster scales increase, communications overhead increases exponentially, considerably diminishing the advantage of parallel training.

\subsubsection{Performance vs Resource consumption
}
Distributed machine learning systems have the potential to improve performance and reliability, but they also introduce challenges in balancing accuracy and efficiency while managing resources. This is because achieving better performance often requires more resources, which can lead to increased costs [LR112]. Therefore, it is necessary to find ways to improve efficiency while maintaining accuracy, such as subsampling or using a minibatch of the dataset and low-precision numerical representations, which are standard techniques used to achieve this [LR122]. However, these techniques also sacrifice some accuracy for faster results.

Resource-constrained devices, such as IoT devices, are limited in their ability to perform local computations [LR122]. Therefore, offloading computation or storage to remote computers can be beneficial, but it also comes with a cost in speed [LR122]. Real-time distributed machine learning systems, such as those used in autonomous vehicles, have unique challenges in balancing consistency and latency while maintaining accuracy.

Various solutions, such as hardware accelerators, parallelisation techniques, and cloud-based delivery models, can be employed to overcome these challenges [LR120]. Architectural approaches, algorithm-level optimisations, gradient compression, and low-level network infrastructure optimisations are some techniques used to improve distributed machine learning systems' performance [LR114].

Specialised resource schedulers, such as Heuristic algorithms, Elastic ML frameworks, Decentralised data-parallel DL systems, and Hybrid Communication, enable parallel computation and inter-worker communication while reducing computation time and balancing workload across available machines [LR119].

Efficient learning algorithms, system-level optimisation, and model compression are other techniques used to improve the efficiency of pre-trained models [LR118]. Researchers have also developed various techniques and models, such as Megatron-LM, Mesh-Tensorflow, ZeRO optimiser, GPipe, and Tera Pipe, that use these methods to accelerate large-scale pre-training while reducing memory and computational requirements [LR118].

However, these solutions come with challenges, such as compatibility issues and longer deployment cycles. The long-term success of distributed machine learning also relies on addressing open challenges such as performance, fault tolerance, privacy, and portability [LR120]. Therefore, it is essential to carefully consider the accuracy-efficiency tradeoff and select appropriate techniques and tools to achieve efficient and effective distributed machine learning.

\subsubsection{Flexibility and Stability }

Navigating the challenges of machine learning testing and maintenance involves balancing flexibility and stability. Flexibility enables ML systems to adjust to shifting requirements and environments, improving maintainability but possibly decreasing stability. In contrast, stability ensures consistent functionality and performance, facilitating maintenance but potentially limiting flexibility.

Interoperability across various ML frameworks and vendor-optimised libraries is essential in the ML domain. The Open Neural Network Exchange (ONNX) is a critical open specification for achieving such interoperability [LR95]. Container technologies like Docker help replicate runtime environments with all dependencies [LR95]. At the same time, the Curious Containers project supports reproducibility by providing applications as container images, loading data directly into the running container, and storing all information in a single file [LR98].

Balancing flexibility and stability in testing and maintenance involves using techniques such as automated deployment across development, QA, and production environments, dedicated CI/CD pipelines, and DevOps integration to ensure secure and efficient model development, deployment, and updates. Standardised ML model packages, metadata schemas, repositories, and serving platforms are necessary to enable discovery and citation and provide on-demand model inference. Furthermore, reproducibility is crucial for scientific rigour and robust applications, requiring tracking and storing relevant model information, including ML source code, datasets, and computation environments.

Several ML platforms and frameworks, such as ONNX [LR95], Docker [LR95], and DistRDF2ML [LR104], tackle these challenges by providing solutions for orchestrating ML pipelines, managing data storage, access control, pipeline execution, job scheduling, and performance monitoring. These platforms offer generic solutions for different ML tasks in hybrid environments and enable developers to describe ML workflows through pipeline configuration or software development kits (SDKs) [LR53]. AutoML, FBP, and MDSD approaches also provide flexibility and abstraction in ML application development [LR80, LR81].

In conclusion, ML practitioners must balance flexibility and stability to manage testing and maintenance effectively. This involves employing various techniques and tools, such as automated deployment, CI/CD pipelines, DevOps integration, standardised model packages, metadata schemas, and reproducibility practices.

\subsubsection{Scalability vs Cost }

When deciding between scalability and cost, it is important to weigh the advantages of improved scalability, such as enhanced performance and reduced downtime, against the potentially higher costs associated with certain methods like cloud computing. In large-scale online services, issues such as cold start and sparse data can arise, necessitating a scheduling system to manage new training data and a monitoring framework to detect and handle failures. While techniques like warm starting and transfer learning can help mitigate these challenges, configuring hyperparameters and understanding large datasets can still be time-consuming. [LR101]

Scalable system architectures and application frameworks are complicated and specific to certain use cases, lacking a general and standard deployment approach. Issues such as data scaling, model size, and several models must be addressed, with challenges like cold start and sparse data, tradeoffs in distributed ML models, balancing efficiency and effectiveness in ML workflows, continuous ML model retraining, and difficulties in designing ML serving infrastructure. Active research areas include data parallelism and model parallelism. [LR95, 96]

Distributed machine learning offers benefits such as saving time and energy, increasing reliability, and avoiding the need to collect large volumes of data on a single machine. Various approaches, such as distributing data or distributing models, can be used to achieve distributed machine learning, which can be performed in a centralised or decentralised architecture. However, implementing distributed machine learning algorithms and frameworks can be difficult, especially when deploying them on edge devices. Traditional distributed high-performance infrastructures like Apache Hadoop and Spark may not be adapted to edge devices due to limited computing and storage capacities, data privacy, energy savings, and limited bandwidth for communication. [LR112]

Training deep neural networks can be time-consuming due to the computation required. Deploying DNN training tasks on multiple nodes can result in frequent communication requirements and communications overhead, which can become a critical bottleneck in distributed training. While high-performance hardware accelerators like GPUs and TPUs can decrease the computation overhead, communications overhead remains unchanged. This creates a tradeoff between functionality, usability, memory, and compute/communication efficiency. As cluster scales increase, communications overhead increases exponentially, considerably diminishing the advantage of parallel training. [LR116, 114]

Implementing distributed machine learning can offer benefits like saving time and energy and avoiding the need for large volumes of data on a single machine. However, there are tradeoffs in terms of increased costs and challenges, such as communication overhead, portability, and privacy. Various techniques, such as synthetic data generation and transfer learning, can help overcome these challenges. Cloud providers offer machine learning as a service, making it more accessible to a broader audience. Tools like Amazon SageMaker can provide scalable, elastic model training on large data streams. [LR124]

\begin{table*}[htbp]
\centering
\caption{Considerations for these Trade-offs}
\label{tab:strategies}
\begin{tabular}{@{}p{3.5cm}p{14cm}@{}}
\toprule
\textbf{Trade-off} & \textbf{Considerations} \\
\midrule
Complexity vs Simplicity & 
\begin{itemize}
    \item Find the right balance between complexity and simplicity. Consider the application's specific needs, available resources, and team expertise.
    \item Manage the complexity of AI components. Choose appropriate models (deep learning vs simpler models).
    \item Consider tradeoffs between retrained and incremental modelling.
    \item Be aware of training ML models' costs and environmental impacts.
\end{itemize} \\
\addlinespace
Speed vs Accuracy & 
\begin{itemize}
    \item Faster models may sacrifice accuracy. Consider tradeoffs between training time and model accuracy.
    \item Use simpler models as a starting point for faster deployment.
    \item Increase computational resources to improve training speed.
    \item Employ techniques like hyperparameter optimisation and distributed machine learning.
\end{itemize} \\
\addlinespace
Scalability vs Performance & 
\begin{itemize}
    \item Balancing workload across machines in distributed systems.
    \item Utilise scalable ML algorithms and frameworks.
    \item Optimise system-level performance of large-scale models.
    \item Address communication overhead and end-to-end software support.
\end{itemize} \\
\addlinespace
Performance vs Resource Consumption & 
\begin{itemize}
    \item Balance accuracy and efficiency. Use techniques like subsampling and low-precision numerical representations.
    \item Offload computation or storage to remote computers.
    \item Utilise hardware accelerators, parallelisation techniques, and cloud-based models.
    \item Optimise distributed machine learning systems for resource efficiency.
\end{itemize} \\
\addlinespace
Flexibility and Stability & 
\begin{itemize}
    \item Balance flexibility and stability in testing and maintenance.
    \item Employ automated deployment, CI/CD pipelines, and DevOps integration.
    \item Use standardised model packages, metadata schemas, and reproducibility practices.
    \item Employ ML platforms and frameworks for managing ML workflows.
\end{itemize} \\
\addlinespace
Scalability vs Cost & 
\begin{itemize}
    \item Weigh the advantages of improved scalability against potentially higher costs.
    \item Address issues like cold start, sparse data, and efficient resource management.
    \item Utilise distributed machine learning and edge computing techniques.
    \item Consider cloud-based machine learning services.
\end{itemize} \\
\bottomrule
\end{tabular}
\end{table*}

\section{RQ 10. Finding  balance in achieving Scalability and Maintainability}

Achieving scalability and maintainability for a machine learning (ML) system simultaneously can be challenging. However, several strategies can help in overcoming some of the tradeoffs. Table XVII provides a well-defined list of strategies for brevity. 

\subsubsection{Modular design and Microservices architecture
}
John et al. [LR28] emphasise the importance of modular design and microservices in managing machine learning applications, mainly through MLOps and machine learning platforms. They provide a framework for adopting MLOps in companies, including a maturity model outlining four stages, validated through case studies. Granlund et al. [LR52] propose the use of integration mechanisms for ML/AI to create multi-organisational systems, using Continuous Delivery for Machine Learning (CD4ML) {SM37} as a concrete example of MLOps pipelines, and discuss various machine learning platforms that have been developed to address the end-to-end lifecycle management of ML applications such as TensorFlow Extended (TFX), ModelOps, and Kubeflow.

Zhou et al. [LR53] also discuss machine learning platforms that provide solutions for orchestrating ML pipelines and managing data storage, access control, pipeline execution, job scheduling, and performance monitoring. They highlight the importance of a generic solution for different ML tasks in hybrid environments, allowing developers to describe ML workflows through pipeline configuration or SDKs.

Docker is commonly used for automation in ML-based software project deployment, with functionalities related to data management, interactive development, task scheduling, and model management [LR82]. Amazon SageMaker is a scalable, elastic system for model training on large streams of data that addresses challenges like incremental training and model freshness, predictability of training costs, elasticity and support for pausing and resuming training jobs, handling ephemeral data, and automating hyperparameter optimisation and model tuning [LR124].

The Open Neural Network Exchange (ONNX) is an essential open specification in achieving interoperability across different ML frameworks and vendor-optimised libraries [LR95]. Within Facebook, ONNX transfers PyTorch research models to a production environment in Caffe2. Container technologies such as Docker support replication of a runtime environment with all its dependencies in a container image filesystem. The Curious Containers project supports reproducibility by providing applications as container images, loading data directly into the running container, and storing all information in a single file [LR98].

\subsubsection{Code organisation:
}

Flow-Based Programming (FBP) and Model-Driven Software Development (MDSD) are software development approaches that provide flexibility and abstraction to software systems design. These approaches have been applied to simplify the development of machine learning applications using the principles of FBP and MDSD. Researchers have modelled the Java APIs of Spark ML as composable components and then connected these components to create a machine learning flow, enabling non-programmers to create an ML program graphically at a higher level of abstraction [LR81]. Although there is limited research on supporting graphical ML programming at a higher level of abstraction, there are several products in the market that support high-level ML programming, such as WEKA, Azure Machine Learning Studio, KNIME, Orange, BigML, mljar, RapidMiner, Streamanalytix, Lemonade, and Streamsets [LR81].

\subsubsection{Automation and Continous integration 
}

The benefits of adopting DevOps tools within ML projects, such as improved code sharing, integration speed, and issue resolution, are highlighted by Rzig et al. [LR74]. The study also emphasises the importance of automatically synchronising DevOps configuration files to reduce maintenance overhead and the need for co-evolution analysis of functional code and DevOps configuration files to facilitate early bug detection. It is recommended that data scientists and ML developers adopt DevOps tools within their projects to improve code sharing, integration speed, and issue resolution, as the benefits of DevOps outweigh the associated costs  [LR74].

To build a deployable model, infrastructure must be created to make it consumable and supportable, which can be achieved by reusing data and models [LR20]. Implementing MLOps environments enables continuous integration and delivery through a CI/CD pipeline and continuous retraining through a CT pipeline. This leads to a safe and repeatable path for ML model development, deployment, and updating [LR28]. Utilising MLOps for a transition towards fully automated monitoring of models requires CI/CD integration, CT pipeline, certification of models, governance and security controls, model explainability, auditing, reproducible workflow and models, and mechanisms to perform end-to-end QA test and performance checks  [LR28].

In hyperparameter optimisation (HPO), different algorithms have various strengths, and automated processes such as Google Vizier, Amazon SageMaker, and SigOpt have been developed to address the challenges  [LR21]. However, open-source projects like Optunity, Tune, and Auptimizer offer more user-friendly APIs but have limitations in customisation and integration. Ultron-AutoML is a framework designed for reliability, efficiency, and scalability to ensure the timely and cost-effective completion of a user's HPO job. It offers the user ease of use, flexibility, and customisation [LR24].

Automated Machine Learning (AutoML) is an emerging field that searches for the best-performing model for a given task, dataset, and evaluation metric, reducing the human resources and expertise required to develop machine learning models and decreasing the time-to-market for these models [LR80]. Various AutoML software is available but they still require an understanding of the underlying concepts. The Machine Learning Bazaar is a new framework that simplifies the development of machine learning and automated machine learning software systems with ML primitives, a unified API, and a specification for data processing and ML components  [LR51]. Meta-learning approaches have been developed as frameworks, including Auto-WEKA, Auto-Sklearn, and SmartML, implemented as microservice architecture for increased scalability and maintainability [LR60].

In summary, Automation and Continuous integration are crucial in ML projects. Utilising DevOps tools within ML projects can improve code sharing, integration speed, and issue resolution. MLOps environments enable continuous integration and delivery, providing a safe and repeatable path for ML model development, deployment, and updates. Automated processes like hyperparameter optimisation can be achieved through open-source projects like Auptimizer or frameworks like Ultron-AutoML.

\subsubsection{Cloud-based infrastructure
}
Cloud infrastructure has become an essential factor in adopting and applying machine learning (ML) technology. Major cloud providers such as Google, Microsoft, Amazon, and IBM offer a range of options to make it easier for businesses and individuals to take advantage of machine learning technology. These options include providing pre-packaged machine-learning software that can be deployed on virtual machines and offering machine learning as a service [LR120].

Cloud providers also support standard distributed machine learning systems and libraries such as TensorFlow and MXNet. This means that users can use existing machine learning technology and tools they are already familiar with [LR120]. Moreover, cloud-based delivery models reduce the entry barrier to designing innovative applications incorporating machine learning techniques. This means that even small businesses or individuals can start using machine learning without investing in expensive infrastructure or hiring specialised staff [LR120].

However, there are limitations to the AutoML libraries and systems that are mentioned, such as ATM, Vizier, Rafiki, Google AutoML, DataRobot, and Azure Machine Learning Studio. They focus on specific subsets of ML use cases and are designed as proprietary applications that do not support community-driven integration of innovations. It is essential to consider these limitations when choosing an AutoML solution [LR51][LR62][LR64].

To support the entire lifecycle of ML models, frameworks such as Flame, TFX, KubeFlow, MLflow, H2O, Skymind Intelligence Layer, Uber's Michelangelo, Facebook's FBLearner Flow, DataRobot, Polyaxon, Comet, Atalaya, Amazon's SageMaker, and Microsoft Azure Machine Learning are available [LR48]. These platforms provide end-to-end solutions for building and deploying ML applications without writing code via high-level, declarative abstractions [LR70].

Choosing the correct algorithm for a specific use case, known as the Algorithm Selection Problem, can take time and effort for non-experts. Software architectures and systems such as KOI and Looper ML platforms provide simple APIs and architecture that cover the entire ML lifecycle, including data collection, model training, deployment, and inference, enabling broad adoption of machine learning models in software systems [LR60].

Deploying ML-based software projects on GitHub can be challenging, but Docker is being used for automation in ML-based software project deployment. Docker is used for logging and monitoring, cloud-based development, interactive development, model management, and continuous integration and deployment (CI/CD) [LR82].

To evaluate and compare machine learning libraries and services, CompareML is a new approach that allows practitioners to determine the most appropriate machine learning algorithm and provider best suited to their data. Currently, no existing solutions can compare machine learning algorithms from different providers while requiring no depth of machine learning knowledge from the user, which is what CompareML aims to provide [LR84].

Cirrus is a machine learning framework that automates the management of data centre resources for ML workflows by utilising serverless infrastructures, such as AWS Lambdas and S3. It overcomes the resource constraints of serverless computing by using an ultra-lightweight worker runtime, streaming training mini-batches, and a stateless worker architecture. AWS SageMaker is also designed to reduce the training time and cost for customers regardless of the number and types of machines used. It uses multiple EC2 instance types and supports modern hardware such as GPUs [LR91][LR124].

In conclusion, cloud infrastructure is essential in leveraging machine learning technology, and there are a variety of platforms and tools available to support the entire ML lifecycle, from data collection to retraining deployed models.

\subsubsection{Collaboration and Communication 
}
Research has explored creating ML platforms with DevOps capabilities and running ML pipelines, identifying potential performance bottlenecks such as GPU utilisation, and automating the ML process from data preprocessing to application runtime monitoring to allow users to focus on application development. Effective feedback loops from monitoring systems to previous phases are necessary when new training data arrives, or model performance degradation is detected [LR53]. However, incorporating MLOps in a multi-organisational setting presents unique challenges related to integration, data privacy, security, and regulatory compliance, requiring custom solutions to maintain adherence to governance, auditing, and regulations [LR52].

Collaboration challenges between data scientists and software engineers can be addressed by investing in interdisciplinary teams, documenting responsibilities and interfaces, considering engineering work as a key contribution, and investing in process and planning. Formalised collaboration practices, including creating a shared understanding of the project's goals and scope, establishing clear communication channels between data scientists and software engineers, and documenting decisions and outcomes, are recommended to address these challenges. A new process model called CRISP-ML(Q) is introduced to develop machine learning applications with a quality assurance methodology, covering six phases from defining scope to maintaining the deployed application. The proposed model offers iterative paths to redefine objectives, stop or postpone the project, and is aligned with risk-based thinking [LR85].

In addition, incorporating data distribution awareness in testing and enhancement phases has improved results by up to 21.5\%. Best practices for communicating ML model quality within big teams have also been explored, and the introduction of MLCask, a version control system, tracks changes in ML pipelines and facilitates collaboration [LR38, LR50]. Furthermore, while technical aspects of developing ML components have received significant research attention, human factors such as coordination and documentation of responsibilities and interfaces and planning for system operation and evolution have been neglected, highlighting the importance of successful collaboration between software engineering and data science teams for developing ML-enabled systems [LR75].

To ensure human safety and health, developing safe and trustworthy AI systems is necessary. Assurance cases have been proposed to support quality assurance and certification of AI applications, but a detailed list of concrete criteria for these cases still needs to be provided. The development of safe AI systems requires a good understanding of the components of trustworthy artificial intelligence, and risk management for systems that use AI must be carefully adapted to the new problems associated with this technology  [LR73]. Additionally, surveys reveal that practical ML projects fail to meet their sponsors' or clients' expectations. This highlights the need for practical guidelines through standards and development process models specific to ML applications. The Japanese industry consortium QA4AI was founded to address these needs [LR85].

Collaborative and communicative efforts are necessary for successful ML project development, from interdisciplinary teams to formalised collaboration practices and process models. Additional considerations such as data privacy, security, and regulatory compliance are necessary for incorporating MLOps in a multi-organisational setting. Assurance cases and risk management are essential for developing safe and trustworthy AI systems. Practical guidelines through standards and development process models specific to ML applications are needed to meet clients' expectations, and cloud-based AI platforms must address practitioners' needs for complexity and flexibility.

\subsubsection{Documentation}

When developing AI, it is essential to prioritise transparency and accountability throughout the entire data development lifecycle to ensure responsible use. This involves treating datasets as potentially problematic, documenting assumptions, conducting independent peer reviews, evaluating datasets with expertise and judgment, utilising visualisation tools, performing sensitivity studies, and comprehending dataset limitations [LR16].

\subsubsection{Data and Model Management tools }

The need for effective data and model management tools has grown considerably in recent years. One such tool is Shuffler, which is designed for data annotation and labelling of images. It uses a relational database and SQL query for storing and manipulating annotations, and it supports loading, modifying, and storing images in various formats. Shuffler also enables users to perform multiple basic manipulation operations and to create custom functions and operations. It supports object detection, semantic segmentation, and object matching tasks, making it a versatile tool for image annotation and labelling  [LR3,4].

Another valuable tool for data management is JUNEAU. This tool aids in indexing, searching, and reusing tabular data like CSV and relational datasets. It acts like a backend data lake management subsystem and integrates relational and key-value stores to capture and index any external files loaded by the notebook. It also captures intermediate data computational steps within the cells and notebooks. By replacing the backends and extending the user interface of Jupyter notebooks, JUNEAU provides a seamless data management experience for users  [LR4].

In addition to data management tools, machine learning experiment management tools have also gained popularity. These tools, such as MLFlow, NeptuneML, and WandB, aim to improve the efficiency of ML pipelines by reducing manual intervention and addressing challenges in managing machine-learning-specific assets such as reproducibility and traceability. Versioning tools like DVC, Pachyderm, ModelDB, and Quilt Data have also become popular in machine learning projects. These tools enable users to specify data and model pipelines and manage model experiments, similar to infrastructure-as-code. They aim to reduce manual intervention and improve the efficiency of ML pipelines in open-source projects  [LR14].

Experiment management tools fall under machine-learning lifecycle management tools, which provide functionalities to store, track, and version assets from different experiment runs. Examples include MLFlow, Neptune, KubeFlow, and ModelDB. Some experiment management tools specialise in supporting deep learning. These tools are critical in ensuring the reproducibility of machine learning experiments and maintaining traceability in the development process. They offer a comprehensive solution for managing machine learning experiments and assets  [LR68].

Overall, developing data and model management tools is crucial for ensuring efficient and effective machine-learning projects. These tools help users to manage data and models effectively, reduce manual intervention, and ensure reproducibility and traceability in the development process. While several existing tools are available, the field is continually evolving, and there is a need to assess and improve the current tool landscape while developing new tools to meet the growing demands of machine learning projects.

\subsubsection{Testing and Debugging tools}

This study evaluates the efficacy of unit test generation techniques for machine learning (ML) libraries using testing and debugging tools like EVOSUITE and Randoop[LR36]. These tools help assess the quality of unit test suites in ML libraries based on metrics such as code coverage and mutation score.

Zhu et al. [LR37] propose a functional regression testing approach for ML systems that utilises production data, perturbation, and clustering to detect software bugs. This method has been applied to an ML-based spelling checker and has shown promising results.

Berend et al. [LR38] investigate the impact of data distribution awareness in deep learning (DL) systems, focusing on coverage-guided testing. Their research explores data mutations, coverage criteria, and test case effectiveness for enhancing robustness.

Various studies identify challenges in testing ML programs and suggest potential research directions [LR32,39]. Tools like DeepXplore and Themis have been developed to detect bugs and discrimination in ML systems [LR39].

Pastor et al. [LR32] extensively reviews current testing practices, including error detection in ML models and numerical-based testing techniques for ML code implementations.

The dpEmu framework [LR41] allows for modelling data faults in ML systems but remains a prototype requiring further research. Meanwhile, TensorFI [LR42] is a flexible, easy-to-use framework that evaluates the resilience of TensorFlow-based ML applications.

CALLISTO [LR44] is a tool designed to improve ML system quality by addressing issues in training data. Panichella et al. [LR45] discuss mutation testing in deep learning and recommend further research.

Dutta et al. [LR78] conducted a large-scale study on seed usage in testing ML projects, suggesting alternative strategies to address randomness and test flakiness. Their research provides valuable insights into using seeds in ML project testing.

Testing ML systems presents significant challenges due to their stochastic behaviour and training-induced faults. Potential solutions include developing automated oracles [LR86], creating common frameworks and benchmarks, generating realistic inputs using input mutation and search-based approaches, measuring test adequacy criteria through neuron activation metrics, and characterising decision boundaries.

Fault injection techniques and tools like Ares, TensorFI, and PyTorchFI help evaluate ML application reliability [LR71]. TensorFlow Data Mutator (TF-DM) is an open-source data mutation tool that targets various data faults and improves model robustness and resilience, which is crucial for safety-critical systems such as self-driving cars [LR71].

\subsubsection{Model governance and monitoring}

Model governance and monitoring are crucial aspects of the machine learning lifecycle that ensure AI systems' trustworthiness, reproducibility, and long-term performance. As the importance of machine learning continues to grow in scientific research and industry, there is a growing need for standardisation and comprehensive infrastructure to support model development, publication, sharing, and evaluation. Several tools and frameworks have been developed to address these challenges.

To ensure reproducible AI systems, relevant model information such as the ML source code, data sets, and computation environment should be tracked and stored [LR85, 88]. Additionally, tool-based approaches such as version control and metadata handling can also help ensure reproducibility.

One tool that can be used for model publication, sharing, and evaluation is DLHub, which allows for the publication of models with descriptive metadata, persistent identifiers, and flexible access control [LR33]. Another tool that can be used to improve productivity and developer awareness in machine learning projects is MSR4ML, which uses Mining Software Repositories (MSR) techniques to automatically identify and trace the usage of artefacts in Git-based ML projects [LR34].

Model risk assessment is another critical consideration in developing and deploying machine learning models, involving a specialised team and the model owner reviewing documentation to ensure compliance with regulations and standards [LR20,27]. To address this, various AI governance frameworks have been proposed, such as the three components of trustworthy AI from the European Commission's High-Level Expert Group on Artificial Intelligence (AI HLEG) [LR22,73,88].

Monitoring the behaviour of deployed machine learning models is necessary to ensure long-term performance and detect model degradation [LR18,27]. Various methods and tools are available for model monitoring, such as MLOps, which can be used to achieve semi-automated or fully automated model monitoring [LR28]. Fully automated monitoring requires CI/CD integration, CT pipeline, certification of models, governance and security controls, model explainability, auditing, reproducible workflow and models, and mechanisms to perform end-to-end QA tests and performance checks [LR18,28].

Introducing new processes and activities can be more effective than technical frameworks to manage the evolution of log data in a DevOps environment optimised for machine learning. Machine-learning management systems such as Overton can streamline the model creation, deployment, and monitoring processes [LR25]. Additionally, experiment management tools like MLFlow, NeptuneML, and WandB can help manage machine-learning-specific assets like reproducibility and traceability [LR17,68]. However, further research and development are needed to improve engineering processes and fully address the complexities of managing machine learning models [LR17,68].

\subsubsection{Scalable infrastructure
}
Scalable infrastructure is critical for implementing decentralised algorithms that minimise data movement, ensure load balance and fault tolerance, respect application and system requirements, and use simulation tools for testing IoT systems and networks. A bagging-based ensemble learning algorithm for image classification that uses weak learners trained on edge devices and aggregated in higher levels can generate a more accurate global model [LR112].

Various techniques have been proposed to handle the complexity and size of deep learning models. Hardware components such as GPUs, FPGAs, TPUs, and neuromorphic hardware are commonly used to accelerate training and inference. Deep learning frameworks like cuDNN and NCCL incorporate low-level libraries that enable efficient implementation of parallelisation strategies [LR103]. Parallelisation techniques like data, model, pipeline parallelism, and hybrid parallelism can train large deep-learning models on distributed infrastructure. Synchronising parameters and communication among parallel workers is critical in distributed deep learning, and various approaches, such as synchronous, bounded asynchronous, and asynchronous training, can be used for parameter synchronisation.

Several techniques related to scheduling in deep learning have been proposed, including heuristic algorithms, elastic ML frameworks, decentralised data-parallel DL systems, and specialised resource schedulers. Cloud providers offer various options for executing machine learning tasks, such as pre-packaged ML software and machine learning service, reducing the burden of entry into designing innovative applications that facilitate machine learning techniques. Amazon SageMaker is a system for scalable, elastic model training on large data streams, available as part of Amazon's cloud offerings. SageMaker uses a streaming model to meet requirements like linear update time, pause/resume capabilities, and elasticity, exhibiting close-to-linear scalability for both runtime and compute costs of model training [LR124].

Researchers have explored various methods to improve the efficiency of large-scale pre-trained models, including system-level optimisation, efficient learning algorithms, and model compression. Different scheduling approaches have also been proposed, such as single-tenant scheduling, multi-tenant scheduling, and scheduling for model architecture and hyper-parameter search [LR119]. Various tools include IBM Fabric for Deep Learning (FfDL), Djinn, Tonic, Ray, and Ease. ml, have been developed to support these approaches.

The SHMEM-ML library is specifically designed for distributed array computations and machine learning model training and inference, utilising the HPC software stack performance to speed up machine learning workflows. SHMEM-ML uses OpenSHMEM as its communication layer, which allows high-performance networking across hundreds or thousands of distributed processes. Additionally, SHMEM-ML interoperates with the broader Python machine learning software ecosystem, built on top of Apache Arrow, allowing data sharing with other libraries without creating copies [LR110].

Overall, these solutions and techniques enable developing and deploying large deep learning models with higher quality and efficiency, making it easier to scale and apply deep learning to various domains and applications [LR117, 118, 119].

\begin{table*}[htbp]
\centering
\caption{Strategies for achieving Scalability and Maintainability}
\label{tab:strategies}
\begin{tabular}{@{}p{3.5cm}p{14cm}@{}}
\toprule
\textbf{Strategy} & \textbf{Description} \\ \midrule
Modular design and Microservices architecture & 
\begin{itemize}
    \item Emphasise the importance of modular design and microservices in managing ML applications.
    \item Adoption of MLOps and machine learning platforms.
    \item Use integration mechanisms for ML/AI and Continuous Delivery for Machine Learning (CD4ML).
    \item Machine learning platforms for orchestrating ML pipelines and managing data storage, access control, and performance monitoring.
    \item Generic solutions for different ML tasks in hybrid environments.
    \item     Docker for automation in ML-based software project deployment.
    \item Open Neural Network Exchange (ONNX) for interoperability across different ML frameworks.
\end{itemize} \\

Code organisation & 
\begin{itemize}
    \item Flow-Based Programming (FBP) and Model-Driven Software Development (MDSD) for simplifying ML application development.
    \item High-level ML programming tools like WEKA, Azure Machine Learning Studio, KNIME, Orange, BigML, mljar, RapidMiner, Streamanalytix, Lemonade, and Streamsets.
\end{itemize} \\

Automation and Continuous integration & 
\begin{itemize}
    \item DevOps tools improve code sharing, integration speed, and issue resolution.
    \item MLOps environments enable continuous integration and delivery and continuous retraining .
    \item Automated hyperparameter optimisation with tools like Google Vizier, Amazon SageMaker, and SigOpt.

\end{itemize} \\

Cloud-based infrastructure & 
\begin{itemize}
    \item Major cloud providers offer pre-packaged ML software and machine learning services.
    \item Support for standard distributed ML systems and libraries.
    \item Platforms like Flame, TFX, KubeFlow, MLflow, H2O, Skymind Intelligence Layer, Uber's Michelangelo, Facebook's FBLearner Flow, DataRobot, Polyaxon, Comet, Atalaya, Amazon's SageMaker, and Microsoft Azure Machine Learning for end-to-end ML application building and deployment.
    \item Tools for algorithm selection, software architecture, and system quality.
\end{itemize} \\

Collaboration and communication & 
\begin{itemize}
    \item Interdisciplinary teams, formalised collaboration practices, and process models for successful collaboration between data scientists and software engineers.
    \item AI assurance cases, risk management, and practical guidelines for safe and trustworthy AI systems.
    \item Tools for collaboration, communication, and quality assurance in ML projects.
    \item Cloud-based AI platforms and benchmarks for improved collaboration and communication.
\end{itemize} \\

Documentation & 
\begin{itemize}
    \item Prioritising transparency and accountability in AI development through data lifecycle management practices.
\end{itemize} \\

Data and Model Management Tools & 
\begin{itemize}
    \item Tools like Shuffler and JUNEAU for data annotation, labelling, indexing, searching, and reusing tabular data.
    \item ML experiment management tools (MLFlow, Neptune, KubeFlow, ModelDB) and versioning tools (DVC, Pachyderm, Quilt Data) for managing ML pipelines and experiments.
    \item Tools for data and model management to ensure reproducibility, traceability, and efficient ML project management.
\end{itemize} \\

Testing and Debugging tools & 
\begin{itemize}
    \item Techniques and tools for unit test generation, functional regression testing, coverage-guided testing, bug detection, and mutation testing in ML systems.
    \item Frameworks and approaches for testing ML programs and evaluating ML application reliability.
    \item Tools for evaluating ML system quality, handling randomness and test flakiness, and characterising decision boundaries in ML models.
    \item Fault injection techniques and tools for ML application reliability assessment.
\end{itemize} \\

Model governance and monitoring & 
\begin{itemize}
    \item Tools for model publication, sharing, and evaluation (DLHub).
    \item Mining Software Repositories (MSR) techniques for tracing the usage of artefacts in ML projects (MSR4ML).
    \item AI governance frameworks for model risk assessment and compliance (European Commission's High-Level Expert Group on AI).
    \item MLOps, CI/CD integration, CT pipeline, and certification for automated model monitoring and performance checks.
    \item Machine-learning management systems and experiment management tools for efficient ML project management.
\end{itemize} \\

Scalable infrastructure & 
\begin{itemize}
    \item Techniques and hardware components (GPUs, FPGAs, TPUs, neuromorphic hardware) for handling the complexity and size of deep learning models.
    \item Parallelisation techniques (data, model, and pipeline parallelism, hybrid parallelism) for training large deep learning models on distributed infrastructure.
    \item Scheduling approaches and cloud-based execution options for efficient training and deployment of machine learning tasks.
    \item Use Tools and frameworks for improving the efficiency and scalability of large deep learning models like DeepSpeed2.
\end{itemize} \\
\bottomrule
\end{tabular}
\end{table*}

\section{Discussion}

The paper presents a systematic literature review on the maintainability and scalability challenges in Machine Learning (ML) systems. We extensively analysed 124 papers to identify and consolidate the challenges and potential solutions at different stages of the ML workflow. The study provides valuable insights into the interdependencies between these stages and offers recommendations for addressing the identified challenges.

One of the critical contributions of this study is the comprehensive catalogue of maintainability and scalability challenges in Data Engineering and Model Engineering workflows. By examining the complexities involved in building ML systems and applications within the current ecosystem of frameworks and tools, the authors highlight the specific challenges that arise during data preprocessing, feature engineering, model training, and deployment. This catalogue serves as a valuable resource for practitioners and organisations, enabling them to better understand the potential hurdles they may encounter in different stages of the ML workflow.

The paper also identifies and consolidates 41 maintainability challenges and 13 scalability challenges, along with potential solutions, tools, and recommendations. These challenges include data quality and governance, model versioning and reproducibility, model interpretability, technical debt management, resource allocation, and scalability bottlenecks. By providing concrete solutions and recommendations, the study equips ML tool developers and researchers with valuable insights to address these challenges effectively.

Furthermore, we have synthesised the interdependencies between different maintainability challenges that impact the overall workflow of ML systems. This mapping helps understand how addressing one challenge can ripple effects on other workflow stages. It highlights the need for a holistic approach to maintainability rather than treating it as an isolated concern within each stage. This perspective is crucial for developing sustainable ML systems that can adapt to evolving requirements and datasets.

The study also explores the tradeoffs between maintainability and scalability in ML systems. As datasets and models grow in size and complexity, maintaining maintainability and scalability becomes challenging. The authors discuss the delicate balance between these two aspects and the potential tradeoffs that must be considered. This analysis provides practitioners and researchers with valuable insights into finding an optimal solution that balances both maintainability and scalability, depending on the specific requirements of their ML systems.

Overall, the paper contributes significantly to understanding ML systems' maintainability and scalability challenges. By consolidating the challenges, providing potential solutions, and offering insights into interdependencies and tradeoffs, the study provides a foundation for future research in this area. The identified challenges and recommendations can guide practitioners and organisations in developing maintainable and scalable ML systems and applications, thereby avoiding potential pitfalls and improving their ML solutions' overall quality and longevity.

\subsection{Implications for Practitioners and Researchers}

\subsubsection{Implications for ML Tool Developers and Practitioners}

For ML tool developers and practitioners, the implications of this study are twofold - focusing on maintainability and scalability in ML systems.

Maintainability is crucial for practitioners to ensure long-term value and prevent performance decline in ML models. The identified maintainability challenges and solutions provide valuable insights for ML tool developers to design tools and frameworks that facilitate version control, monitoring, documentation, automated testing, continuous integration, and model governance. By incorporating these features into ML tools, practitioners can easily manage and maintain ML models, improve collaboration, reduce downtime, and make informed decisions based on performance trends.

Scalability is another critical consideration for ML tool developers and practitioners. ML systems must efficiently handle large datasets, computational requirements, and peak loads. The implications of scalability highlighted in the study can guide tool developers in designing frameworks and architectures that support parallel processing, distributed storage, cloud services, model compression, caching, load balancing, and automated scaling. Practitioners can leverage these scalable ML tools to handle large-scale data processing, reduce costs, improve performance, and ensure high availability during peak traffic.

By addressing maintainability and scalability challenges, ML tool developers and practitioners can build robust and efficient ML systems that deliver long-term value, maintain high performance, and adapt to changing requirements.

\subsubsection{Implications for Researchers}

For researchers in the field of ML, this study provides valuable implications and actionable insights to enhance further the understanding of maintainability and scalability challenges in different stages of the ML workflow.

The catalogue of challenges and solutions presented in the study serves as a foundation for future research. Researchers can delve deeper into specific challenges and investigate advanced techniques, algorithms, and methodologies to overcome these challenges. For example, researchers can explore novel approaches for model versioning, monitoring, data quality assurance, technical debt management, or resource allocation to improve the maintainability of ML systems. Similarly, they can focus on techniques for distributed computing, model compression, load balancing, or automated scaling to enhance the scalability of ML systems.

The tradeoffs and balance between maintainability and scalability identified in the study also provide a rich area for further research. Researchers can delve into the nuances of these tradeoffs and develop frameworks or decision-making models that can help practitioners make informed choices based on their specific requirements and constraints. This research can contribute to developing guidelines and best practices for achieving an optimal balance between maintainability and scalability in ML systems.

Additionally, researchers can contribute by conducting empirical studies and experiments to validate the proposed solutions and recommendations. By evaluating the effectiveness of different approaches in real-world scenarios, researchers can provide empirical evidence to support the adoption of specific techniques or methodologies for addressing maintainability and scalability challenges.

Overall, this study provides researchers with a roadmap for further investigations in the domain of ML system maintainability and scalability, thereby advancing the knowledge and understanding in this crucial area.

\subsection{Guidelines and Actionable Insights}

The paper provides practical guidelines and actionable insights for practitioners and researchers to achieve maintainability and scalability in ML systems. These actional insights can guide ML models' development, deployment, and maintenance. 

To achieve maintainability, practitioners can focus on the following:

1. Model versioning: Implement tools like Git or DVC to track different versions of the model and associated data and configurations.

2. Monitoring: Utilise tools like TensorFlow Model Analysis or MLflow to monitor model performance, detect data drift, and identify potential issues.

3. Documentation: Maintain detailed documentation of the model, including data and configurations used, performance metrics, and updates/modifications made.

4. Automated testing: Implement automated testing to ensure changes to the model do not break existing functionality.

5. Continuous integration: Adopt continuous integration practices to automatically build and test the model with each change, catching and addressing problems early in development.

6. Model governance: Establish model governance policies and procedures, including a model change approval process and a model retirement policy, to ensure controlled and transparent modifications and retirements.

7. Quality assessment: Use established quality assessment tools to evaluate the reliability and accuracy of the studies or models included in the ML system.

8. Team collaboration: Encourage collaboration and communication among team members, fostering knowledge sharing and feedback for continuous improvement.

On the other hand, to achieve scalability, practitioners can focus on the following:

1. Parallel processing: Utilise techniques such as using multiple GPUs or distributed computing clusters to distribute the training process across multiple machines, reducing training time.

2. Distributed storage: Employ distributed storage systems like HDFS or Amazon S3 to handle large amounts of data without running out of storage space on a single machine.

3. Cloud services: Leverage cloud service providers like Amazon Web Services, Microsoft Azure, or Google Cloud to access scalable computational resources at a reasonable cost.

4. Model compression and quantisation: Apply techniques like model compression and quantisation to reduce computational requirements, making running models on smaller, less expensive machines feasible.

5. Horizontal and vertical scaling: Implement horizontal scaling by adding more machines to handle increased traffic or vertical scaling by adding more resources to a single machine to handle increased computational demands.

6. Caching and load balancing: Utilise caching and load balancing techniques to distribute traffic across multiple machines and improve system scalability.

7. Monitoring: Deploy monitoring tools to track the performance of the ML system, identify bottlenecks, and make informed decisions on scaling.

8. Automated scaling: Implement automated scaling mechanisms to adjust the resources the system uses dynamically in response to changes in workload or traffic.

By following these guidelines, practitioners and researchers can develop maintainable and scalable ML systems that optimise model performance, facilitate collaboration, reduce downtime, make better decisions, handle large datasets, and meet computational requirements efficiently.

\section{ Conclusion }

This study underscores the pivotal role of maintainability and scalability in designing and implementing ML systems. Our systematic review offers a holistic catalogue of the challenges, intricacies, and potential solutions in these areas, derived from both Data Engineering and Model Engineering workflows. With the identification of 54 distinct challenges and their interdependencies, along with synthesised strategies and tradeoffs, this research is a valuable resource for practitioners and researchers. Leveraging these insights can empower stakeholders to navigate the complexities of building ML systems, ensuring their sustainability and efficiency. We hope that this foundation aids in steering future research, fostering a deeper understanding, and promoting the development of ML applications that are both scalable and maintainable.

\section{ Reference }

\begin{enumerate}

\item S. Amershi et al., "Software Engineering for Machine Learning: A Case Study," 2019 IEEE/ACM 41st International Conference on Software Engineering: Software Engineering in Practice (ICSE-SEIP), 2019, pp. 291-300, doi: 10.1109/ICSE-SEIP.2019.00042.

\item Sculley, D., et al.: Hidden technical debt in machine learning systems. In: Cortes, C., Lawrence, N.D., Lee, D.D., Sugiyama, M., Garnett, R. (eds.) Advances in Neural Information Processing Systems 28, pp. 2503–2511. Curran Associates, Inc. (2015)

\item Lwakatare L.E., Raj A., Bosch J., Olsson H.H., Crnkovic I. (2019) A Taxonomy of Software Engineering Challenges for Machine Learning Systems: An Empirical Investigation. In: Kruchten P., Fraser S., Coallier F. (eds) Agile Processes in Software Engineering and Extreme Programming. XP 2019. Lecture Notes in Business Information Processing, vol 355. Springer, Cham. https://doi.org/10.1007/978-3-030-19034-714

\item Anders Arpteg, Björn Brinne, Luka Crnkovic-Friis, and Jan Bosch. 2018. Software engineering challenges of deep learning. In Euromicro Conference on Software Engineering and Advanced Applications (SEAA). IEEE, 50–59.

\item Foutse Khomh, Bram Adams, Jinghui Cheng, Marios Fokaefs, and Giuliano Antoniol. 2018. Software Engineering for Machine-Learning Applications: The Road Ahead. IEEE Software 35, 5 (2018), 81–84

\item Zhiyuan Wan, Xin Xia, David Lo, and Gail C Murphy. 2019. How Does Machine Learning Change Software Development Practices? IEEE Transactions on Software Engineering (2019).
\item Alex Serban, Koen van der Blom, Holger Hoos, and Joost Visser. 2020. Adoption and Effects of Software Engineering Best Practices in Machine Learning. In Proceedings of the 14th ACM / IEEE International Symposium on Empirical Software Engineering and Measurement (ESEM) (ESEM '20). Association for Computing Machinery, New York, NY, USA, Article 3, 1–12. DOI:https://doi.org/10.1145/3382494.3410681

\item A. Munappy, J. Bosch, H. H. Olsson, A. Arpteg and B. Brinne, "Data Management Challenges for Deep Learning," 2019 45th Euromicro Conference on Software Engineering and Advanced Applications (SEAA), 2019, pp. 140-147, doi: 10.1109/SEAA.2019.00030.
\item "IEEE Standard for Software Maintenance" in IEEE Std 1219-1998 , vol., no., pp.1-56, 21 Oct. 1998, doi: 10.1109/IEEESTD.1998.88278.
\item Malhotra, Ruchika and Anuradha Chug. “Software Maintainability Prediction using Machine Learning Algorithms.” (2012)

\item R. Malhotra and A. Chug, ‘‘Software maintainability: Systematic literature review and current trends,’’ Int. J. Softw. Eng. Knowl. Eng., vol. 26, no. 8, pp. 1221–1253, 2016.
\item  T. Mikkonen, J. K. Nurminen, M. Raatikainen, I. Fronza, N. Mäkitalo, and T. Männistö, ‘Is Machine Learning Software Just Software: A Maintainability View’, in Software Quality: Future Perspectives on Software Engineering Quality, vol. 404, D. Winkler, S. Biffl, D. Mendez, M. Wimmer, and J. Bergsmann, Eds. Cham: Springer International Publishing, 2021, pp. 94–105. doi: 10.1007/978-3-030-65854-08.
\item  ‘Designing Machine Learning Systems [Book]’.https://www.oreilly.com/library/view/designing-machinelearning/9781098107956/ (accessed Mar. 17, 2023).
\item  ‘The Importance of Scalability In Software Design’. https://www.conceptatech.com/blog/importance-of-scalability-in-software-design (accessed Mar. 17, 2023).
\item  ‘ml-ops.org’. https://ml-ops.org/ (accessed Mar. 17, 2023).
\item B. Kitchenham, “Procedures for performing systematic reviews,” Keele, UK, Keele University, vol. 33, no. TR/SE-0401, 2004.
\item R. F. Paige, J. Cabot, and N. A. Ernst, “Foreword to the special
the section on negative results in software engineering,” Empirical
Software Engineering, vol. 22, no. 5, pp. 2453–2456, 2017
\item M. Kuhrmann, D. M. Fernandez, and M. Daneva, “On the ´
Pragmatic Design of Literature Studies in Software Engineering:
An Experience-based Guideline,” Empirical Software Engineering,
vol. 22, pp. 2852–2891, 2017.
\item  T. Saracevic, “Evaluation of Evaluation in Information Retrieval,”
in Proceedings of 18th Annual International Conference on RDIR, 1995,
pp. 138–146.
\item H. Edison, X. Wang and K. Conboy, "Comparing Methods for Large-Scale Agile Software Development: A Systematic Literature Review," in IEEE Transactions on Software Engineering, vol. 48, no. 8, pp. 2709-2731, 1 Aug. 2022, doi: 10.1109/TSE.2021.3069039.
\item S. Wang et al., "Machine/Deep Learning for Software Engineering: A Systematic Literature Review," in IEEE Transactions on Software Engineering, vol. 49, no. 3, pp. 1188-1231, 1 March 2023, doi: 10.1109/TSE.2022.3173346.
\item T. Dyba and T. Dingsøyr, “Empirical studies of agile software development: A systematic review,” Inf. Softw. Technol., vol. 50, no. 9–10, pp. 833–859, 2008.
\item Landis, J.R. and Koch, G.G., 1977. The measurement of observer agreement for categorical data. biometrics, pp.159-174.
\item Krippendorff, Klaus. Content analysis: An introduction to its methodology. Sage Publications, 2018.
\item Hallgren, K.A., 2012. Computing inter-rater reliability for observational data: an overview and tutorial. Tutorials in quantitative methods for psychology, 8(1), p.23.
\item Malinda Dilhara, Ameya Ketkar and Danny Dig, "Understanding Software-2.0: A Study of Machine Learning Library Usage and Evolution", ACM Trans. Softw. Eng. Methodol, vol. 30, 4, no. 55, pp. 42, October 2021.
\item C. Wohlin, P. Runeson, P.A. da Mota Silveira Neto, E. Engström, I. do Carmo Machado, E.S. de Almeida; On the reliability of mapping studies in software engineering
J. Syst. Software, 86 (2013), pp. 2594-2610
\item Webster, J. and Watson, R.T., 2002. Analysing the past to prepare for the future: Writing a literature review. MIS quarterly, pp.xiii-xxiii.
\item Wohlin, C., 2014, May. Guidelines for snowballing in systematic literature studies and a replication in software engineering. In Proceedings of the 18th international conference on evaluation and assessment in software engineering (pp. 1-10).
\item Silver, Christina , Lewins, Ann. (2014). Using Software in Qualitative Research: A Step-by-Step Guide. 10.4135/9781473906907. 
\item E. d. S. Nascimento, I. Ahmed, E. Oliveira, M. P. Palheta, I. Steinmacher and T. Conte, "Understanding Development Process of Machine Learning Systems: Challenges and Solutions," 2019 ACM/IEEE International Symposium on Empirical Software Engineering and Measurement (ESEM), Porto de Galinhas, Brazil, 2019, pp. 1-6, doi: 10.1109/ESEM.2019.8870157.
\item Nadia Nahar, Shurui Zhou, Grace Lewis, and Christian Kästner. 2022. Collaboration challenges in building ML-enabled systems: communication, documentation, engineering, and process. In Proceedings of the 44th International Conference on Software Engineering (ICSE '22). Association for Computing Machinery, New York, NY, USA, 413–425. https://doi.org/10.1145/3510003.3510209
\item Alex Serban, Koen van der Blom, Holger Hoos, and Joost Visser. 2020. Adoption and Effects of Software Engineering Best Practices in Machine Learning. In Proceedings of the 14th ACM / IEEE International Symposium on Empirical Software Engineering and Measurement (ESEM) (ESEM '20). Association for Computing Machinery, New York, NY, USA, Article 3, 1–12. https://doi.org/10.1145/3382494.3410681
\item K. Shivashankar and A. Martini, "Maintainability Challenges in ML: A Systematic Literature Review," 2022 48th Euromicro Conference on Software Engineering and Advanced Applications (SEAA), Gran Canaria, Spain, 2022, pp. 60-67, doi: 10.1109/SEAA56994.2022.00018.
\item Ask Berstad Kolltveit and Jingyue Li. 2023. Operationalising machine learning models: a systematic literature review. In Proceedings of the 1st Workshop on Software Engineering for Responsible AI (SE4RAI '22). Association for Computing Machinery, New York, NY, USA, 1–8. https://doi.org/10.1145/3526073.3527584
\item Chao Liu, Cuiyun Gao, Xin Xia, David Lo, John Grundy, and Xiaohu Yang. 2021. On the Reproducibility and Replicability of Deep Learning in Software Engineering. ACM Trans. Softw. Eng. Methodol. 31, 1, Article 15 (January 2022), 46 pages. https://doi.org/10.1145/3477535
\item Giray, G., 2021. A software engineering perspective on engineering machine learning systems: State of the art and challenges. Journal of Systems and Software, 180, p.111031.

\end{enumerate}

\renewcommand\refname{Literature Review Papers }
\makeatletter
\renewcommand\@bibitem[1]{\item\if@filesw \immediate\write\@auxout
    {\string\bibcite{#1}{L\the\value{\@listctr}}}\fi\ignorespaces}
\def\@biblabel#1{[LR#1]}
\makeatother

\end{document}